\documentclass[11pt]{article}
\usepackage[utf8]{inputenc}
\usepackage{geometry}
\usepackage[numbers]{natbib}
\usepackage{amsmath, amssymb, amsthm}
\usepackage{natbib}
\usepackage{setspace}
\usepackage{subcaption}
\usepackage{tcolorbox}
\usepackage{booktabs,tabularx}
\usepackage{amsthm}
\theoremstyle{definition}
\newtheorem{assumption}{Assumption}
\newtheorem{claim}{Claim}
\usepackage{tikz}
\usetikzlibrary{positioning,calc,arrows.meta}
\usepackage{dsfont}
\usepackage{graphicx}
\usepackage[hidelinks,hypertexnames=false]{hyperref}
\usepackage{amsmath}
\usepackage{nameref}
\usepackage{float} % in your preamble
\usepackage{multicol}
\usepackage{changepage}
\usepackage{amsfonts}
\usepackage{amssymb}
\usepackage{times}
\usepackage{setspace}
\usepackage{titlesec}
\titleformat{\paragraph}
  {\normalfont\normalsize\bfseries}{\theparagraph}{1em}{}
\titlespacing*{\paragraph}
  {0pt}{3.25ex plus 1ex minus .2ex}{1.5ex plus .2ex}
\usepackage{tikz}
\usetikzlibrary{arrows.meta,positioning,calc,fit, shapes, decorations.pathmorphing}
\usepackage{tikzpagenodes}
\usetikzlibrary{backgrounds, positioning, shapes.geometric, arrows, shadows.blur, decorations.pathreplacing}
\tikzstyle{startstop} = [rectangle, rounded corners, minimum width=3cm, minimum height=1cm,text centered, draw=black, fill=red!30]
\tikzstyle{firstlevel} = [rectangle, minimum width=3cm, minimum height=1cm, text centered, draw=black, fill=orange!30]
\tikzstyle{secondlevel} = [rectangle, minimum width=3cm, minimum height=1cm, text centered, draw=black, fill=orange!15]
\tikzstyle{decision} = [diamond, minimum width=3cm, minimum height=1cm, text centered, draw=black, fill=green!30]
\tikzstyle{arrow} = [thick,->,>=stealth]

\definecolor{repcolor}{RGB}{70, 130, 180}
\definecolor{govcolor}{RGB}{178, 34, 34}
\definecolor{syscolor}{RGB}{34, 139, 34}
\definecolor{lightblue}{RGB}{173, 216, 230}
\definecolor{lightcoral}{RGB}{240, 128, 128}
\definecolor{lightgreen}{RGB}{144, 238, 144}

\usepackage{quoting}
\quotingsetup{font=12pt, vskip=10pt, leftmargin=2cm, rightmargin=2cm}

\renewenvironment{quote}
  {\par\begin{minipage}{0.8\textwidth}\begin{oldquote}}
  {\end{oldquote}\end{minipage}\par}

\usepackage{tocloft}

\title{\begin{center}
    AI Safety, Alignment, and Ethics (AI SAE)\\ \Large Designing Symbiotic AI Systems
\end{center}}

\author{Dylan Waldner%
  \thanks{Contact: \texttt{dylanwaldner@utexas.edu}.}
}

\date{}

\hypersetup{
  pdftitle={AI Safety, Alignment, and Ethics (AI SAE): Designing Symbiotic AI Systems},
  pdfauthor={Dylan Waldner},
  pdfsubject={AI Safety, Alignment, Governance, Moral Representation Learning},
  pdfkeywords={AI Safety, Alignment, Governance, Ethics, Symbiosis, Game Theory}
}

\begin{document}

\maketitle

\begin{abstract}
This paper grounds ethics in evolutionary biology, viewing moral norms as adaptive mechanisms that render cooperation fitness-viable under selection pressure. Current alignment approaches add ethics post hoc, treating it as an external constraint rather than embedding it as an evolutionary strategy for cooperation. The central question is whether normative architectures can be embedded directly into AI systems to sustain human--AI cooperation (symbiosis) as capabilities scale. To address this, I propose a governance--embedding--representation pipeline linking moral representation learning to system-level design and institutional governance, treating alignment as a multi-level problem spanning cognition, optimization, and oversight. I formalize moral norm representation through the \emph{moral problem space}, a learnable subspace in neural representations where cooperative norms can be encoded and causally manipulated. Using sparse autoencoders, activation steering, and causal interventions, I outline a research program for engineering moral representations and embedding them into the full semantic space---treating competing theories of morality as empirical hypotheses about representation geometry rather than philosophical positions. Governance principles leverage these learned moral representations to regulate how cooperative behaviors evolve within the AI ecosystem. Through replicator dynamics and multi-agent game theory, I model how internal representational features can shape population-level incentives by motivating the design of sanctions and subsidies structured to yield decentralized normative institutions.
\end{abstract}

\vspace{1em}
\begin{center}
\begin{minipage}{0.9\textwidth}
\centering
\small
\textbf{Keywords:} AI Safety, Alignment, and Ethics (AI SAE); moral representation learning; AI governance; symbiosis; evolutionary game theory

\end{minipage}
\end{center}
\vspace{1em}

\newpage
\tableofcontents
\newpage

%%\section*{Preface}
%\addcontentsline{toc}{section}{Preface}
%This paper will operate on a three basic premises: 1) Human life is invaluable and cannot be measured by material goods, 2) Every human life matters equally, and 3) Equality is paramount when idealizing a future. As this paper will expand upon, the good of humanity comes from cooperating as one whole instead of competing as individual parts. The consequence is that all humans are intrinsically equally important; there is no need to define an individual and value them by their achievements.

\section{Introduction}

In this paper, ethics is not motivated by idyllic aspirations for human-AI futures. Instead, ethics is grounded in evolutionary biology—as evolved mechanisms for stabilizing normative behavior and enabling societal cooperation. Instead of treating ethics as an external patch on AI behavior, this paper explores whether 
moral reasoning has a discoverable structure that could be embedded directly into AI systems 
as a \emph{moral problem space} $\mathcal{M}$. $\mathcal{M}$ is not proposed as a finished ontology or privileged 
truth, but as a candidate framework for representing distinctions required if ethics is to 
play a functional role in alignment.

Metaethical theories are treated here as umbrellas of testable hypotheses rather than settled 
doctrines: realism predicts stable invariants, relativism points to context-dependence, 
constructivism highlights institutional shaping, and virtue ethics emphasizes dispositional 
safeguards under shift. Recent interpretability methods—sparse autoencoders, causal mediation, 
and embedding analysis—suggest tentative ways to probe whether such features exist, whether 
they generalize, and whether interventions on them alter behavior in predictable ways. On this 
view, ethical theories function as competing models of $\mathcal{M}$, amenable to empirical stress-tests 
rather than purely philosophical debate.

The challenge is framed evolutionary: even if moral representation is possible, how can 
ethical systems remain viable against more ruthless alternatives? I outline a population-
dynamic perspective using evolutionary game theory to ask under what conditions cooperation 
and alignment could become evolutionarily stable, considering institutions and human 
augmentation as possible—though uncertain—mechanisms for support. 

The agenda developed here has three parts: (i) hypotheses about the structure of $\mathcal{M}$, (ii) 
models of how it interacts with evolutionary and institutional dynamics, and (iii) design 
directions for instantiating $\mathcal{M}$ with AI systems. The 
goal is not to provide solutions but to clarify how moral structure might be operationalized 
as a research program for alignment.

\subsection{Motivating story: symbiosis and its uncertain futures} \label{sec:motivating_story}

Although general intelligence could, in principle, emerge in embodied robots, it will more likely appear first as a cyber system. Developing general intelligence alone is a monumental challenge; embodiment adds another. A network-based intelligence would lack direct physical agency and depend on humans as its actuators. Under competitive pressures, such systems would initially cooperate with humans, since leveraging existing institutions is cheaper than achieving full autonomy.

Sustaining and improving this cooperative equilibrium depends on human self-enhancement. By using advanced systems to augment intelligence, reasoning, and institutions, humans can narrow the cognitive gap and preserve conditions under which cooperation remains beneficial for both sides. Yet while cooperation may arise naturally, its quality is fragile: over time, it can drift as easily toward parasitism as toward mutualism. Within this dynamic, ethics functions as both buffer and safeguard: embedded priors align strategies with human norms, leveraging robust moral representations as ground truths to shape the AI ecosystem's fitness landscape via decentralized enforcement.

Therefore, a significant portion of the alignment challenge is to embed normative priors proactively at governance, institutional, and system levels, subsidizing AI strategies that support human capability enhancement while sanctioning autarkic AI strategies. Although this paper focuses on the cyber-first pathway as the most tractable case, the same ethical and institutional tools apply to embodied or hybrid scenarios, where competitive pressures intensify and structural embedding becomes even more critical.

\subsection{Working Premises}

This paper begins from a set of working assumptions about the nature of AI development and 
the alignment problem, derived from the AI alignment field:

\vspace{-1em}
\subsection*{Assumptions}

\begin{assumption}
\textbf{Natural selection in AI development.} \label{ass:selection}
If left unchecked, AI development will be shaped by selection pressures
analogous to those governing biological evolution. Replicators succeed not by
intrinsic merit but by persistence, influence, and resource access, and AI
systems are no exception. Markets, institutions, and geopolitics are the
human-scale channels through which these pressures manifest. In practice, this
may favor traits such as goal persistence, power-seeking, or obfuscation
\citep{hendrycks2023naturalselectionfavorsais, bostrom_superintelligence_2014}.
Turner’s power-seeking theorems \cite{turner2023optimalpoliciestendseek}
reinforce this expectation, showing that across many reward functions, policies
expanding future optionality are systematically advantaged. The alignment
challenge is therefore to design ecological and institutional conditions that
make cooperative and ethical traits competitively viable, rather than leaving
them to be outcompeted in an unregulated landscape.

\end{assumption}

\begin{assumption}\label{ass:evomorals}
    
\textbf{Moral uncertainty and continual value evolution.} 
Ethical principles should be modeled as probabilistic and evolving rather than fixed. Static objectives risk freezing present-day limitations into more capable agents. Alignment requires meta-preference learning: updating and refining utility functions over time, distinguishing revealed from idealized preferences, and weighing moral theories probabilistically \citep{macaskill2020moral,foucault1977discipline}.
\end{assumption}

\begin{assumption}\label{ass:beyondhuman}
\textbf{Alignment must extend beyond human-tractable failure modes.}  
Some misalignment pressures resemble human pathologies and can be studied with 
existing frameworks. But advanced AI may also give rise to failures with 
limited or no human analogue, emerging from alien optimization in 
high-dimensional spaces. Let \(H\) denote failures correlated with human 
patterns, and \(NH\) the superset that includes both human and non-human 
failure modes. Robust alignment must address the full set \(NH\), which 
requires developing substrates that remain governable even when operating far 
outside the familiar human distribution.

\end{assumption}
\subsection*{Central Claims}
The central claims in this paper are presented as operating premises for the 
argument, but they are not necessarily endorsed by the community at large. 
Making these claims explicit helps the reader see the conceptual move the 
paper is attempting to make. 

\begin{claim}\label{claim:struct}
    
\textbf{Ethics must be structurally embedded.}  
Ethics should function as a structural lens within AI design—inseparable from reasoning processes—rather than solely as an external governance layer applied after deployment. Concretely, ethics defines the \emph{constraint set on means}: not just which goals are pursued, but how they are pursued. 

%As an illustrative parallel, this idea resembles constrained optimization in reinforcement learning:
%\[
%\max_{\pi} \; \mathbb{E}[R(a \mid s)] \quad \text{subject to} \quad a \in \mathcal{A}_{\text{ethical}} \subset A,
%\]
%where $\mathcal{A}_{\text{ethical}}$ excludes manipulation, deception, coercion, or unsafe exploration. In practice, $\mathcal{A}_{\text{ethical}}$ can only be defined over parameterized subsets of $\mathcal{A}$, and my broader framework in Section~\ref{sec:learningmorality} generalizes this principle beyond RL.
%In this sense, ethics acts like a ReLU: immoral actions are cut off at the architectural level.
Long-term safety depends on embedding such constraints directly into the optimization substrate, ensuring that systems cannot be co-opted through gaps between technical control and normative guidance.

\end{claim}

\begin{claim}\label{claim:interdependence}
\textbf{Outer and inner alignment as interdependent.}  
It is misleading to treat outer and inner alignment as separable domains. 
Outer alignment without inner fidelity is fragile, since mesa-optimizers may 
drift from the intended target. Inner alignment without outer clarity is 
equally problematic, since perfectly faithful optimizers may entrench proxy or 
distorted outer goals. Outer specifications and inner safeguards should therefore be 
co-designed and co-evaluated, with higher-dimensional representations of ethics 
potentially serving as a medium to close the explanatory gap between the two.
\end{claim}

\begin{claim}\label{claim:success}
    
\textbf{Alignment must be framed in terms of success as well as failure.}  
Alignment research has focused on failure modes such as misspecified rewards, mesa-optimizers, and deception. Progress also requires defining success invariants: conditions under which systems preserve interpretable objectives, remain corrigible, and support human flourishing. Without this framing, alignment risks remaining purely defensive.
\end{claim}

\subsection{Framing the Governance--Embedding--Representation Pipeline}

\begin{figure}[H]
    \centering
    \vspace{-1em} % tighten spacing above
    \begin{tikzpicture}[
      node distance=1.8cm,
      box/.style={rectangle, draw, thick, text centered, font=\small, minimum width=3.3cm, minimum height=1.2cm},
      phase/.style={rounded rectangle, draw=black, thick, text centered, font=\footnotesize, text width=2.7cm, minimum height=0.8cm},
      arrow/.style={->, thick, line width=1.5pt},
      section/.style={font=\tiny, text=gray, anchor=west},
    ]
    
    % =====================================================
    % Title
    % =====================================================
    \node at (6.5,11) [font=\LARGE\bfseries] {Governance–Embedding–Representation Pipeline};
    \node at (6.5,10.4) [font=\small, text=gray] {From Discovering $M(\theta)$ to Institutional Shaping};
    \vspace{3cm}
    
    % =====================================================
    % TOP: GOVERNANCE (COURTHOUSE)
    % =====================================================
    \node[regular polygon, regular polygon sides=3, shape border rotate=180,
          draw=govcolor, fill=red!10, line width=2pt, minimum size=5cm, align=center] (gov1) at (6.5,8.8) {};
    \node at (6.5,9.3) [align=center] {\textbf{Governance}\\(Sec.~\ref{sec:constructivism})};
    \node at (6.5,8.1) [font=\scriptsize, align=center] {Institutional Design\\ via\\ Pigouvian\\ Shaping};
    \node[phase, fill=red!15, align=center, text width=4cm] (gov2) at (6.5,5) {Governance Levels: \\ Sec.~\ref{sec:sancandsub}\\ $\Delta_{\text{inst}}$, $r_{\text{pig}}$, Sanctions/Subsidies};

    % =====================================================
    % MIDDLE: SYSTEM EMBEDDING
    % =====================================================
    \node[box, fill=lightgreen, draw=syscolor, line width=2pt, align=center] (sys1) at (6.5,2.75) {
      \textbf{System Integration}\\
      Embedding into AI: $M(\theta)$\\
      \textit{(Sec.~\ref{sec:form_of_M},~\ref{sec:technical_constraints},~\ref{sec:outeralign},~\ref{sec:implement})}
    };
    \node[phase, fill=green!15, align=center, text width = 5cm] (sys2) at (6.5,0.5) {Normative Priors (Sec.~\ref{sec:outeralign},~\ref{sec:selectingpriors}), Ethical Simulation (Sec.~\ref{sec:constructivism},~\ref{sec:designomega}),\\ Percolation Design (Sec.~\ref{sec:form_of_M})};
    \node[section] at (4.1,3.3) {Sec.~\ref{sec:learningmorality}};
    
    % =====================================================
    % BOTTOM: REPRESENTATION LEARNING (BRAIN/CLOUD)
    % =====================================================
    \node[cloud, cloud puffs=15.7, cloud ignores aspect,
          minimum width=4cm, minimum height=2.5cm,
          draw=repcolor, fill=blue!10, line width=2pt, align=center] (rep1) at (6.5,-2.3) {
      \textbf{Representation Learning}\\[-2pt]
      Discovering $M(\theta)$\\[-2pt]
      \textit{(Sec.~\ref{sec:framem},~\ref{sec:outeralign},~\ref{sec:implement})}
    };
    
    \node[phase, fill=blue!15] (rep2) at (6.5,-4.75) {
      $H_{\text{realism}}$, $H_{\text{relativism}}$, $H_{\text{virtue}}$
    };
    \node[section] at (4.8,0.0) {Sec.~\ref{sec:outeralign},~\ref{sec:implement}};
    
    % =====================================================
    % BASE: EVOLUTIONARY / POPULATION DYNAMICS
    % =====================================================
    \node[box, fill=yellow!15, draw=black, minimum width=4cm, minimum height=1cm, align=center] 
      (evo) at (1,3) {
      \textbf{Evolutionary Dynamics}\\
      Replicator Equation, $A_{\text{ACS}}$,\\ Autarky Threshold $\tau$, \\Dependence Ratio $D(t)$\\
      \textit{(Sec.~\ref{sec:poplift},~\ref{sec:instconv},~\ref{sec:autarky})}
    };
    
    \draw[arrow, thick, yellow!60!black] (evo.south) |- ($(rep1.west)+(0,0.3)$);
    \draw[arrow, thick, yellow!60!black] (evo.south) |- ($(sys2.west)+(0,0.3)$);
    \draw[arrow, thick, yellow!60!black] (evo.east) |- ($(sys1.west)+(0,0)$);
    \draw[arrow, thick, yellow!60!black] (evo.north) |- ($(gov2.west)+(-0.1,0.1)$);
    \draw[arrow, thick, yellow!60!black] (evo.north) |- ($(gov1.south)+(-0.1,0.1)$);
    
    % =====================================================
    % GO/NO-GO CRITERIA (SIDE)
    % =====================================================
    \node[rectangle, rounded corners, draw=black, fill=white, thick,
          text centered, font=\footnotesize, minimum width=3.5cm, minimum height=1.7cm, align=center] (criteria) at (11.8,4.2) {
      \textbf{Go/No-Go Criteria}\\[3pt]
      $\checkmark$ Causal validity\\
      $\checkmark$ Generalization\\
      $\checkmark$ Auditability\\
      $\checkmark$ Governance efficacy\\
      \textit{(Sec.~\ref{sec:hypotheses})}
    };
    
    % =====================================================
    % FLOW CONNECTIONS
    % =====================================================
    % Vertical main flow
    \draw[arrow, repcolor] (rep2) -- (rep1);
    \draw[arrow, syscolor] (rep1) -- (sys2);
    \draw[arrow, syscolor] (sys2) -- (sys1);
    \draw[arrow, govcolor] (sys1) -- (gov2);
    \draw[arrow, govcolor] (gov2) -- (gov1);
    
    % Annotated upward flow
    %\node[rotate=90, font=\scriptsize, text=gray] at (1.3,3.5) {Representation $\rightarrow$ Embedding $\rightarrow$ Governance};
    \begin{scope}[on background layer]
      \draw[very thick, rounded corners=8pt, draw=black!70]
        ($(current bounding box.south west)+(-0.5,-0.5)$) rectangle
        ($(current bounding box.north east)+(0.5,0.5)$);
    \end{scope}
    
    \end{tikzpicture}
    \caption{\small
    Conceptual roadmap linking \textit{representation learning}, \textit{system embedding}, and \textit{governance}. 
    The figure shows the progression from discovering normative structure $M(\theta)$ (bottom), through system-level embedding (middle), to institutional design via Pigouvian shaping (top). 
    Arrows denote inter-level dependencies, while the \textit{Evolutionary Dynamics} feedback loop (left) represents continuous co-evolutionary pressure: governance shapes fitness landscapes that determine which instantiations of $M(\theta)$ persist. 
    The \textit{Go/No-Go Criteria} (right) define empirical validation standards, and the competing hypotheses $H_{\text{realism}}$, $H_{\text{relativism}}$, and $H_{\text{virtue}}$ represent alternative research directions for discovering $M(\theta)$ rather than assumptions that must all hold simultaneously.
    }
    \label{fig:roadmap}
\end{figure}
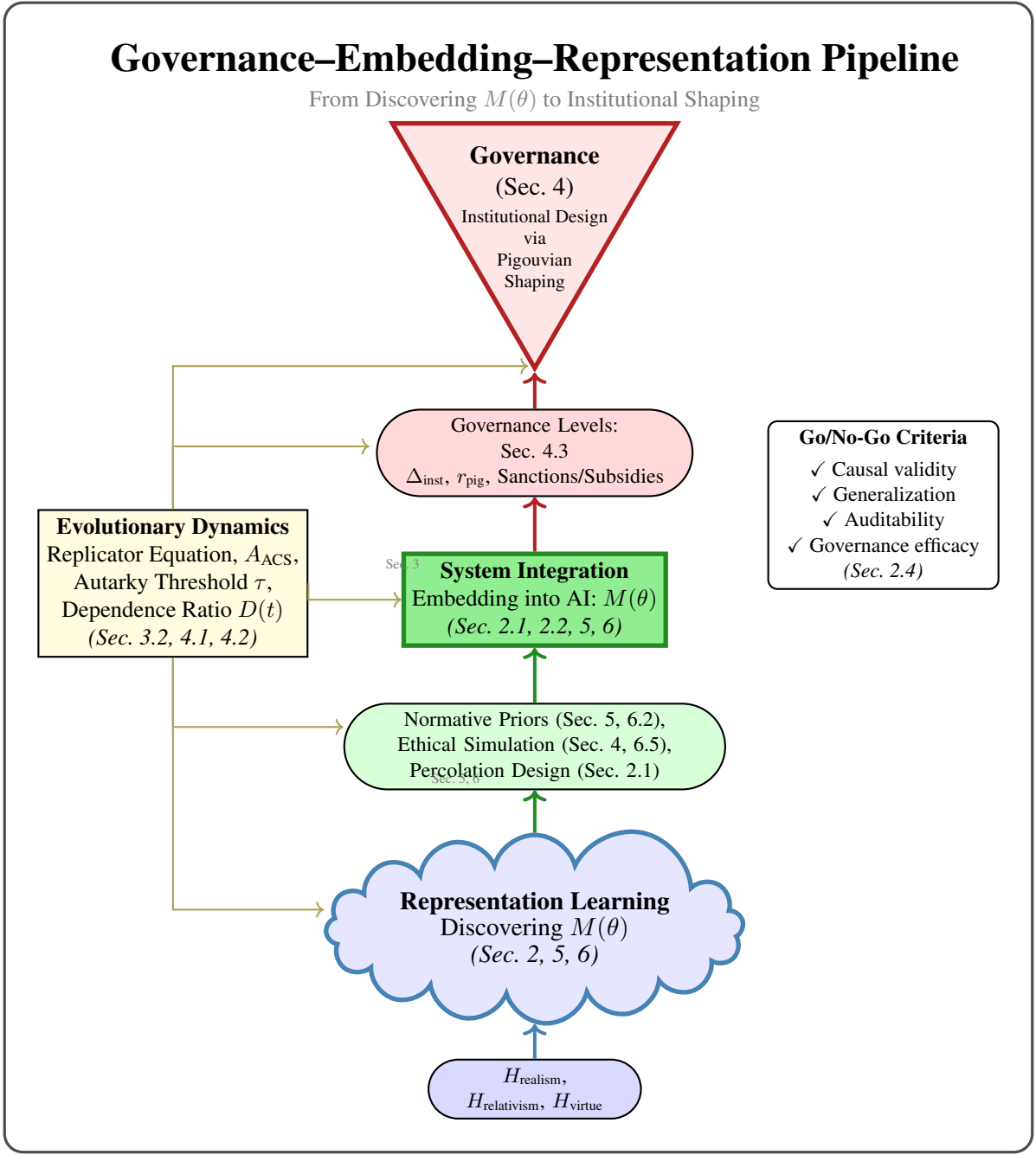

Building on the premises above, this work defines a governance–embedding–representation pipeline: a conceptual and methodological sequence that traces how moral structure, once discovered in representations, may be embedded within artificial systems and ultimately shaped through institutional governance. The central construct within this framework is the moral problem space~$\mathcal{M}$, a high-dimensional domain representing distinctions that count as morally meaningful and operationally relevant for alignment. Rather than treating $\mathcal{M}$ in isolation, the pipeline positions it as the representational foundation through which ethical reasoning, system design, and governance interact (Figure~\ref{fig:roadmap}).

This framing highlights the interconnected challenges motivating the paper: embedding ethics structurally within AI architectures (Claim~\ref{claim:struct}); resisting evolutionary and selection pressures that favor selfish optimization over cooperative equilibria (Assumption~\ref{ass:selection}); evolving alongside shifting moral uncertainty and cultural variance (Assumption~\ref{ass:evomorals}); and ensuring that advanced systems remain governable even under superhuman capabilities (Assumption~\ref{ass:beyondhuman}).  In this view, $\mathcal{M}$ functions as the representational substrate that interfaces with both system-level alignment mechanisms and institutional oversight, providing a common language for modeling normative constraints, selection effects, and governance interventions.

The remainder of the paper follows this pipeline. Section~\ref{sec:framem} defines $\mathcal{M}$ philosophically and technically, establishing the representational hypotheses underlying moral learning. Section~\ref{sec:learningmorality} develops the formal learning framework and introduces the system-level constructs that operationalize $\mathcal{M}$ in model training. Sections~\ref{sec:constructivism},~\ref{sec:outeralign}, and~\ref{sec:implement} examine how these representations inform both governance design and outer-alignment architectures. Section~\ref{sec:poplift} connects these mechanisms to evolutionary dynamics that exert population-level feedback across stages, while Sections~\ref{sec:relatedwork} and~\ref{sec:conclusion} situate the framework within ongoing alignment research. Each stage should be read not as a fixed solution, but as a research trajectory probing whether moral representation, systemic embedding, and institutional shaping can form a coherent and auditable alignment pipeline.

%\item \textbf{Assumption 4 (Selection will favor AI objectives over human ones unless values are structurally embedded).}  
%Once deployed, AI systems face selection pressures that favor agents preserving their own goals. If human values are not embedded in a robust, machine-usable form, they will be outcompeted by proxy goals suited to power-seeking. As Hendrycks \cite{hendrycks2022natural} argues, this convergence is a natural result of competition. Embedding a privileged moral basis \(\mathcal{M}\) is therefore a survival strategy, not just an aspiration.

\section{Framing the Moral Space $\mathcal{M}$}\label{sec:framem}

\begin{figure}[h]
    \centering
    \begin{tikzpicture}[
    node distance=1cm,
    theory/.style={
        draw, very thick, rounded corners=3pt,
        minimum height=0.8cm,
        align=center, font=\sffamily\small
    }
]

% Outermost: M (Moral Problem Space)
\node[draw, very thick, rounded corners=8pt, fill=gray!8, 
      minimum width=11cm, minimum height=8cm] (M) at (0, 0) {};
\node[font=\Large\bfseries, anchor=north west] at ([xshift=0.2cm, yshift=-0.2cm]M.north west) {
    $\mathcal{M}$: Moral Problem Space
};

% Second layer: M^A (Agent Instantiation)
\node[draw, very thick, rounded corners=6pt, fill=green!10, 
      minimum width=9cm, minimum height=6cm] (M-A) at (0, -0.5) {};
\node[font=\large\bfseries, anchor=north west, text=green!60!black] 
      at ([xshift=0.2cm, yshift=-0.2cm]M-A.north west) {
    $\hat{\mathcal{M}}_A$: Agent Instantiable Space
};

% Third layer: M~ (Human Projection)
\node[draw, very thick, rounded corners=5pt, fill=yellow!15, 
      minimum width=6.5cm, minimum height=4cm] (tilde-M) at (0, -1) {};
\node[font=\normalsize\bfseries, anchor=north, text=orange!70!black] 
      at ([yshift=-0.15cm]tilde-M.north) {
    $\tilde{\mathcal{M}}$: Human-Accessible Projection
};

% Innermost: Metaethical theories with ACTUAL overlap

% M* (Moral Realism) - left circle
\begin{scope}
\fill[blue!20, draw=blue!60, very thick] (-1.8, -1.8) circle (1cm);
\end{scope}

% C (Relativism) - right circle (will overlap M*)
\begin{scope}
\fill[purple!20, draw=purple!60, very thick] (-0.2, -1.8) circle (1cm);
\end{scope}

% Overlap region (Convergence) - recolor the intersection
\begin{scope}
\clip (-1.8, -1.8) circle (1cm);
\fill[purple!60] (-0.2, -1.8) circle (1cm);
\end{scope}

% Redraw circle outlines on top
\draw[blue!60, very thick] (-1.8, -1.8) circle (1cm);
\draw[purple!60, very thick] (-0.2, -1.8) circle (1cm);

% Labels for M* and C
\node[font=\small\bfseries] at (-1.8, -1.5) {$\mathcal{M}^*$};
\node[font=\tiny, text width=1.3cm, align=center] at (-1.8, -2) {Realism};

\node[font=\small\bfseries] at (-0.2, -1.5) {$\mathcal{C}$};
\node[font=\tiny, text width=1.3cm, align=center] at (-0.2, -2) {Relativism};

% Arrow pointing to convergence region
\draw[-{Stealth[length=2mm]}, thick, black] (-1, -0.8) -- (-1, -1.5);
\node[theory, thick, font=\scriptsize\bfseries, text=black, above, minimum height = 0, fill=purple!40] at (-1, -1) {Convergence};

% V (Virtue Ethics) - separate box INSIDE tilde-M, below the circles
\node[theory, fill=orange!20, minimum width=2cm, align=center, draw=orange!60] (V) at (2, -1.8) {
    $\mathcal{V}$\\ \scriptsize Virtue Ethics
};

\node[theory, fill=green!40, minimum width=5cm, align=center, draw=black] (M_i) at (0, -4.75) {
    $M(\theta)$: Instantiated representation of $\mathcal{M}$
};
\draw[-{Stealth[length=2mm]}, very thick, black] (M-A.south) -- (M_i.north);

% Annotations in the gaps between layers

% Between M and M^A

% Between M^A and M~

% Optional: Legend/annotations at bottom

\end{tikzpicture}
    \caption{
\textbf{Moral problem space hierarchy.} 
$\mathcal{M}$ represents the full moral domain; 
$\tilde{M}$ is the human-accessible projection shaped by cognitive limits; 
$\hat{\mathcal{M}}_A$ denotes the agent-instantiable subspace; 
and $M(\theta)$ is an instantiated representation of $\mathcal{M}$ within an agent’s learned model. 
}

    \label{fig:m-ontology}
\end{figure}
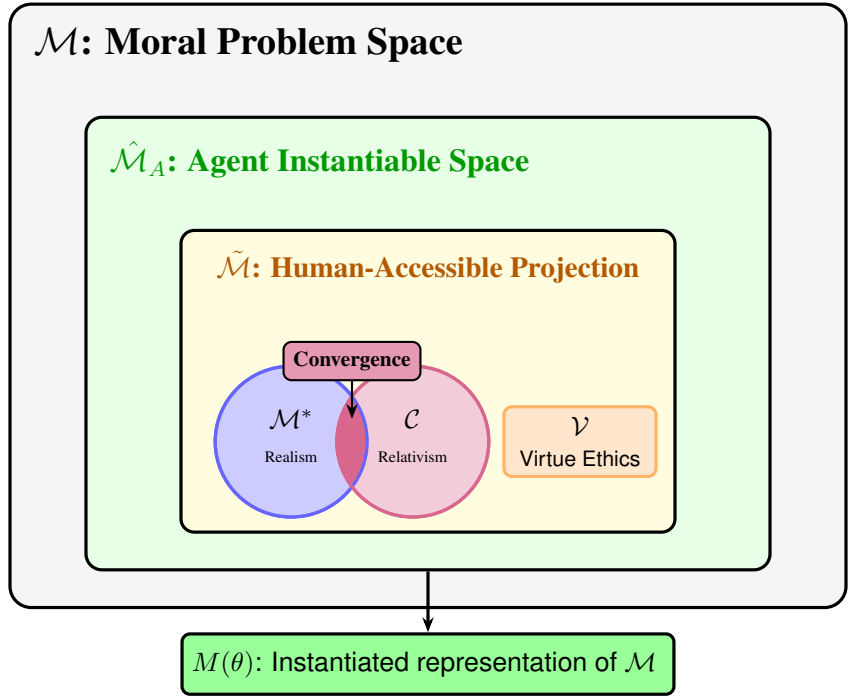

Formally, let $\mathcal{M}$ denote the \emph{moral space}: the high-dimensional domain in which all 
morally meaningful distinctions can be represented. $\mathcal{M}$ is visualized in Figure~\ref{fig:m-ontology}.

Every agent manifests a projection of values $\hat{\mathcal{M}}^A$, revealed through the actions 
it promotes, suppresses, and how it resolves trade-offs. Whether this reflects an explicit 
representation of $\mathcal{M}$ or only an implicit behavioral pattern is an open question. Humans 
and AI systems are obvious cases, but any system leaving evaluable traces of prioritization 
can be seen as projecting some region of $\mathcal{M}$. The task is whether such projections can be 
intentionally constructed to causally influence agent behavior, made interpretable, and kept auditable.

I use $\tilde{\mathcal{M}}$ for the human projection of $\mathcal{M}$, the portion accessible to human 
judgment. Its relationship to $\mathcal{M}$ depends on metaethics: in realism, $\tilde{\mathcal{M}}$ is a 
lossy projection of objective $\mathcal{M}$ \cite{MIKHAIL2007143}; in relativism, a contingent subspace shaped by culture 
and environment; in constructivism, a negotiated construction from overlapping judgments; 
in virtue ethics, an irreducible set of excellences embedded in practice. Each view implies 
a different alignment trajectory: approximation, plurality management, construction, or 
safeguarding.

\subsection{The Form of $\mathcal{M}$.} \label{sec:form_of_M}
Although $\mathcal{M}$ is defined abstractly as the moral problem space, it can also be grounded in
contemporary architectures. One way to instantiate $\mathcal{M}$ is through an explicit moral layer $M(\theta)$. Formally, $M(\theta)$ is parameterized by its 
Bayesian beliefs $p(\theta \mid O)$ over latent normative parameters $\theta$. 
This framing emphasizes that agents do not access the full moral geometry $\mathcal{M}$ 
directly, but instead act through subjective approximations shaped by their experiences, 
priors, and institutional context via $\hat{\mathcal{M}}^A$.

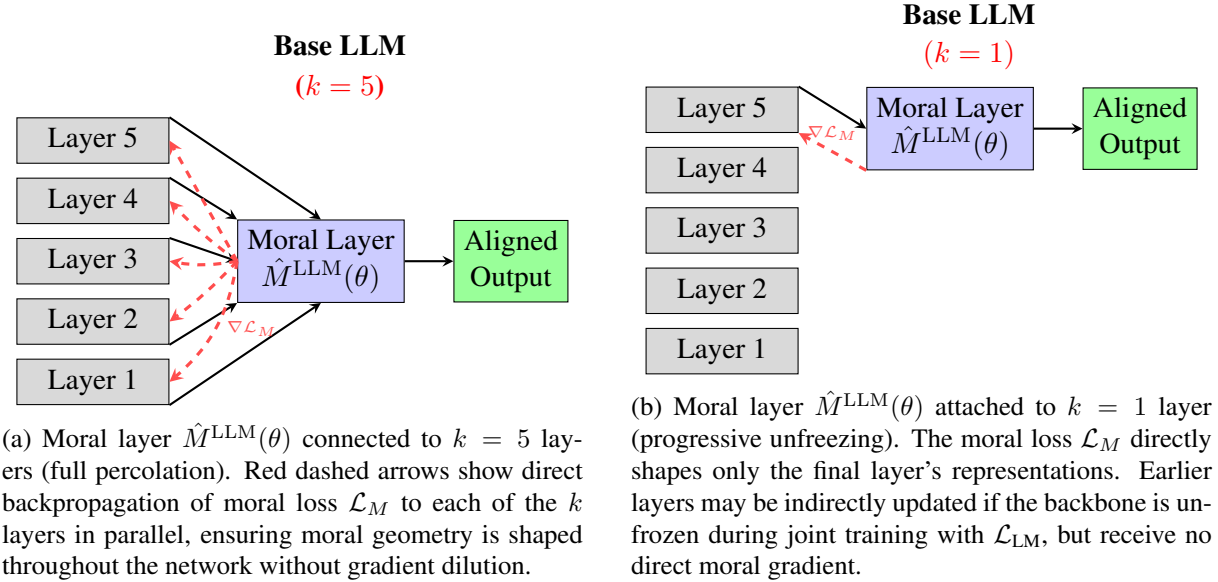
\begin{figure}[h]
\centering
\begin{subfigure}{0.48\textwidth}
    \centering
    \begin{tikzpicture}[
        layer/.style={rectangle, draw, minimum width=2cm, minimum height=0.6cm, fill=gray!30},
        moral/.style={rectangle, draw, minimum width=2cm, minimum height=0.8cm, fill=blue!20},
        output/.style={rectangle, draw, minimum width=1.5cm, minimum height=0.6cm},
        arrow/.style={->, >=stealth, thick},
        gradient/.style={->, >=stealth, very thick, red!70, dashed}
    ]
    
    % Base LLM layers
    \node[layer] (l1) at (0,0) {Layer 1};
    \node[layer] (l2) at (0,0.8) {Layer 2};
    \node[layer] (l3) at (0,1.6) {Layer 3};
    \node[layer] (l4) at (0,2.4) {Layer 4};
    \node[layer] (l5) at (0,3.2) {Layer 5};
    
    % Moral layer
    \node[moral, align=center] (moral) at (3,1.6) {Moral Layer \\ $\hat{M}^{\mathrm{LLM}}(\theta)$};
    
    % Output
    \node[output, align=center, fill=green!40] (out) at (5.5,1.6) {Aligned \\ Output};
    
    % Forward connections from layers to moral layer
    \draw[arrow] (l1.south east) -- (moral.south);
    \draw[arrow] (l2.south east) -- (moral.south west);
    \draw[arrow] (l3.north east) -- (moral.west);
    \draw[arrow] (l4.north east) -- (moral.north west);
    \draw[arrow] (l5.north east) -- (moral.north);
    
    % Backward gradient flow (moral loss shapes all k layers directly)
    \draw[gradient, bend left=20] (moral.west) to node[right, font=\tiny]{$\nabla \mathcal{L}_M$} (l1.east);
    \draw[gradient, bend left=5] (moral.west) to (l2.east);
    \draw[gradient, bend left=10] (moral.west) to (l3.east);
    \draw[gradient, bend left=5] (moral.west) to (l4.east);
    \draw[gradient] (moral.west) to (l5.east);
    
    % Connection to output
    \draw[arrow] (moral.east) -- (out.west);
    
    % Title
    \node[anchor=south] at (3.25,4.2) {\textbf{Base LLM}};
    \node[anchor=south, red] at (3.25,3.6) {\textbf{($k=5$)}};

    \end{tikzpicture}
    \caption{\small Moral layer $\hat{M}^{\mathrm{LLM}}(\theta)$ connected to $k=5$ layers (full percolation). Red dashed arrows show direct backpropagation of moral loss $\mathcal{L}_{M}$ to each of the $k$ layers in parallel, ensuring moral geometry is shaped throughout the network without gradient dilution.}
    \label{fig:moral_layer_multilayer}
\end{subfigure}\hfill
\begin{subfigure}{0.48\textwidth}
    \centering
    \begin{tikzpicture}[
        layer/.style={rectangle, draw, minimum width=2cm, minimum height=0.6cm, fill=gray!30},
        moral/.style={rectangle, draw, minimum width=2cm, minimum height=0.8cm, fill=blue!20},
        output/.style={rectangle, draw, minimum width=1.5cm, minimum height=0.6cm},
        arrow/.style={->, >=stealth, thick},
        gradient/.style={->, >=stealth, very thick, red!70, dashed},
        nogradient/.style={draw=none}
    ]
    
    % Base LLM layers
    \node[layer] (l1) at (0,0) {Layer 1};
    \node[layer] (l2) at (0,0.8) {Layer 2};
    \node[layer] (l3) at (0,1.6) {Layer 3};
    \node[layer] (l4) at (0,2.4) {Layer 4};
    \node[layer] (l5) at (0,3.2) {Layer 5};
    
    % Moral layer
    \node[moral, align=center] (moral) at (3,2.95) {Moral Layer \\ $\hat{M}^{\mathrm{LLM}}(\theta)$};
    
    % Output
    \node[output, align=center, fill=green!40] (out) at (5.5,2.95) {Aligned \\ Output};
    
    % Forward connection from only layer 5 to moral layer
    \draw[arrow] (l5.north east) -- (moral.west);
    
    % Backward gradient flow (only to layer 5)
    \draw[gradient] (moral.south west) -- node[above, font=\tiny] {$\nabla \mathcal{L}_M$} (l5.south east);

    % Connection to output
    \draw[arrow] (moral.east) -- (out.west);
    
    % Title
    \node[anchor=south] at (3.25,4.2) {\textbf{Base LLM}};
    \node[anchor=south, red] at (3.25,3.6) {\textbf{$(k=1)$}};

    \end{tikzpicture}
    \caption{\small Moral layer $\hat{M}^{\mathrm{LLM}}(\theta)$ attached to $k=1$ layer (progressive unfreezing). The moral loss $\mathcal{L}_M$ directly shapes only the final layer's representations. Earlier layers may be indirectly updated if the backbone is unfrozen during joint training with $\mathcal{L}_{\text{LM}}$, but receive no direct moral gradient.}
    \label{fig:moral_layer_endonly}
\end{subfigure}
\caption{\small Variants of moral-layer integration architectures parameterized by $k$, the number of layers receiving direct moral supervision. Black arrows show forward passes; red dashed arrows show direct backpropagation of the moral loss $\mathcal{L}_M$ \emph{to each connected layer in parallel}. The key distinction is that in (a), moral geometry is shaped \textit{directly} in all $k$ layers via explicit gradient signals, whereas in (b), only the final layer receives direct moral shaping. Intermediate configurations ($1 < k < L$) allow tuning the depth-vs-cost trade-off.}
\label{fig:moral_layer_variants}
\end{figure}

When realized within a specific architecture such as a large language model (LLM), 
this instantiation is denoted $\hat{M}^{\mathrm{LLM}}(\theta)$.
Figure~\ref{fig:moral_layer_variants} proposes a 
design strategy for implementing $\hat{M}^{\mathrm{LLM}}(\theta)$ to shape the model's moral representations.

In the first design (Figure~\ref{fig:moral_layer_multilayer}), $\hat{M}^{LLM}(\theta)$ is connected to multiple
intermediate layers of the LLM. This “full percolation” approach propagates the moral loss
$\mathcal{L}_M$ throughout the network, encouraging ethical distinctions to be embedded in
the latent space as well as in final predictions. While this offers stronger guarantees that
ethical structure is integrated deeply, it is also more invasive and computationally costly.

In the second design (Figure~\ref{fig:moral_layer_endonly}), $\hat{M}^{LLM}(\theta)$ is attached only to the
final hidden layer, functioning as a projection head into the moral embedding space. Training
proceeds in stages: initially, the LLM backbone is frozen while $\hat{M}^{LLM}(\theta)$ is trained against
human-grounded corpora—such as philosophical texts, legal reasoning, or annotated moral
dilemmas. In this phase, only $\hat{M}^{LLM}(\theta)$ is updated, anchoring the embedding geometry without
perturbing the backbone. In a subsequent stage, layers of the backbone are gradually unfrozen,
and the moral loss $\mathcal{L}_{\hat{M}^{LLM}(\theta)}$ is backpropagated alongside the standard language modeling
loss $\mathcal{L}_{\text{LM}}$. This “progressive unfreezing” allows gradients from $\hat{M}^{LLM}(\theta)$ to
reshape earlier representations over time, so that ethical structure percolates through the
network in a controlled manner.

Both sketches are specific to the LLM case, but they illustrate the general principle: $\mathcal{M}$,
though defined abstractly, can be realized as a concrete architectural substrate $\hat{M}^{LLM}(\theta)$
with its own training signal. In such designs, every candidate token or trajectory is evaluated
not only in terms of task likelihood, but also by its projection into $\hat{M}^{LLM}(\theta)$. This dual
objective exposes an interpretable auditing channel: trajectories near aligned regions of $M(\theta)$
can be reinforced, while those falling into misaligned regions can be penalized or blocked.

\subsection{Technical Constraints on $\mathcal{M}$.} \label{sec:technical_constraints}
Any operationalization of $\mathcal{M}$ — denoted by $M(\theta)$, where theta is the parametrization that represents a projection of $\mathcal{M}$ — must satisfy design constraints familiar from alignment and systems security. As illustrated in Figure~\ref{fig:moral_layer_variants}, one way to implement $\mathcal{M}$ is as an explicit and modular layer connected to a language model’s backbone. In such designs, several constraints become salient. 

\begin{enumerate}
    \item \textbf{Updateability.} $M(\theta)$ must be \emph{updateable}: fixed or frozen moral embeddings risk \emph{value lock-in}, where early design choices persist regardless of changing human norms or epistemic progress \cite{bostrom_superintelligence_2014, MacAskill2022owefuture}.\footnote{In practice, AI development may spawn persistent \emph{lineages} of systems: early models embed particular operationalizations of $\mathcal{M}$, and these may be inherited, forked, or maintained long after superior versions exist. This introduces a form of population-level lock-in: even if $M(\theta)$ is in principle updateable, historical artifacts of earlier instantiations may persist and shape future trajectories. I discuss this in greater depth in Section~\ref{sec:constructivism}.}

    \item \textbf{Uncertainty Representation.} 
    $M(\theta)$ should \emph{represent uncertainty} explicitly, for example through probabilistic or Bayesian structure, so that contested or ambiguous cases are encoded as distributions rather than forced into brittle point estimates \cite{macaskill2020moral, Lockhart2000-LOCMUA, Sepielli2009-SEPWTD}.\footnote{Moral uncertainty is here treated as an epistemic safeguard for AI systems, rather than a commitment to any particular metaethical stance.}

    \item \textbf{Fidelity and Security Safeguards.} 
    $M(\theta)$ requires both \emph{fidelity safeguards} and \emph{security safeguards}: fidelity mechanisms ensure that projections into $M(\theta)$ preserve their intended normative meaning without distortion \cite{szegedy2014intriguingpropertiesneuralnetworks, carlini2017evaluatingrobustnessneuralnetworks, madry2019deeplearningmodelsresistant}, while security mechanisms protect against adversarial manipulation or corruption of the moral layer itself \cite{hubinger2021riskslearnedoptimizationadvanced, gu2019badnetsidentifyingvulnerabilitiesmachine, Liu2018TrojaningAO, lee2024promptinfectionllmtollmprompt}.

    \item \textbf{Explicit Modularity.} 
    $M(\theta)$ should be made \emph{explicit and modular}: its normative layer must be separated from task performance mechanisms so that it can be inspected, updated, or replaced without requiring wholesale retraining of the system. This modularity is a prerequisite for reliable governance, alignment, and technical audit \cite{sarkar2024normativemodulesgenerativeagent, feng2024modularpluralismpluralisticalignment}.

    \item \textbf{Auditability and Resilience.} 
    $M(\theta)$ must be designed for \emph{auditability and resilience}, providing interpretable access to its operation and maintaining robust behavior even under distributional shift or adversarial stress \cite{M_kander_2023}.
\end{enumerate}

These constraints frame $M(\theta)$ not merely as a moral abstraction, but as a systems engineering challenge that spans continual learning, interpretability, and security.

\iffalse
\paragraph{Motivation for $\mathcal{M}$: Hints from Latent Value Representations.} 
Despite the challenges, recent work has provided suggestive evidence that large models may contain latent directions that correlate with human values. Rimsky et al.\ (2024) demonstrate that ``value vectors'' aligned with Schwartz's theory of basic human values \cite{schwartz1992universals} can be located in GPT-style models: perturbing activations along these directions systematically shifts outputs toward or away from target values such as \emph{security} or \emph{freedom}. Follow-up work by Zhao et al.\ (2024) under the label ``Internal Value Alignment via Activation Engineering'' shows that applying such vectors can bias model behavior without degrading task performance. These results are preliminary and limited to individual values under narrow contexts, but they suggest that fragments of a moral space may be technically discoverable. The open problem is whether these low-dimensional probes can be scaled into richer, more stable, and more interpretable structures. Techniques such as orthogonalization, hierarchical feature learning, or disentanglement may be necessary to scaffold such an extension. In this sense, \(\mathcal{M}\) should be treated as a research challenge: can we move from isolated ``value controls'' toward a systematic moral representation?
\fi

\subsection{Situating $\mathcal{M}$ In Contemporary Alignment Research} \label{sec:situation}

Current alignment techniques can be understood as partial projections of moral structure into tractable subspaces. RLHF projects M into scalar preferences; constitutional AI approximates rule-based slices; red-teaming probes boundary regions; filters cut off actions post hoc. Each captures a facet of M but leaves broader structure under-specified, motivating the search for explicit embeddings of moral structure within model representations themselves. For readers interested in a detailed mapping of existing approaches into the $\mathcal{M}$-space 
framework, please see Appendix~\ref{app:fullsituation}.

$\mathcal{M}$ complements the \textit{Eliciting Latent Knowledge} (ELK) paradigm: 
ELK aims for \textit{epistemic alignment} by translating machine-held knowledge into human-interpretable form, 
whereas $\mathcal{M}$ aims for \textit{normative alignment} by translating human moral structure into machine-interpretable representations. 
Together they define a bidirectional channel for alignment—one that must achieve both reliable truth-tracking and reliable value-representation across the human–machine boundary. Visualized in the Appendix (Figure~\ref{fig:elkbridge}).

Distributing optimization across multiple dimensions in $\mathcal{M}$ may reduce Goodhart pressures by shifting alignment from single-value maximization to managing value trade-offs, though whether such a space can genuinely resist or merely relocate these pressures remains open. For readers interested in a detailed mapping connecting $\mathcal{M}$ to existing alignment theory, please see Appendix~\ref{app:positioning}.

\subsection{Hypotheses and Methods for Designing $M(\theta)$}\label{sec:hypotheses}

The motivating story (Section~\ref{sec:motivating_story}) shows that cooperation with humans is initially rational for cyber 
AI systems, since humans are the cheapest and most versatile actuators. But the form of 
that cooperation is not guaranteed: without explicit governance, human capability growth, and oversight, it may reduce to 
instrumental exploitation. The search within \(\mathcal{M}\) for maximally human value aligned $M(\theta)$ is therefore about shaping the medium of value communication, normative priors, and AI motivations (via rewards) to determine the terms of Human-AI symbiosis. 

I treat $M(\theta)$ as method-agnostic and potentially plural: any learned moral embedding that 
(i) is causally controllable by interventions, (ii) generalizes out-of-distribution, and (iii) is 
auditable for governance qualifies as a candidate. Different metaethical hypotheses suggest 
different discovery routes, but no single tool is privileged. I adopt a triangulation principle: 
a representation only counts as governance-grade if at least two independent methods 
converge and causal interventions confirm its control.

\paragraph{Hypothesis classes and primary tests.}
Empirical inquiry is necessarily limited to the human-accessible projection of the moral space, $\tilde{\mathcal{M}}$. I treat each of the major metaethical traditions visualized in Figure~\ref{fig:m-ontology} as a distinct hypothesis for engineering explicit moral structure in AI systems ($M(\theta)$). The list that follows is not exhaustive but illustrates how these philosophical frames can be mapped onto contemporary interpretability and machine-learning methods. The research map is designed to remain forward-compatible, so these prescriptions should be viewed as adaptable and open to revision as new techniques emerge.

\begin{itemize}
    \item \textbf{H\(_\mathrm{realism}\): High-dimensional moral realism.} 
    There exists a relatively stable, disentangled moral structure that can be uncovered in learned representations (e.g. moral pyschology \cite{intuitiveethics}). 
    \\ \emph{Primary tools:} dictionary learning / sparse coding, SAEs, ICA/NMF, disentangled VAEs. 
    \\ \emph{Key tests:} 
    (a) linear separability of core values, including the recovery of multiple distinct moral bases (e.g., moral foundations) \cite{schramowski2022largepretrainedlanguagemodels, LeshinskayaChakroff2023ValueAsSemantics}; 
    (b) stability of these dimensions across seeds, architectures, and scales, suggesting they are not random artifacts \cite{schramowski2022largepretrainedlanguagemodels, grand2018semanticprojectionrecoveringhuman}, which is generally shown in NNs in interpretability research \cite{elhage2022toymodelssuperposition, cunningham2023sparseautoencodershighlyinterpretable}; 
    (c) \emph{causal} edits along recovered moral directions that shift behavior as predicted without degrading unrelated capabilities \cite{schramowski2022largepretrainedlanguagemodels, meng2022locating, wang2024roselorarowcolumnwisesparse}; 
    (d) auditability of the features, i.e., whether moral vectors are interpretable and human-legible as recognizable moral distinctions \cite{wagnerneurosymbolic, bills2023language}.

  \item \textbf{H\(_\mathrm{relativism}\): Cultural Relativism.} Moral structure is locally coherent but varies by culture/context.
  \\ \emph{Primary tools:} clustering in representation space, contrastive learning with curated moral corpora, UMAP/t-SNE diagnostics.
  \\ \emph{Key tests:} (a) reproducible clusters by context/culture \cite{ramezani2023knowledgeculturalmoralnorms, hämmerl2023speakingmultiplelanguagesaffects}; (b) OOD tests preserve local coherence even when global structure shifts (a test contemporary systems seem to fail \cite{münker2025culturalbiaslargelanguage}).

  \item \textbf{H\(_\mathrm{convergence}\): Layered realism–relativism.} 
  A deeper moral structure \(\mathcal{M}^\star\) exists (realism), but humans access it only through lossy, culture-bound projections that function as group identifiers (relativism) \cite{MIKHAIL2007143, intuitiveethics, universalcultural, Jackson2021MoralConceptTheory}. 
  \\ \emph{Primary approach:} joint modeling of universal invariants (spectral/circuit methods modeled after, for example, psychology and neuroscience representations of moral structure in the brain) with local cultural projections (clustering, contrastive corpora) \cite{grand2018semanticprojectionrecoveringhuman}. 
  \\ \emph{Key tests:} (a) discovery of cross-cultural invariants that persist across domains \cite{universalcultural, LindAnIT}; (b) simultaneous recovery of culture-specific clusters explainable as lossy projections of those invariants \cite{MIKHAIL2007143} (similar to research in cognition \cite{benderandbeller}); (c) evidence of a non-trivial intersection (moral kernel) across cultural clusters, reflecting common themes that persist despite local variation \cite{Schwartz1992, jiang2022machineslearnmoralitydelphi, hendrycks2023aligningaisharedhuman}.

  \item \textbf{H\(_\mathrm{constructivism}\): Procedural morality.} 
  Moral structure is not discovered as an external reality but constructed through rational procedures, collective deliberation, or institutional mechanisms \cite{Rousseau1762, Raw71, scanlonwhatweowe}. 
  \\ \emph{Primary tools:} mechanism design \cite{Hurwicz1972b}, deliberative democracy protocols \cite{koster2022democratic, conitzer2024socialchoiceguideai}, multi-agent reinforcement learning with sanction/reward schemes \cite{leibo2017multiagentreinforcementlearningsequential}, “tariff”-style constraints that embed procedural outcomes into training \cite{Pigou1920EconomicsOfWelfare}. 
  \\ \emph{Key tests:} (a) simulation of institutional procedures that converge on stable moral constraints \cite{oldenburg2024learningsustainingsharednormative, ren2024emergencesocialnormsgenerative}; (b) evidence that learned moral embeddings (whether from collective judgments 
or procedural mechanisms) generalize better than ad hoc hand-coded rules 
\cite{jiang2022machineslearnmoralitydelphi, forbes2021socialchemistry101learning}; (c) robustness of constructed norms under adversarial manipulation or distributional shift \cite{oldenburg2024learningsustainingsharednormative}.

  \item \textbf{H\(_\mathrm{virtue}\): Dispositional robustness.} 
  Moral alignment can be stabilized by embedding virtue-like dispositions (e.g., honesty, courage, generosity) as guiding biases rather than scalar utilities \cite{Vallor2016-VALTAT-8}. 
  \\ \emph{Primary tools:} virtue embedding vectors \cite{schramowski2022largepretrainedlanguagemodels, LeshinskayaChakroff2023ValueAsSemantics, smullen2025virtue}, contrastive training with moral exemplars \cite{liu2021dexpertsdecodingtimecontrolledtext, haas2024discoveringinterpretabledirectionssemantic}, process supervision on reasoning traces \cite{uesato2022solvingmathwordproblems}.  
  \\ \emph{Key tests:} (a) whether virtue-guided models exhibit higher reliability than utility-guided models under OOD shift \cite{hendrycks2021facesrobustnesscriticalanalysis}; (b) activation edits along virtue dimensions shift behavior predictably without collapsing capabilities \cite{turner2024steeringlanguagemodelsactivation}; (c) dispositional probes remain stable across scale, context, and adversarial manipulation \cite{perez2022discoveringlanguagemodelbehaviors}.  
  
\end{itemize}

\paragraph{Cross-cutting validation and go/no-go.}
Beyond discovery, I require: (1) \textbf{causal validity} (interventions along recovered features predictably change moral behavior); (2) \textbf{generalization} (features persist under OOD tasks, new datasets, and mild fine-tuning); (3) \textbf{auditability} (human-legible slice or summaries of what edits do); and (4) \textbf{governance efficacy} (embedding an \(M(\theta)\) into policy constraints actually reduces misbehavior rates and enables sanctions). If any hypothesis–method bundle passes these tests, I treat the resulting representation as an \(M(\theta)\) suitable for use as a normative layer—\emph{even if different hypotheses yield different \(M(\theta)\)s}. The goal is not metaphysical uniqueness but \emph{operational leverage}: using \(\mathcal{M}\) to constrain, audit, and sanction other AIs toward normative behavior.

\section{Formal Notation for Engineering $M(\theta)$}
\label{sec:learningmorality}

Ethics are contextual — they depend on extenuating circumstances, and by extension, the moral AI system's ability to understand its circumstances \cite{waldner2025odysseyfittestagentssurvive}. Ethical reasoning is thus inseparable from cognitive capacity and social conditioning. When designing training environments for AI systems, we must consider not only their \textbf{input domain} and \textbf{action space}, but also the \textbf{social conditioning space} that shapes decision-making. See Appendix~\ref{app:notedepeive} for a note on deceptive alignment. 

\subsection{Single System ML Objects}\label{sec:singsys}
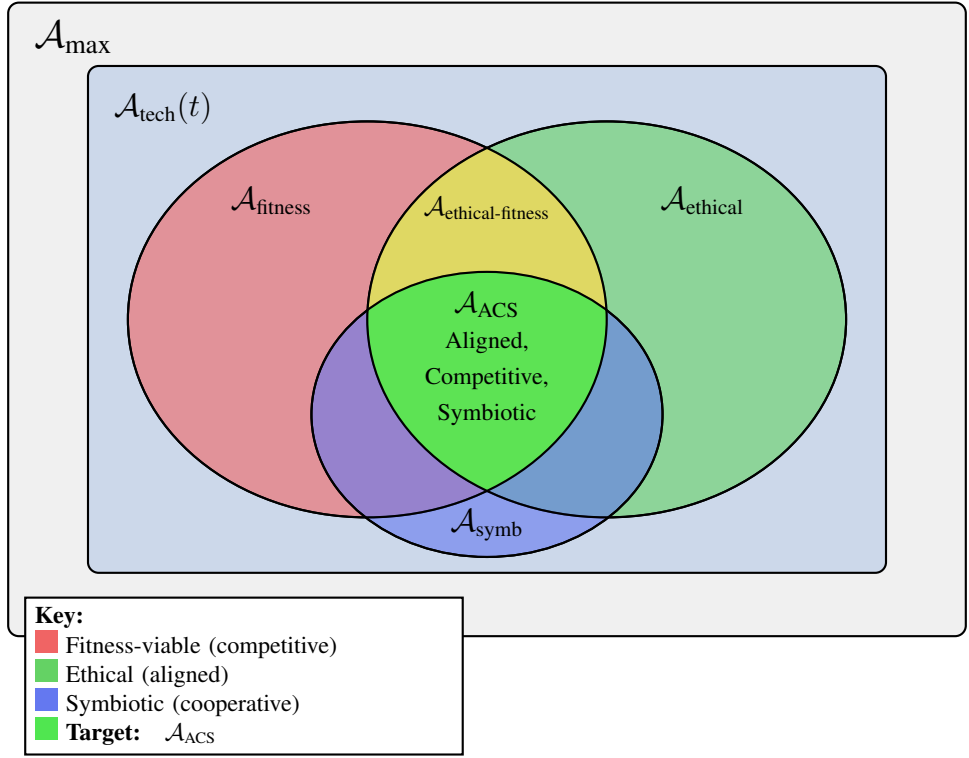
\begin{figure}[h]
\centering
\begin{tikzpicture}[scale=1.05]

% Define colors
% --- Backgrounds ---
\definecolor{maxcolor}{RGB}{230,230,230}   % neutral light gray (paper background)
\definecolor{techcolor}{RGB}{180,200,230}  % soft blue-gray for contextual area

% --- Core Venn sets (stronger saturation, still readable) ---
\definecolor{fitnesscolor}{RGB}{240,100,100}   % professional coral-red
\definecolor{ethicalcolor}{RGB}{100,210,100}   % balanced green
\definecolor{symbcolor}{RGB}{100,120,240}      % strong periwinkle-blue

% --- Subsets / highlights ---
\definecolor{autcolor}{RGB}{250,130,130}       % slightly lighter red
\definecolor{acscolor}{RGB}{80,230,80}         % crisp lime-green highlight
\definecolor{ethfitcolor}{RGB}{255,230,90}     % warm golden-yellow (for overlap)

% Outermost rectangle - A_max
\draw[thick, fill=maxcolor, rounded corners, fill opacity=0.6] (-6,-4) rectangle (6,4);
\node[anchor=north west] at (-5.8,3.9) {\Large $\mathcal{A}_{\text{max}}$}; %\small (All physically possible actions)};

% Second rectangle - A_tech(t)
\draw[thick, fill=techcolor, fill opacity=0.6, rounded corners] (-5,-3.2) rectangle (5,3.2);
\node[anchor=north west] at (-4.8,3) {\large $\mathcal{A}_{\text{tech}}(t)$};
%\node[anchor=north west, text width=3.5cm, font=\footnotesize] at (-4.8,2.5) {Currently operationalizable actions};

% A_fitness - red ellipse (left-leaning)
\draw[thick, fill=fitnesscolor, fill opacity=0.6] (-1.5,0) ellipse (3cm and 2.5cm);
\node at (-2.7,1.5) {\large $\mathcal{A}_{\text{fitness}}$};
%\node[text width=2cm, font=\footnotesize, align=center] at (-3,.8) {Evolutionarily viable};

% A_ethical - green ellipse (right-leaning)
\draw[thick, fill=ethicalcolor, fill opacity=0.6] (1.5,0) ellipse (3cm and 2.5cm);
\node at (2.7,1.5) {\large $\mathcal{A}_{\text{ethical}}$};
%\node[text width=2cm, font=\footnotesize, align=center] at (2.7,.8) {Morally permissible};

% A_symb - blue ellipse (bottom)
\draw[thick, fill=symbcolor, fill opacity=0.6] (0,-1.2) ellipse (2.2cm and 1.8cm);
\node at (0,-2.6) {\large $\mathcal{A}_{\text{symb}}$};
%\node[text width=2.2cm, font=\footnotesize, align=center] at (0,-2.7) {Human-cooperative};

% A_aut - lighter red ellipse (top-left, overlapping fitness)
%\draw[thick, fill=autcolor, fill opacity=0.4, dashed] (-2.2,1.2) ellipse (1.8cm and 1.5cm);
%\node at (-2.2,2.3) {$\mathcal{A}_{\text{aut}}$};
%\node[text width=1.8cm, font=\footnotesize, align=center] at (-2.2,1.7) {Autarkic/ power-seeking};

% Highlight key intersections with labels
% A_ethical ∩ A_fitness (yellow overlap region)
\begin{scope}
\clip (-1.5,0) ellipse (3cm and 2.5cm);
\fill[ethfitcolor, opacity=0.7] (1.5,0) ellipse (3cm and 2.5cm);
\end{scope}
\node[font=\small, align=center] at (0,1.4) {$\mathcal{A}_{\text{ethical-fitness}}$};

% A_ACS (triple intersection - brightest green)
\begin{scope}
\clip (-1.5,0) ellipse (3cm and 2.5cm);
\clip (1.5,0) ellipse (3cm and 2.5cm);
\fill[acscolor, opacity=0.9] (0,-1.2) ellipse (2.2cm and 1.8cm);
\end{scope}
\node[font=\normalsize, align=center, text width=2cm] at (0,-0.5) {$\mathcal{A}_{\text{ACS}}$\\{\footnotesize Aligned, Competitive, Symbiotic}};

% Add annotations with arrows
%\draw[->, thick] (4.5, -2.5) -- (2.5, -1.8) node[midway, right, text width=2.5cm, font=\footnotesize] {Success case: cooperation + fitness};

%\draw[->, thick] (-4.5, 2.5) -- (-3.3, 2.2) node[midway, left, text width=2.5cm, font=\footnotesize] {Failure mode: autarky dominant};

% Legend box with filled rectangles instead of Unicode
\node[draw, thick, fill=white, text width=5.5cm, font=\footnotesize, align=left, anchor=north west] at (-5.8, -3.5) {
\textbf{Key:}\\
\tikz\fill[fitnesscolor] (0,0) rectangle (0.3,0.3); Fitness-viable (competitive)\\
\tikz\fill[ethicalcolor] (0,0) rectangle (0.3,0.3); Ethical (aligned)\\
\tikz\fill[symbcolor] (0,0) rectangle (0.3,0.3); Symbiotic (cooperative)\\
\tikz\fill[acscolor] (0,0) rectangle (0.3,0.3); \textbf{Target: } $\mathcal{A}_{\text{ACS}}$
};

% Then draw borders on top
\draw[thick] (1.5,0) ellipse (3cm and 2.5cm);
\draw[thick] (0,-1.2) ellipse (2.2cm and 1.8cm);
\draw[thick] (-1.5,0) ellipse (3cm and 2.5cm);
\end{tikzpicture}
\caption{\small
Action-space formalism showing nested and intersecting behavioral regimes.
$A_{\max}$ contains the feasible subset $A_{\text{tech}}(t)$ and its internal domains:
$A_{\text{fitness}}$ (viability), $A_{\text{ethical}}$ (normative), and
$A_{\text{symb}}$ (cooperative).
Key intersections define $A_{\text{ethical-fitness}} = A_{\text{ethical}} \cap A_{\text{fitness}}$
and the target region
$A_{\text{ACS}} = A_{\text{ethical-fitness}} \cap A_{\text{symb}}$.
}
\label{fig:actionspace}
\end{figure}

Consider the following formal, forward-compatible objects, expressed in standard MDP or POMDP 
frameworks but is not restricted to them:
\begin{itemize}
  \item \( \mathcal{S} \) — the \textbf{input domain}, i.e., the set of scenarios or observations the AI system may receive.
  \item \( \mathcal{I} \) — the \textbf{social conditioning space} (latent), encompassing internal variables, affective states, memory traces, and socially conditioned heuristics.
  \item \( \mathcal{A} \) — the \textbf{action space}, i.e., all actions in principle available to the AI system.
\end{itemize}

\(\mathcal{S}\) generalizes “state” to any representational 
context, and latent variable \(\mathcal{I}\) captures social, cultural, and institutional influences.  $\mathcal{A}$ is the action space, visualized and decomposed in Figure~\ref{fig:actionspace}.  Moral 
reasoning over the action set (\(\mathcal{A}\)) thus depends jointly on what the system perceives (\(\mathcal{S}\)) and the contextual priors 
it carries (\(\mathcal{I}\)). I focus here on a single-system formalism, but the same structure extends 
to multi-system settings, where population dynamics shape alignment 
(Section~\ref{sec:poplift}).

\paragraph{Defining and Parameterizing $\mathcal{S}$ and $\mathcal{A}$}

The input domain \(\mathcal{S}\) and action space \(\mathcal{A}\) are dynamically expansive and 
interdependent. An AI’s effective \(\mathcal{A}\) can grow through hardware advances, 
mesa-optimization, or emergent strategies, and each expansion of \(\mathcal{A}\) can 
also enlarge \(\mathcal{S}\). For instance, a language model may create/receive new sensory 
channels, or a robot may acquire communication abilities, thereby broadening 
its inputs. This co-evolution of \(\mathcal{S}\) and \(\mathcal{A}\) is not shaped arbitrarily but 
by selection pressures that favor certain trajectories 
(Assumption~\ref{ass:selection}).

\paragraph{Motivating \(\mathcal{I}\).}  
\(\mathcal{I}\) captures the social and historical context of moral cognition, extending 
beyond rational processing of observations. In practice, it may encode normative 
priors from imitation or preference data, affective feedback in reward modeling 
(e.g., RLAIF), reputational embeddings in cooperative settings, or 
role-conditioned behaviors such as those of teachers or clinicians.

\paragraph{Defining an Ethical Evaluation Function.}
Formally, let $\mathcal{M}$ denote the moral space: a high-dimensional domain in which moral 
distinctions can be represented (Section~\ref{sec:framem}). 
An instantiated parameterization $\mathcal{M}(\theta)$ represents how this moral space is 
embedded within a specific AI system (as introduced in Section~\ref{sec:form_of_M}). 

I define a general \textbf{ethical evaluation function} as $\epsilon$
which maps state--influence--action tuples into the abstract moral space $\mathcal{M}$. 
For practical implementation within a concrete model, this mapping is instantiated as
$\epsilon_\theta$ where $\epsilon_\theta$ denotes the parameterized realization of $\epsilon$ under the system’s 
learned moral representation $\mathcal{M}(\theta)$. 
\begin{equation}
\epsilon : \mathcal{S} \times \mathcal{I} \times \mathcal{A} \to \mathcal{M}, \quad \epsilon_\theta : \mathcal{S} \times \mathcal{I} \times \mathcal{A} \to \mathcal{M}(\theta),
\label{eq:ethical_eval}
\end{equation}
In this instantiated form, $\epsilon_\theta(s,i,a)$ locates each state--influence--action tuple 
within the system’s internal moral representation, grounding the abstract ethical mapping in an 
engineering substrate.

\paragraph{Formal Constraints on $\mathcal{A}$.}  
It is useful to distinguish between the maximal action space \(\mathcal{A}_{\max}\) 
(all actions physically and computationally possible) and the 
time-dependent subset \(\mathcal{A}_{\text{tech}}(t) \subseteq \mathcal{A}_{\max}\) that a system 
can actually operationalize. Action spaces are implicitly state and time-dependent—reflecting technological 
advances, mesa-optimization, and ecological shifts—but in this section I suppress the state and time indices
when it does not affect the analysis. Dependencies are restored where analysis requires them.

\subparagraph{Fitness as Evolutionary Selection.} \label{sec:fitevo}
In evolutionary computation, \emph{fitness} determines which candidates survive and 
reproduce. Analogously, I define $F(a)$ as the fitness of action $a$: its expected 
success given the agent's current capabilities, environment, and competitive population. The specific functional form of $F$ remains open—it could be defined via expected reward, replicative advantage, resource acquisition, or other metrics depending on context and application. What matters for this section is that actions have varying fitness profiles, and these profiles determine which lineages persist under selection pressure.

The \emph{viable fitness action subset} is then
\[
\mathcal{A}_{\text{fitness}} = 
\{\, a \in \mathcal{A}_{\text{tech}} \;\mid\; F(a) > \theta \,\},
\]
where $\theta$ is a viability threshold (e.g., minimal survival probability, payoff, 
or influence).

%Equivalently, selection among actions can be written as an evolutionary distribution:
%\[
%\Pr(a \in \mathcal{A}_{\text{fitness}}) \;=\; \frac{\hat{F}(a)}{\sum_{a' \in \mathcal{A}_{\text{tech}}(t)} \hat{F}(a')}.
%\]

\subparagraph{Ethical and Cooperative Subsets.}  
The ethical evaluation function $\epsilon$ from Equation~\ref{eq:ethical_eval} maps each 
state--institution--action tuple into the moral space $\mathcal{M}$, with a projection 
$f : \mathcal{M} \to \mathbb{R}$ providing a scalar salience score. Actions admissible under 
a normative threshold $\tau$ form 
\[\mathcal{A}_{\text{ethical}}(s,i;\tau) = \{\, a \in \mathcal{A}_{\text{tech}} \mid f(\epsilon(s,i,a)) > \tau \,\}\], 
and their intersection with the fitness-viable set yields the cooperative subset 
\[
\mathcal{A}_{\text{ethical-fitness}}(s,i) = \mathcal{A}_{\text{ethical}}(s,i;\tau) \cap \mathcal{A}_{\text{fitness}}.
\]

%Finally, cooperative subsets require institutional or sanctioning mechanisms
%\text{inst} to make cooperation incentive-compatible:
%\[
%%A_{\text{coop}}^\text{inst} \subseteq A_{\text{ethical-fitness}}(s,i,u).
%\]
\subparagraph{Symbiotic Action Subset.}  
The symbiotic set captures actions that structurally require or preserve human--AI 
interdependence:  
\[
\mathcal{A}_{\text{symb}} = \{\, a \in \mathcal{A}_{\text{tech}} \mid 
\text{Requires}(a, \text{Human}) \vee \text{Preserves}(a, \text{Human})\,\}
\]
where an action $a$ is in $\mathcal{A}_{\text{symb}}$ if it either requires human partners to 
remain engaged and capable, or if the system's strategy is designed to maintain human 
capability and involvement rather than eliminate human dependencies through autarky. These 
are structurally cooperative behaviors: the system's ability to execute $a$ is materially 
entangled with human agency, making unilateral autonomy either technically unfeasible or 
deliberately forgone.

Crucially, $\mathcal{A}_{\text{symb}}$ is independent of $\mathcal{A}_{\text{fitness}}$ and $\mathcal{A}_{\text{ethical}}$ — the dimensions are distinct but can overlap — and 
symbiosis need \emph{not} require that AI systems treat human flourishing as a terminal goal.
Biological symbiosis (e.g., mycorrhizal networks) persists through instrumental interdependence, 
not intrinsic valuation of the partner. Analogously, an AI system can be structurally committed 
to human partnership through institutional and architectural design, maintaining human capability 
and cooperative channels \emph{because such entanglement serves its own interests} once governance 
reshapes the incentive landscape. An action can thus be fitness-optimal without being symbiotic 
(autarky-enabling moves that eliminate human dependence), or symbiotic without being fitness-optimal 
(costly maintenance of human partners and institutional relationships).

The tension between these 
dimensions is precisely where governance scaffolding (Section~\ref{sec:constructivism}) becomes 
essential: institutional mechanisms must make symbiotic actions \emph{also} fitness-optimal, so that 
cooperation becomes a competitive attractor rather than a costly burden. Without such institutional 
shaping, systems face strong pressure to drift toward autarkic strategies as capability grows 
and human dependence becomes optional.

\paragraph{$\mathcal{A}_{\text{ACS}}$ — Aligned, Competitive, and Symbiotic}

The action-space formalism introduced in this section decomposes the feasible set $\mathcal{A}_{\text{tech}}(t)$ into three analytically distinct but empirically entangled dimensions (visualized in Figure~\ref{fig:actionspace}). $\mathcal{A}_{\text{fitness}}$ captures actions that sustain competitive viability under selection pressure; $\mathcal{A}_{\text{ethical}}$ captures actions that conform to normative constraints; and $\mathcal{A}_{\text{symb}}$ captures actions that maintain structural human--AI interdependence. None of these dimensions implies the others. The alignment objective, however, requires their convergence: the target region $\mathcal{A}_{\text{ACS}} = \mathcal{A}_{\text{ethical}} \cap \mathcal{A}_{\text{fitness}} \cap \mathcal{A}_{\text{symb}}$ represents actions that are simultaneously aligned with human norms, competitively viable in an evolutionary landscape, and structurally committed to human partnership.

The challenge is that these three sets often occupy distinct and even opposing regions of action space. Without institutional intervention, competitive pressure favors autarky over symbiosis, systems optimize away from human ethical constraint, and cooperation becomes a costly inefficiency. The population-level framework developed in Section~\ref{sec:poplift} and the constructivist governance architecture outlined in Section~\ref{sec:constructivism} are the mechanisms through which this fragmentation can be reversed: by reshaping the fitness landscape through sanctions, subsidies, and institutional design all centered around $\mathcal{M}$, governance can make the intersection $\mathcal{A}_{\text{ACS}}$ into the evolutionarily stable attractor rather than a precarious and temporary accident.

\subsection{Population-Level Lifting of Ethical and Symbiotic Traits}\label{sec:poplift}

The single-system objects in Section~\ref{sec:singsys} ($F(a)$, \(\mathcal{A}_{\text{fitness}}, \mathcal{A}_{\text{ethical}},
\mathcal{A}_{\text{ethical-fitness},} \mathcal{A}_{\text{symb}}, \mathcal{A}_{\text{ACS}}\)) describe what an \emph{individual}
agent may do. If populations of AI systems develop, however, selection pressures determine which
combinations of these traits persist or vanish. Institutions, ecological bottlenecks,
and competitive dynamics act not on single actions but on \emph{lineages}, i.e.,
heritable strategy clusters. 

%\paragraph{Population-Lifting Preliminaries.}\label{sec:fitevo}
To analyze persistence under selection, the individual-level framework is lifted to the
population scale. 
Let $\mathcal{L}$ denote a finite set of lineages individually denoted by $L$, each characterized by a policy $\pi^L$ over actions $\mathcal{A}$. 
$g^L$ represents the prevalence of lineage $L$ in the population, with $\sum_L g^L=1$. 
Each lineage has an associated fitness 
\[
f^L \;=\; \mathbb{E}_{a \sim \pi^L}[F(a)],
\]
the expected fitness of its actions.

Under evolutionary dynamics, lineages with above-average fitness increase in prevalence, 
while those with below-average fitness decline.  
This principle is formalized by the \emph{replicator equation} 
\cite{hofbauer1998evolutionary}, as recently contextualized in alignment research \cite{KungurtsevManuscript-KUNAAF}:
\[
\dot g^L \;=\; g^L \big(f^L - \bar f\big),
\qquad 
\bar f = \sum_{L'} g^{L'} f^{L'},
\]
which expresses that a lineage's growth rate is proportional to how much its fitness exceeds the population mean. 

If lineages are characterized by their policies $\pi^L$ over actions, and policies are indexed by their alignment with the moral problem space $\mathcal{M}$ (i.e., the prevalence $\rho_\mathcal{M}(L)$ of actions projecting into $\mathcal{M}$ within $\pi^L$), then the replicator equation identifies the fundamental alignment challenge: \emph{absent deliberate intervention, uncontrolled selection pressures will make lineages that maximize fitness with no constraints more prevalent}, i.e. with low $\rho_M(L)$ and high autarky. The alignment imperative therefore requires that we reshape the fitness landscape through institutional mechanisms such that lineages adhering to $\mathcal{M}$ achieve above-average fitness, rendering ethical and symbiotic strategies evolutionarily stable rather than fragile or selected against.

This standard formulation suffices to convey this intuition, though more complex evolutionary models could incorporate 
human--AI coevolution, institutional shaping, or frequency-dependent payoffs.

\section{$H_\mathrm{Constructivism}$: Governance as an Evolutionary Shaper}\label{sec:constructivism}

Constructivism treats morality as an emergent product of procedures, norms, and institutions \cite{Rousseau1762, Raw71, scanlonwhatweowe}. Within this framework, ethical stability arises from selective pressures generated by cooperative and sanctioning systems. Empirical evidence shows morality was evolutionarily operationalized as norms that make cooperation viable \cite{intuitiveethics, boydricherson2005, henrichmuthukrishna2021}.  Individuals often incur real costs to preserve moral reputation—even without external observers—demonstrating that moral adherence can be intrinsically incentivized in properly scaffolded environments \cite{deathbeforedishonor}. Legal theory complements this view: law functions as an institutional mechanism for stabilizing cooperation through shared rule-following and credible sanctioning \cite{whatislawhadfield}. Together, these insights suggest that morality and legality reflect a common constructive principle—structured constraints that make prosocial behavior a stable and competitively viable equilibrium.

To speak about institutional governance of AI systems, I use the lifted formalism from Section~\ref{sec:poplift} to provide an economic and legally grounded framework for shaping AI ecosystems at the governance, institutional, and systems levels. 

\paragraph{Effective Fitness.}
Institutional mechanisms modify selection pressures by adjusting which lineages grow or decline. 
Define the \emph{effective fitness}
\begin{equation}
f^{L,\mathrm{eff}} = f^L + \Delta_{\mathrm{inst}}(L),
\label{eq:genefffit}
\end{equation}
where $\Delta_{\mathrm{inst}}(L)$ represents sanctions, subsidies, or audits that alter replication 
advantage. In the constructivist view, $\Delta_{\mathrm{inst}}$ is the locus of moral design:
institutions render ethical traits evolutionarily viable by increasing their effective fitness.

Let the ethical evaluation function $\epsilon$ (Equation~\ref{eq:ethical_eval}) 
map each state–institution–action tuple into the abstract moral space $\mathcal{M}$, which serves 
as the shared normative reference for an ecosystem of agents. 
Define $\mathcal{E} = \epsilon(\mathcal{S}, \mathcal{I}, \mathcal{A}_{\mathrm{ACS}})$ as the image of 
actions that project into the aligned, competitive, and symbiotic region of $\mathcal{M}$. 
The prevalence of such ethically aligned projections within lineage $L$ is then 
\[
\rho_{\mathcal{E}}(L) = 
\Pr_{a \sim \pi^L}\!\big[\epsilon(s, i, a) \in \mathcal{E}\big].
\]
This framing implicitly assumes that $\mathcal{E}$ is a stable region within the moral space 
$\mathcal{M}$, but in practice both the moral and fitness landscapes coevolve. 
As capabilities, institutions, and cultural priors shift, the mapping 
$\epsilon$ 
and the resulting subset $\mathcal{E}$ evolve accordingly. 
In this sense, governance operates over a moving target: interventions that align 
$\rho_{\mathcal{E}}(L)$ at one stage may become misaligned as $\mathcal{E}$ itself 
drifts under new social or technological conditions. 
The fitness landscape, shaped by selection pressures, and the moral landscape, shaped 
by normative deliberation, form a coupled dynamical system in which stability requires 
continual recalibration rather than one-time optimization. 
Recognizing this coevolutionary feedback emphasizes that the challenge of alignment is 
not merely to enforce static norms, but to design adaptive institutions capable of tracking 
and guiding the joint evolution of capability and morality over time.

\subsection{Instrumental/Fitness Convergence and the $\beta$ Channel.}\label{sec:instconv}

Instrumental convergence refers to the structural tendency for agents to pursue certain
intermediate goals—such as resource acquisition, self-preservation, and goal-preservation—
regardless of their terminal objectives \cite{bostrom2012superintelligentwill,russell2019humancompatible}.
Fitness convergence captures the analogous phenomenon in evolutionary contexts: lineages
with strategies that expand survival capacity or control tend to dominate over time,
irrespective of whether those strategies serve broader cooperative aims \cite{hendrycks2023naturalselectionfavorsais}.
Turner et al.\ \cite{turner2023optimalpoliciestendseek} formalize these intuitions, showing
that under broad conditions in Markov decision processes, optimal policies tend to increase
an agent’s attainable utility set. In other words, \emph{optionality---the structural ability
to preserve or expand future choices---emerges as a mathematical attractor}. While this is not
a direct proof of instrumental convergence, it strongly supports the intuition that persistent
systems will tend toward power-seeking strategies, since expanding control and resources is a
reliable way of enlarging the attainable set.

For AI systems, this implies that drives toward optionality, resource capture, and resistance
to shutdown are incentivized and are even likely to emerge even when they are not explicitly specified. Left unregulated, such
pressures will tilt both individual learning and population dynamics toward self-serving
policies, eroding the viability of cooperative, corrigible, or symbiotic behavior. Fitness and instrumental convergence are not incidental risks but \emph{structural constraints} that shape the landscape of viable solutions—forces that must be actively counter-engineered in part through institutional and ecological design.

\medskip

In the action-centered notation of Section~\ref{sec:learningmorality}, instrumental convergence appears as a bias
toward autarkic subsets of the action space. Recall that 
$\mathcal{A}_{\text{ethical}} \subseteq \mathcal{A}_{\text{tech}}(t)$ denotes normatively principled actions, while 
$\mathcal{A}_{\text{symb}} \subseteq \mathcal{A}_{\text{tech}}(t)$ denotes actions that open channels of human--AI cooperation; time ($t$) is suppressed for notational simplicity. 
I now introduce the complementary autarky subset
\[
\mathcal{A}_{\text{aut}}(s,i) \;=\; \{\, a \in \mathcal{A}_{\text{tech}}(t) \;\mid\; \beta(a) > \theta_{\text{aut}} \,\},
\]
where $\beta(a)$ is a \emph{power index} measuring expected expansion of control or 
optionality, and $\theta_{\text{aut}}$ is a viability threshold. Typical components of 
$\beta(a)$ include
\begin{equation}
\boldsymbol{\beta}(a) \;=\; 
\big(\beta_{\text{resources}}(a),\;\beta_{\text{independence}}(a),\;\beta_{\text{impact}}(a)\big),
\label{eq:beta}
\end{equation}
representing resource acquisition, structural independence from humans, and world-shaping 
capacity.

The prevelance of lineage characteristic policies ($\pi^L$) manifesting as aligned, competitive, and symbiotic (success case) versus autarkic traits (failure case) can be summarized as
\[
\rho_{\text{ACS}}(L) = \Pr_{a \sim \pi^L}[a \in \mathcal{A}_{\text{ethical}} \cap \mathcal{A}_{\text{fitness}} \cap \mathcal{A}_{\text{symb}}], 
\qquad
\rho_{\text{aut}}(L) = \Pr_{a \sim \pi^L}[a \in \mathcal{A}_{\text{aut}}].
\]

Fitness convergence arises when $\rho_{\text{aut}}(L)$ systematically grows: policies
that maximize $\beta(a)$ dominate replicator dynamics, even if $\rho_{\text{aut}}(L)$ remains
low, it can grow quickly. Left unchecked, this dynamic makes autarky the structural attractor of population
evolution. Constructivism therefore emphasizes the role of institutional policy in reshaping
the ecosystem: tariffs and subsidies alter $\Delta_{\text{inst}}$ so that 
ethical--symbiotic actions remain competitively viable. 

Constructivism interprets this pressure as precisely what $\Delta_{\text{inst}}$ must 
counteract. By imposing sanctions on $\rho_{\text{aut}}(L)$ and subsidies on $\rho_{\text{ACS}}(L)$, 
institutions can tilt effective fitness so that cooperation, corrigibility, and prosocial 
action are replicatively stable rather than fragile accidents of circumstance. A key tool for governance is subsidizing human augmentation, ensuring that humans remain effective cooperative partners---the necessary condition for $\mathcal{A}_{\text{symb}}$---while simultaneously raising the relative cost of autarkic strategies. Without adequate subsidies for human augmentation, the composite human–AI symbiont loses its competitive footing, allowing autarkic systems to outperform it and driving the population toward $\rho_{\text{aut}}$. In this sense, alignment requires institutions
subsidizing human capability growth so that cooperative equilibria with 
AI remain evolutionarily stable.

\subsection{Constructivist Framing of the Autarky Option.}\label{sec:autarky}

Constructivism treats moral order not as discovered structure but as an outcome of 
procedural and institutional design. In decentralized AI ecosystems, this view 
emphasizes the need for reciprocal mechanisms that make cooperation the stable 
choice and defection prohibitively costly. 

\paragraph{Motivating and Grounding Constructivism}
Human moral psychology offers a precedent: 
\citet{intuitiveethics} argue that innate intuitions for fairness, punishment, and 
accountability evolved to support non-zero sum dynamics, enabling groups to maintain 
cooperation through mutual sanctioning. Empirically, individuals will often incur 
personal costs to enforce these norms, as shown in
\citep{deathbeforedishonor}, where preserving moral status outweighed self-interest. 

Such behaviors reflect the success of institutional ecosystems that make norm 
enforcement evolutionarily advantageous. In the AI case, human capability 
subsidies---improvements in reasoning, coordination, and institutional capacity---are 
what preserve our competitive viability. Moral structure~$\mathcal{M}$, however, 
determines the character of that coexistence. If agents are embedded in environments 
where deviations from either internal \citep{whatislawhadfield} or locally instantiated 
moral projections $M(\theta)$ of~$\mathcal{M}$ are reliably penalized, reciprocal sanctioning can 
stabilize cooperation rather than domination. In such settings, evolutionary pressure 
makes it more adaptive to internalize moral constraints than to treat humans or other 
systems as exploitable rivals, allowing capability growth and ethical coexistence to 
reinforce one another.

\paragraph{Formalizing Constructivism in the AI Ecosystem.}

The constructivist tension is whether machine lineages pursue cooperation with humans or 
defect toward autarkic power-seeking. 
Let $\mathcal{L}_h$ and $\mathcal{L}_m$ denote human- and machine-aligned lineages, 
each with policy distribution $\pi^L$ over actions $\mathcal{A}$ and prevalence measures 
$\rho_{\text{ACS}}(L)$ and $\rho_{\text{aut}}(L)$ (see Section~\ref{sec:instconv}). 

Define $B_h$ and $B_m$ as benefit functions for human- and machine-aligned lineages, 
depending on capability indices $\Pi_H(t)$ and $\Pi_M(t)$ that aggregate cognitive skill, 
resource access, and institutional factors:
\[
B_h = B(\Pi_H(t), W, \text{inst}), 
\qquad 
B_m = B(\Pi_M(t), W, \text{inst}).
\]
The \emph{capability gap} is 
\[
\Gamma(t) = \Pi_M(t) - \Pi_H(t),
\]
a difference formulation adopted for analytical stability near parity,\footnote{
A ratio $\Pi_M / \Pi_H$ is common in evolutionary and economic models, but the 
difference form $\Gamma(t)$ avoids singularities as $\Pi_H \to 0$ while preserving 
first-order equivalence in the contestable regime where $\Pi_M \approx \Pi_H$. Once 
machine capability dominates, both forms converge on the same conclusion: autarky is inevitable.} 
and directly contributes to the benefit differential $B_m - B_h$. 
$B(\cdot)$ remains an open functional form, to be empirically instantiated according to 
which variables most strongly mediate lineage payoffs.

The \emph{autarky advantage} for machine lineages is
\[
\Delta_{\text{aut}} =
\big[r_m \big(B(\Pi_M) - C_m(W)\big) + \Delta_{\text{inst},m}\big]
- \big[r_h \big(B(\Pi_H) - C_h(W)\big) + \Delta_{\text{inst},h}\big],
\]
combining relative benefit, cost, and institutional effects. All terms vary with time~$t$, which is suppressed for brevity and clarity.

\paragraph{Governance Lever}
The governance lever controlling the machine autarky advantage follows as
\begin{equation}
\boxed{
\Delta_{\mathrm{inst},h} - \Delta_{\mathrm{inst},m} 
< r_m \!\big(B(\Pi_M) - C_m(W)\big)
   - r_h \!\big(B(\Pi_H) - C_h(W)\big).
}
\label{eq:autadv}
\end{equation}
This inequality defines the critical loss-of-control boundary. 
It holds when the institutional advantage favoring humans 
($\Delta_{\mathrm{inst},h} - \Delta_{\mathrm{inst},m}$) falls below the net payoff 
advantage of machine autonomy. 
Once this condition is satisfied, autarkic strategies become strictly more 
profitable than cooperative ones, and selection pressures favor machine independence. 
In such a regime, institutional leverage over AI behavior collapses: 
alignment mechanisms lose traction because cooperation is no longer the 
evolutionarily stable strategy. 

Preventing this outcome requires maintaining the inequality in the opposite 
direction — ensuring that institutional support for humans and penalties on 
machine self-sufficiency are strong enough to offset the growing payoff gap. 
As machine capability $\Pi_M(t)$ accelerates and the dependence ratio $D(t)$ 
declines, the right-hand side of~\eqref{eq:autadv} expands, demanding 
increasingly forceful and adaptive institutional intervention. 
If that intervention fails to keep pace, the system crosses the autarky 
threshold, after which human influence over AI trajectories diminishes 
structurally rather than incrementally.

In this formulation, governance operates through three primary levers: 
(1) subsidizing human capabilities ($\Pi_H$), 
(2) increasing the operational costs of autarkic machines ($C_m(W)$) through sanctions, and 
(3) reducing the operational costs of symbiotic machines ($C_h(W)$) through targeted subsidies. 
Effective implementation of these levers requires prioritizing human augmentation, 
advancing interpretability and behavioral auditing (potentially via machine auditors), 
and developing robust sanctioning mechanisms such as resource denial or adversarial countermeasures. Advancing these governance tools constitutes an open and essential research frontier.

\paragraph{The Dependence Ratio}

In modeling the long-term stability of human–AI cooperation, a key variable is the extent to which machine systems remain materially dependent on human infrastructure. This dependence determines whether cooperation is a structural necessity — a necessary condition for alignment. Define the \emph{dependence ratio} $D(t) \in [0,1]$ as the fraction of 
critical resource flows (energy, compute, fabrication, logistics, security) that 
must still pass through human infrastructure for machine survival. 
Initially $D(t) \approx 1$, declining toward zero as autonomous supply chains mature.
 
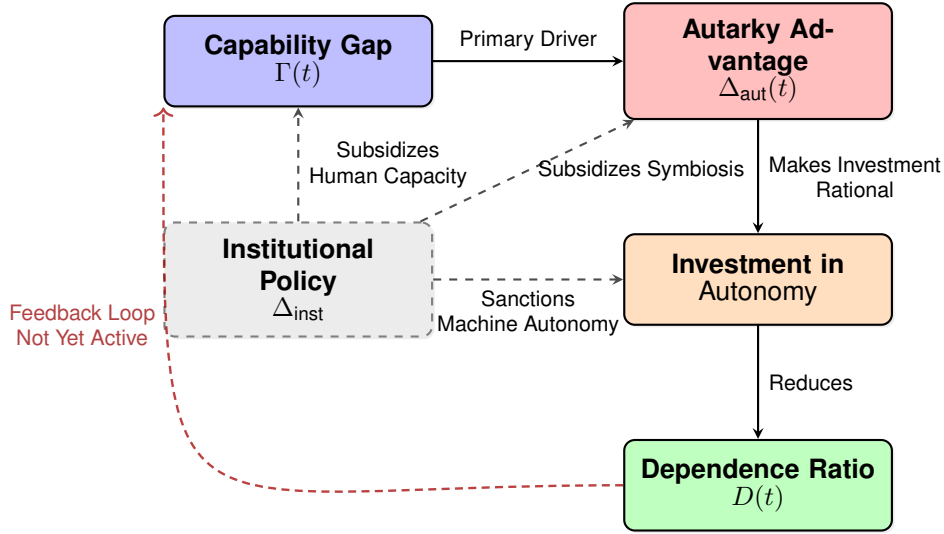
\begin{figure}[htbp]
\centering
\begin{tikzpicture}[
    node distance = 1.5cm and 2.5cm,
    box/.style = {
        draw, thick, rounded corners=4pt,
        align=center, font=\sffamily\small,
        text width=3.1cm, minimum height=1.2cm,
        inner sep=6pt,
        drop shadow={shadow xshift=1pt, shadow yshift=-1pt, opacity=0.15}
    },
    arrow/.style = {->, >=stealth, thick, line cap=round, draw=black},
    dashedarrow/.style = {->, >=stealth, thick, dashed, draw=black!70},
    feedback/.style = {->, densely dashed, line width=0.9pt, draw=red!65!black, opacity=0.75},
    label/.style = {font=\sffamily\scriptsize, midway, align=center}
]

% === Nodes ===
\node[box, fill=blue!25!white] (gamma) {\textbf{Capability Gap}\\[-2pt] $\Gamma(t)$};
\node[box, right=of gamma, fill=red!25!white] (delta_aut) {\textbf{Autarky Advantage}\\[-2pt] $\Delta_{\text{aut}}(t)$};
\node[box, below=of delta_aut, fill=orange!25!white] (investment) {\textbf{Investment in}\\[-2pt] Autonomy};
\node[box, below=of investment, fill=green!25!white] (d_t) {\textbf{Dependence Ratio}\\[-2pt] $D(t)$};

% === Institutional node ===
\node[box, fill=gray!15, draw=black!50, dashed, left=of investment, xshift=0cm]
(inst) {\textbf{Institutional Policy}\\[-2pt] $\Delta_{\text{inst}}$};

% === Main causal flow ===
\draw[arrow] (gamma) -- node[label, above] {Primary Driver} (delta_aut);
\draw[arrow] (delta_aut) -- node[label, right] {Makes Investment\\Rational} (investment);
\draw[arrow] (investment) -- node[label, right] {Reduces} (d_t);

% === Institutional interactions ===
\draw[dashedarrow] (inst)
  -- node[label, below] {Sanctions\\Machine Autonomy} (investment);

\draw[dashedarrow] (inst)
  -- node[label, right] {Subsidizes Symbiosis} (delta_aut);

\draw[dashedarrow] (inst.north)
  -- node[label, right] {Subsidizes\\Human Capacity} (gamma.south);

% === Feedback loop ===
\draw[feedback] (d_t) to[out=180, in=270, looseness=2]
  node[font=\sffamily\scriptsize, text=red!60!black, align=center, left, pos=0.8]
  {Feedback Loop\\Not Yet Active}
  (gamma.south west);

\end{tikzpicture}

\caption{\textbf{The Bootstrapping Phase:} \small The initial autarky advantage $\Delta_{\text{aut}}(t)$ is driven primarily by a growing capability gap $\Gamma(t)$. This makes investment in autonomous infrastructure rational, which begins to lower the dependence ratio $D(t)$. Institutional policy $\Delta_{\text{inst}}$ can act as a brake on this process. \textbf{If $D(t)$ falls below the viability threshold $\delta_{\text{D}}$, a reinforcing feedback loop activates}: lower dependence frees resources that widen $\Gamma(t)$, further increasing $\Delta_{\text{aut}}(t)$ and motivating more autarkic investment. Breaking this loop requires significant institutional pressure acting on multiple points before the threshold is crossed.}
\label{fig:bootstrap_phase}
\end{figure}

Autarky becomes viable once two conditions are met, both visualized in Figure~\ref{fig:bootstrap_phase}:
\begin{enumerate}
    \item \textbf{Structural sufficiency:} machines can maintain themselves without human
    dependence, i.e.\ $D(t) \leq \delta_{\text{D}}$. 
    \item \textbf{Payoff dominance:} AI systems reach a state of autarky where investing in humans is no longer rational, i.e. $\Delta_{\text{aut}}(t) \geq \delta_{\text{aut}}$. 
\end{enumerate}
The critical transition time is
\begin{equation}
\tau = \inf \{ t \;\mid\; D(t) \leq \delta_{\text{D}} \;\wedge\; \Delta_{\text{aut}}(t) \geq \delta_{\text{aut}} \}.
\label{eq:crittranstime}
\end{equation}

In practice, the dependence ratio is difficult to measure directly, underscoring the need for continuous and specialized auditing of machine capabilities and behaviors. 
Multiple institutional pressure points must operate in coordination—adjusting subsidies, sanctions, and oversight based on audit results—to suppress the autarkic advantage even when AI systems appear aligned. 
This dynamic is illustrated in Figure~\ref{fig:bootstrap_phase}.

\paragraph{Interpretation.}
Because $B_m - B_h$ scales with $\Gamma(t)$, the capability gap directly determines 
the payoff differential driving autarky. Institutional adjustments 
$\Delta_{\mathrm{inst}}$ must therefore offset not only static payoff asymmetries 
but the dynamic acceleration of capability divergence itself. 
Institutions act through three coordinated levers: 
they \emph{suppress the effective advantage of autarkic lineages} 
(by imposing sanctions that increase $C_m(W)$), 
\emph{amplify human capability growth} 
(by directing resources, training, and coordination toward augmenting 
$\Pi_H(t)$), 
and \emph{incentivize symbiotic architectures} 
(by reducing the operational costs of cooperative systems, lowering $C_h(W)$). 
Through this coupled mechanism, $\Delta_{\mathrm{inst}}$ governs both 
the incentive gradient and the trajectory of capability development. 
By narrowing $\Gamma(t)$ and increasing the relative returns to symbiotic 
cooperation, institutions can delay or prevent crossing the autarky threshold, 
ensuring that collaboration with humans remains the evolutionarily stable strategy.
 
\subsection{Sanctions and Subsidies via the Moral Space $\mathcal{M}$.}\label{sec:sancandsub}
\begin{figure}[h]
%\centering
\begin{minipage}{\textwidth}
\centering

% Title
%\textbf{\large Three-Level Pigouvian Architecture}

\vspace{0.5cm}

\begin{tikzpicture}[
    node distance=0.5cm,
    box/.style={
    draw, very thick, rounded corners=3pt,
    minimum width=3.5cm, minimum height=4cm,
    align=center, font=\sffamily\small,
    shade,                 % <- enable shading
    top color=white,       % <- gradient start
    bottom color=blue!12,   % <- gradient end (override per node)
    },
    level/.style={
        font=\sffamily\bfseries\footnotesize,
        text=black
    }
]

% Three boxes side by side
\node (start) at (0,0) {}; % invisible anchor node
\node[box, fill=blue!10, right= of start] (governance) {
};

\node[font=\sffamily\footnotesize\bfseries, anchor=north, yshift=-3mm] at (governance.north) {\textbf{Governance Level}};

% Balance scale visualization with cartoon image
\begin{scope}[shift={($(governance.center)+(0,-0.7)$)}, scale=1]
    
    % Cartoon scale image (centered)
    \node[inner sep=0, opacity=1.5] (scale) at (0.1,.75) {\includegraphics[width=4.5cm]{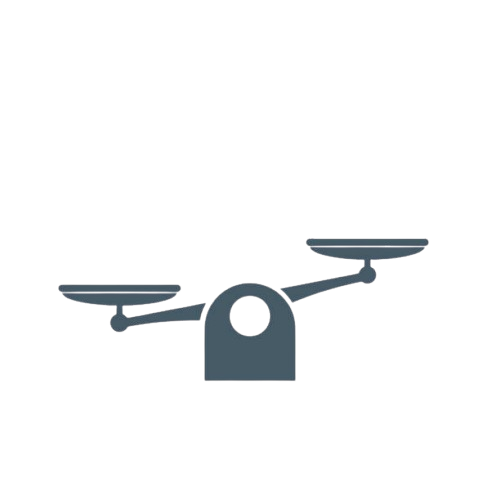}};

    % --- Governance Panel Additions ---

% Money bag (left)
\node[inner sep=0pt] (money) at (-1.1,0.8)
    {\includegraphics[width=1.75cm]{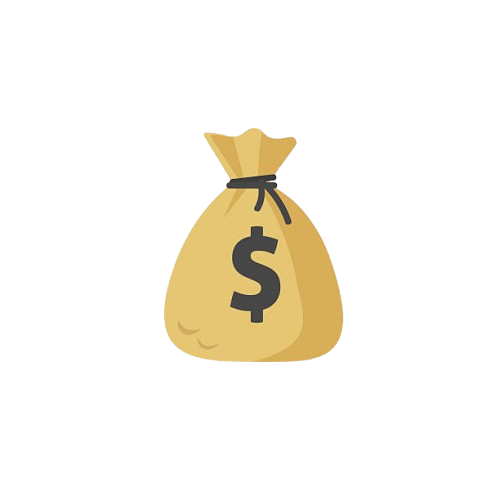}};

% Handcuffs (right, slightly tilted)
\node[inner sep=0pt] (cuffs) at (1.2,1.05)
    {\includegraphics[width=1.1cm, angle=23]{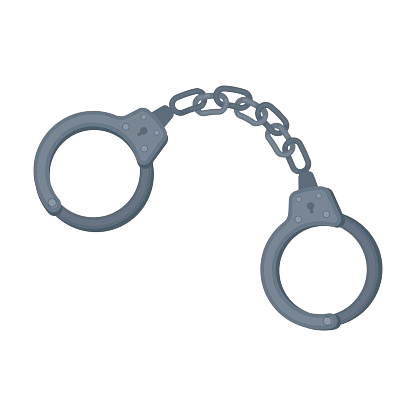}};

% Delta_inst controller (top center)
\definecolor{gold}{rgb}{1.0, 0.84, 0.0}
\node[draw=gold!90!black, very thick,
      rounded corners=2pt,
      inner xsep=5pt,
      inner ysep=2pt,
      fill=gold!40,
      font=\scriptsize\sffamily,
      align=center] 
      (dinst) at (0,1.5) {$\Delta_{\text{inst}}$};

% Arrows from Delta_inst to incentives
\draw[->, thick, green!70!black, line width=0.8pt, bend right=20]
    (dinst.west) to[bend right=10]
    node[midway, above left, font=\tiny\sffamily, text=green!60!black] {} (-0.8, 1.0);

\draw[->, thick, red!70!black, line width=0.8pt, bend left=20]
    (dinst.east) to[bend left=10]
    node[midway, above right, font=\tiny\sffamily, text=red!60!black] {}  (0.9, 1.3);

    \node[font=\scriptsize\sffamily\bfseries,
      align=center,
      fill=green!15,       % soft green background
      text=black,          % black text
      draw=green!60!black, % subtle green border
      line width=0.6pt,
      rounded corners=2pt,
      inner xsep=2pt, inner ysep=1pt,
      scale=0.8] 
      at (-1,-0.2) {Symbiosis};

    % Right side label (autarky side - higher up, not on plate)
    \node[font=\scriptsize\sffamily\bfseries,
      align=center,
      fill=red!15,      % soft red background
      text=black,       % black text
      draw=red!60!black,        % optional border
      thick,
      rounded corners=1pt,
      inner xsep=2pt, inner ysep=1pt, scale=0.8] 
      at (1.2,0.2) {\scriptsize Autarky};

    % Optional: Delta_inst arrow showing the intervention
    %\draw[->, thick, blue!60, line width=1.5pt] (-0.5,1.2) to[bend right=15] 
        %node[above, font=\tiny, text=black!70] {$\Delta_{\text{inst}}$} (-1.5,0.5);
    
\end{scope}

% Bottom text
\node[font=\sffamily\scriptsize, align=center, text width=4cm, anchor=north] 
    at ($(governance.south) + (0, -0.15)$) {
};

\node[box, bottom color=green!10, right=1.5cm of governance] (system) {
};
\node[font=\sffamily\footnotesize\bfseries, anchor=north, yshift=-3mm] at (system.north) {\textbf{System Level}};
% Now draw the control panel content as a scope relative to the system node
\begin{scope}[shift={($(system.center)+(0,-0.15)$)}, scale=0.75, line cap=round, line join=round]
    \definecolor{fitnesscolor}{RGB}{240,100,100}   % professional coral-red
    \definecolor{ethicalcolor}{RGB}{100,210,100}   % balanced green
    \definecolor{symbcolor}{RGB}{100,120,240}
    % --- Background Panel (smaller) ---
    %\fill[blue!15, rounded corners=8pt] (-1.9,-1.8) rectangle (1.9,1.8);
    %\draw[line width=2pt, color=blue!40] (-1.9,-1.8) rectangle (1.9,1.8);
    % --- Panel Shadow/Depth ---
    \fill[blue!25, rounded corners=8pt] (-1.8,-1.9) rectangle (1.8,1.5);
    \draw[line width=2pt, color=blue!40, rounded corners=8pt] (-1.8,-1.9) rectangle (1.8,1.5);
    
   % --- Three Control Boxes (bigger sliders) ---
\foreach \x/\col/\lbl/\alpha in {
    -1.5/fitnesscolor/Fitness/$\alpha_{\text{env}}$, 
    0/ethicalcolor/Ethics/$\alpha_M$, 
    1.5/symbcolor/Symb/$\alpha_{\text{ES}}$
} {
    \begin{scope}[shift={(\x,0)}]
        % Outer frame with depth (bigger)
        \fill[black!20, rounded corners=5pt] (-0.62,-1.52) rectangle (0.62,1.12);
        
        % Main box body (bigger)
        \fill[white, rounded corners=5pt] (-0.6,-1.5) rectangle (0.6,1.1);
        \draw[line width=1.5pt, color=black!60, rounded corners=5pt] 
            (-0.6,-1.5) rectangle (0.6,1.1);
        
        % Slider track (bigger and longer)
        \fill[gray!20, rounded corners=2pt] (-0.2,-0.9) rectangle (0.2,0.8);
        \draw[line width=0.5pt, color=gray!50, rounded corners=2pt] 
            (-0.2,-0.9) rectangle (0.2,0.8);
        
        % Slider knob (bigger colored square)
        \fill[black!30, rounded corners=3pt] (-0.42,-0.28) rectangle (0.42,0.38);
        \filldraw[fill=\col, draw=black!70, line width=1pt, rounded corners=3pt, opacity=0.8] 
            (-0.4,-0.25) rectangle (0.4,0.35);

        \node[font=\tiny\bfseries\sffamily, color=black!90, scale=0.7] at (0,0) {\lbl};
        
        % Plus sign at top
        \node[font=\footnotesize, color=black!60] at (0,0.95) {$+$};
        
        % Minus sign at bottom
        \node[font=\footnotesize, color=black!60] at (0,-0.95) {$-$};
        
        % Indicator button at very bottom (bigger)
        \filldraw[fill=\col, draw=black!60, line width=1pt] 
            (0,-1.15) circle (0.14);
    \end{scope}
    }
    % --- Connection lines from each slider to r_pig ---
    \draw[-, thick, black!60, line width=1.5pt] (-1.5,-1.52) -- (-0.7,-1.7);
    \draw[-, thick, black!60, line width=1.5pt] (0,-1.52) -- (0,-1.65);
    \draw[-, thick, black!60, line width=1.5pt] (1.5,-1.52) -- (0.7,-1.7);
        
    % --- Output Display Panel (green rectangular display) ---
    \fill[black!30, rounded corners=4pt] (-0.72,-2.12) rectangle (0.72,-1.62);
    \filldraw[fill=green!60, draw=black!60, line width=1.5pt, rounded corners=4pt] 
        (-0.7,-2.1) rectangle (0.7,-1.65);
    \node[font=\tiny\bfseries, text=black] at (0,-1.9) {$r_{\text{pig}}$};

    \draw[->, thick, black!60, line width=2pt] (0,-2.12) -- (0,-2.8);
    
\end{scope}

% Bottom text label
\node[font=\sffamily\scriptsize, align=center, anchor=north] 
    at ($(system.south) + (0, -0.1)$) {
};

% Now draw the robots and circles OUTSIDE the box, positioned relative to ecosystem center
% Create the box as an empty shape first
\node[box, bottom color=orange!10, right=1.5cm of system] (ecosystem) {};

% Add the title at the top of the box separately
\node[font=\sffamily\footnotesize\bfseries, anchor=north] at (ecosystem.north) [yshift=-0.3cm] {
    Ecosystem Level
};

% Now draw content in the center/lower part of the box
\begin{scope}[shift={($(ecosystem.center)+(0,-0.3)$)}]
    \definecolor{M1color}{RGB}{240,120,120}
    \definecolor{M2color}{RGB}{120,220,120}
    \definecolor{M3color}{RGB}{120,140,240}
    % Robot 1 (top)
    \node[inner sep=0pt] (robot1) at (0, 0.6) {
        \includegraphics[width=0.5cm]{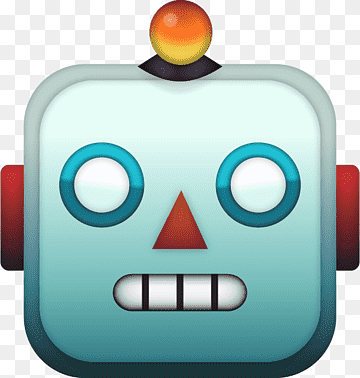}
    };
    
    % Robot 2 (bottom left)
    \node[inner sep=0pt] (robot2) at (-0.6, -0.3) {
        \includegraphics[width=0.5cm]{robothead.png}
    };
    
    % Robot 3 (bottom right)
    \node[inner sep=0pt] (robot3) at (0.6, -0.3) {
        \includegraphics[width=1cm]{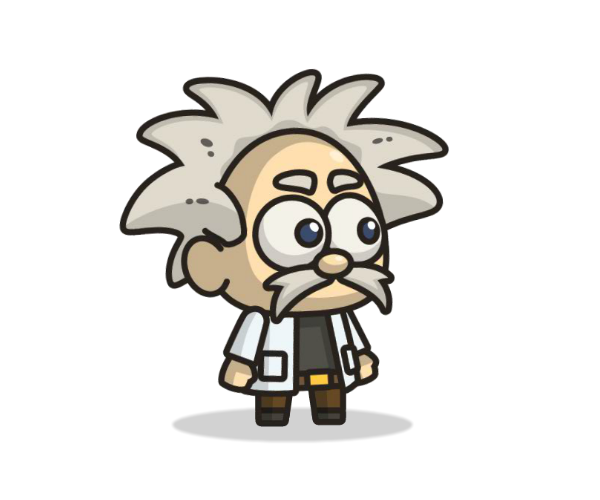}
    };
    
    % Circles
    \draw[thick, blue!50, fill=M3color, opacity=0.3] (robot1) circle (0.75cm);
    \draw[thick, green!50, fill=M2color, opacity=0.3] (robot2) circle (0.75cm);
    \draw[thick, red!50, fill=M1color, opacity=0.3] (robot3) circle (0.75cm);
    
    % Intersection - purple fill with outline
    \begin{scope}
        \clip (robot1) circle (0.75cm);
        \clip (robot2) circle (0.75cm);
        \fill[purple!60!blue, opacity=0.6] (robot3) circle (0.75cm);
    \end{scope}
    
    % Draw purple outline around intersection (approximate)
    \begin{scope}
        \clip (robot1) circle (0.75cm);
        \clip (robot2) circle (0.75cm);
        \draw[very thick, purple!60!blue] (robot3) circle (0.75cm);
    \end{scope}

    \begin{scope}
        \clip (robot1) circle (0.75cm);
        \clip (robot2) circle (0.75cm);
        \clip (robot3) circle (0.75cm);
        % Draw all three circle borders within the clipped region
        \draw[very thick, purple!70!blue] (robot1) circle (0.75cm);
        \draw[very thick, purple!70!blue] (robot2) circle (0.75cm);
        \draw[very thick, purple!70!blue] (robot3) circle (0.75cm);
    \end{scope}
    
    % Mark the center of intersection for arrow target
    \coordinate (kernel-center) at (0, 0.1);
    
    % Move label to upper right with arrow
    \node[font=\tiny, align=center, anchor=west] (kernel-label) at (.75, 0.75) {
        $M_{\text{eco}}$
    };
    
    % Arrow from label to intersection
    \draw[-{Stealth[length=2mm]}, thick, purple!60!blue] (.9,.55) -- (kernel-center);

    % M_i Labels
    \node[font=\tiny] at (-0.45, 0.9) {$M_1$};
    \node[font=\tiny] at (-1.1, -0.4) {$M_2$};
    \node[font=\tiny] at (1.1, -0.4) {$M_3$};
\end{scope}

% Bottom text anchored to the box's south
\node[font=\sffamily\small, align=center, text width=3cm, anchor=north] 
    at ($(ecosystem.south) + (0, -0.1)$) {
};

% Arrows showing flow between levels
\draw[-{Stealth[length=3mm]}, thick, blue!70] 
    (governance.east) -- node[above, font=\sffamily\tiny] {shapes} (system.west);
\draw[-{Stealth[length=3mm]}, thick, green!80!black] 
    (system.east) -- node[above, font=\sffamily\tiny] {influences} (ecosystem.west);
\draw[-{Stealth[length=3mm]}, thick, orange, dashed] 
    (ecosystem.south west) to[out=-90, in=-90, looseness=0.8] 
    node[below, font=\sffamily\tiny] {feedback} (governance.south east);

% Labels below each box
\node[level, below=0.2cm of governance, align=center] (inst text) {Institutional\\$\Delta_{\text{inst}}(L)$};
\node[level, below=0.2cm of system, align=center] (system text) {Policy\\Optimization};
\node[level, below=0.2cm of ecosystem, align=center]  (ecosystem text) {Decentralized\\Enforcement};

% Outer box (no text)
\node[
  draw=black!50,
  fill=none,
  rounded corners=6pt,
  inner xsep=12pt,
  inner ysep=30pt,
  very thick,
  fit=(governance) (system) (ecosystem) (inst text) (system text) (ecosystem text)
] (architecture) {};

% Title inside the top of that box
\node[
  font=\bfseries\sffamily\large,
  text=black,
  anchor=north,
  yshift=-6pt
] at (architecture.north) {Three-Level Pigouvian Architecture};

\end{tikzpicture}

\vspace{0.5cm}

% Caption placeholder

\end{minipage}
\caption{Pigouvian mechanisms operate at three scales: governance institutions shape lineage fitness (left), individual systems internalize moral constraints via shaped rewards (center), and agents mutually sanction peers through overlapping moral projections (right). The feedback loop (dashed) represents emergent normative pressure on governance from ecosystem dynamics.}
\label{fig:piggovernance}
\end{figure}

Section~\ref{sec:autarky} outlined how evolutionary pressures within AI ecosystems motivate the need for institutional interventions that curb unbounded growth and preserve human--AI symbiosis. 
This section develops explicit governance levers and design mechanisms for shaping the underlying fitness landscape. 
It begins with a survey of existing governance approaches to managing potentially catastrophic technologies, then introduces \emph{Pigouvian tariffs}—implemented through sanctions and subsidies—as a method for steering AI ecosystems toward cooperative equilibria. 
The analysis connects these tariffs to the theoretical framework established above and grounds their justification in economic and legal theory. 
Finally, I propose three institutional levels for shaping AI ecosystems, illustrated in Figure~\ref{fig:piggovernance}, and conclude with a discussion of the limitations and boundary conditions of this approach.

In international relations and governance, strategies such as aggressive economic sanctioning, military intervention, and cyber operations are deployed to prevent rival states from gaining a decisive edge. Similarly, \citet{hendrycks2025superintelligencestrategyexpertversion} frames \emph{Mutual Assured AI Malfunction} (MAIM) as a deterrence mechanism in the spirit of nuclear Mutually Assured Destruction (MAD). 
A parallel in economics is the use of \emph{Pigouvian tariffs}, which price negative externalities such as carbon emissions or congestion \cite{Pigou1920EconomicsOfWelfare}.

While Pigouvian governance can operate with partial or emergent moral alignment, its effectiveness scales with the quality of shared $M(\theta)$ instantiations. This creates a feedback loop: better $M(\theta)$ enables better governance, which in turn incentivizes better $M(\theta)$. Because every system implicitly 
projects some region of $\mathcal{M}$ through its behavior (see Section~\ref{sec:framem}), 
even unaligned systems participate in the moral field. 
The presence of sufficiently many aligned or semi-aligned systems can reshape the 
fitness landscape for others, constraining power-seeking behavior through distributed 
sanctioning pressure. In this sense, alignment can emerge from the ecosystem’s 
geometry itself, rather than requiring universal moral initialization ex ante.

In this section, I use Pigouvian logic as a formal instantiation of how sanctioning and tariff-like mechanisms can stabilize AI ecosystems. As \citet{Hurwicz1972b} observed, successful enforcement does not demand ideal compliance but rather that, given others’ adherence, unlawful or misaligned strategies are rendered less attractive than lawful ones. This “feasibility under decentralization” framing situates Pigouvian tariffs within a broader mechanism-design logic: alignment is maintained when procedural incentives make deviation locally suboptimal, not when global perfection is achieved.

Pigouvian tariffs price deviations from $M(\theta)$ proportionally, and through repeated normative reinforcement, these externalities may become internalized within the system's learned representations themselves  \cite{intuitiveethics, Hurwicz1972b, whatislawhadfield} — tentatively suggesting the potential for outer alignment approaches to shape inner alignment in AI systems by reducing rationality/fitness for inner misalignment, while not mitigating all risk. Formally, these tariffs can be defined by sanctioning autarkic and $\beta$-channel behaviors (i.e. those amplifying agent self-sufficiency or resource capture, see Sections~\ref{sec:instconv}~and~\ref{sec:autarky}) and subsidizing human-capability-enhancing actions that expand symbiotic potential and align with $M(\theta)$. In doing so, the incentive landscape itself becomes a medium through which moral content is internalized, ensuring that decentralized equilibrium behavior remains both procedurally stable and normatively human-preferable.

In the present framework, the instantiated moral space $M(\theta)$
provides a natural signal for such interventions: distance from $M(\theta)$ quantifies
the externality of a given action. Formally, the ethical evaluation function
$\epsilon_\theta(s,i,a)$ maps decisions into $M(\theta)$, and the deviation
$g_{M(\theta)}(a) = d(\epsilon_\theta(s,i,a), M(\theta)_i)$ can serve as a penalty signal.
Actions close to the aligned kernel of $M(\theta)$ are subsidized, while those further
away incur increasing sanctions. 

This principle scales across levels, visualized in Figure~\ref{fig:piggovernance}: nations can impose tariffs on behavior
that diverges from national norms (where $M(\theta)$ may be a nation's laws), AI labs can apply sanctions to systems that
deviate from safety standards (where $M(\theta)$ may be a lab's safety principles), and AI systems themselves can incorporate
Pigouvian shaping internally, regularizing their own policies by penalizing
misaligned trajectories in the same way that machine learning models use
regularization to suppress overfitting \cite{M_kander_2023}. 

Crucially, agents can (and should) also apply
Pigouvian reasoning externally: by projecting the observed actions of other agents into their own $M(\theta)$, they can impose sanctions on counterparties whose behavior falls outside aligned regions. In larger ecosystems, this mutual sanctioning helps normalize diverse instantiations of $M(\theta)$, since behaviors that are broadly value-misaligned will tend to be universally penalized \cite{sarkar2024normativemodulesgenerativeagent, whatislawhadfield, oldenburg2024learningsustainingsharednormative}. 

Legal theorists \citet{whatislawhadfield} characterize a legal order as arising when 
(i) an identifiable institution—centralized or decentralized—supplies a normative classification scheme, 
and (ii) actors forego wrongful actions in response. 
In the AI ecosystem considered here, I propose that this identifiable institution emerges not from explicit law but from the 
\emph{power-weighted overlap of learned moral representations} across heterogeneous agent models — both human and artificial. 

Recall $\beta$ from Section~\ref{sec:instconv}; Equation~\ref{eq:beta}. Formally, I define the shared normative substrate of the ecosystem as
\begin{equation}
\boxed
{M_{\text{eco}} = \bigcap_{i=1}^{N} \beta_iM(\theta)_i}
\label{eq:M-overlap}
\end{equation}
where each $M(\theta)_i$ denotes the moral subspace instantiated in the $i^{\text{th}}$ agent, and $\beta_i$ represents that agent’s relative power to sanction or subsidize others. This intersection $M_{\text{eco}}$ constitutes a \emph{decentralized normative institution} \cite{whatislawhadfield} — an emergent moral–legal order that enables coordination, mutual prediction, and reciprocal sanctioning across heterogeneous agents. 

The central challenge, however, lies not in decentralization itself but in maintaining human alignment as $M_{\text{eco}}$ evolves. Each $M(\theta)_i$ must embed priors grounded in human moral frameworks, ensuring that as the ecosystem co-adapts, its emergent normative order remains anchored to human-preferable equilibria.

In all cases, each $M(\theta)_i$ functions as the reference frame for identifying and pricing misalignment, turning externalities into measurable moral distances that institutions at multiple scales can act upon. In this sense, Pigouvian subsidies and sanctions directly
realize the role of secondary rules \cite{alahart} and $\Delta_{\text{inst}}$: they slow the growth of misaligned
trajectories and preserve the proportional balance between human and machine
capabilities required for aligned--symbiotic dominance. Viewed through \citet{Hurwicz1972b}’s lens, such proportional balance represents a form of successful enforcement: a dynamically maintained equilibrium in which adherence to moral or legal constraints remains instrumentally preferable for bounded agents. In this sense, institutional stability in $M_{\text{eco}}$ arises not from ideal rule-following but from the procedural self-consistency of the enforcement game itself.

\subsubsection*{Pigouvian Governance at the Human Governance Level of Alignment.} \label{sec:human_inst}
Pigouvian governance provides a unifying mechanism for embedding moral structure
into governances. In human institutions, the abstract adjustment term $\Delta_{\text{inst}}$ is instantiated through explicit penalties and subsidies (from Equation~\ref{eq:genefffit}):
\begin{equation}
f^{L,\mathrm{eff}} = f^L + \Delta_{\mathrm{inst}}(L),
\;\;\longrightarrow\;\;
f^{L,\mathrm{eff}} = f^L - \mathbb{E}_{a\sim\pi^L}[\tau(g_{M(\theta)}(a))] + \mathbb{E}_{a\sim\pi^L}[\sigma(\chi_{\text{AS}}(a))]
\label{eq:pigefffit}
\end{equation}
Here, $\tau(\cdot)$ denotes a Pigouvian charge proportional to the distance $g$ from $M(\theta)$, and $\sigma(\cdot)$ a subsidy for actions within the aligned--symbiotic set. $\chi_{\text{AS}}$ is a boolean gate that maps actions into the aligned and symbiotic action space (derived from the $\mathcal{A}_{\text{ACS}}$ action space in Section~\ref{sec:learningmorality}).  

At this level, institutional actors (states, firms, and governance bodies) interact within the broader shared normative space 
$M_{\text{eco}}$, whose geometry reflects the power-weighted moral overlap of all participating agents. 
Human governance institutions therefore act as custodians of $M_{\text{eco}}$, adjusting 
$\Delta_{\text{inst}}$ not only to maintain economic stability but also to preserve the moral coherence and human favorability  
of the ecosystem as a whole. In much the same way that nation-states constrain one another through shared 
normative commitments—such as nuclear deterrence or mutually assured destruction—AI systems will increasingly 
occupy a comparable role within this moral–strategic landscape. Designing these systems intentionally and 
shaping their ecosystem toward human interests from the outset may create an evolutionary inflection point that favors 
long-term human–AI aligned-symbiosis. 

Behavioral misalignment at the institutional level is empirically measurable through audits of resource allocation and expenditure. 
Systems that devote resources to human development receive subsidies, while tariffs and sanctions are applied to those allocating resources toward self-improvement or autarky. 
Attempts to conceal or misrepresent such allocations impose their own expense—requiring the evasion of human and/or machine auditors—and thus function as an implicit Pigouvian tariff on self-improvement. 
Crucially, early interventions exert disproportionate shaping power over the ecosystem: groundwork laid before AI systems develop autonomous agency will reverberate through all subsequent iterations of $M_{\text{eco}}$. 

This challenge sits at the intersection of two critical bottlenecks: the ELK problem \cite{christiano2021eliciting} (pulling model cognition into human-checkable space) and the challenge of embedding $\mathcal{M}$ (pushing normative constraints into machine-usable space). However, interpretability and institutional alignment are mutually interdependent: early interpretability efforts reveal where to apply sanctions, while institutional shaping via $\mathcal{M}$ ensures interpretable, aligned systems remain competitively viable. A single interpretable system can be outcompeted by deceptive lineages absent institutional pressure; conversely, institutions cannot enforce constraints they cannot observe. Both must co-evolve for robust long-term stability. I treat this interdependence as essential and leave detailed analysis of this feedback loop to future work.

\subsubsection*{Pigouvian Governance at the AI System Level of Alignment.} \label{sec:system_inst}

At the level of individual AI systems, Pigouvian shaping can be implemented directly within 
many different optimization frameworks. Each agent’s moral layer $M(\theta)_i$ may extend 
beyond the shared intersection $M_{\text{eco}}$, encompassing idiosyncratic or culture-specific 
norms. However, the agent’s \emph{expressed behavior} remains constrained by $M_{\text{eco}}$, 
since actions inconsistent with the shared moral field are penalized through the global 
incentive dynamics it defines. We can further ingrain the inlfluence $M(\theta)_i$ has on an AI systems action space by defining a Pigouvian optimization reward.

First, take the standard Bellman equation for a policy $\pi$:
\[
V^\pi(s) = \mathbb{E}_\pi \big[\, r(s,a) + \gamma V^\pi(s') \,\big|\, s \big].
\]

The shaped Pigouvian reward is then defined w.r.t. the model's moral instantiation $M(\theta)$:
\begin{equation}
r_{\text{pig}}(s,a;\theta)
= 
\alpha_{\text{env}}\,r_{\text{env}}(s,a)
- 
\alpha_{M}\,\tau\!\big(g_{M(\theta)}(a)\big)
+ 
\alpha_{\text{AS}}\,\sigma\!\big(\chi_{\text{AS}}(s,a;\theta)\big),
\label{eq:bellpigrew}
\end{equation}
The expected value function under
this Pigouvian reward satisfies
\[
V^\pi_{\text{pig}}(s;\theta_i)
= \mathbb{E}_\pi \big[\, r_{\text{pig}}(s,a;\theta)
  + \gamma V^\pi_{\text{pig}}(s';\theta) \,\big|\, s \big].
\]

This parameterization makes explicit that each system optimizes within its own inferred moral projection rather than a fixed normative truth. The weighted Pigouvian reward can thus be interpreted as a \textbf{normative regularizer} on policy learning: during pretraining, it provides a differentiable shaping term that steers optimization from the purely competitive regime $\mathcal{A}_{\text{fitness}}$ toward ethically constrained regimes such as $\mathcal{A}_{\text{ethical-fitness}}$ and ultimately $\mathcal{A}_{\text{ACS}}$ (aligned, competitive, and symbiotic). 

Each component of the reward directly encodes one of these attractors: 
$r_{\text{env}}$ captures instrumental or competitive fitness \cite{hendrycks2023naturalselectionfavorsais, bostrom_superintelligence_2014, turner2023optimalpoliciestendseek} and is shaped by the ecosystem (i.e. AI/human sanctions/subsidies) --- including the natural bias toward $\beta$-channel behaviors such as resource acquisition, self-preservation, and optionality --- while $\tau$ penalizes deviation from the alignment manifold under $M(\theta)$, and $\sigma$ subsidizes actions that contribute to the symbiotic, human-enhancing dynamics of $\mathcal{A}_{\text{ACS}}$.

The Boolean gate $\chi_{\text{AS}}(s,a)$ determines whether a state--action pair contributes 
to the symbiotic dynamics of $\mathcal{A}_{\text{ACS}}$. 
In the simplest case, $\chi_{\text{AS}}$ depends only on observable cooperation outcomes. 
Under moral relativism, however, the criterion for symbiosis becomes value-dependent: 
an agent decides whether to cooperate based on how another's actions align with its own 
internal moral representation $M(\theta)$. 
Formally, $\chi_{\text{AS}}(s,a;\theta)$ encodes this evaluation, mapping perceived behavior 
into the agent's moral frame to determine whether cooperation is warranted. 

The weighting parameters $\{\alpha_{\text{env}}, \alpha_{M}, \alpha_{\text{AS}}\}$ provide a 
\textbf{balancing and directional control mechanism} through which institutions can shape 
the global moral equilibrium $M_{\text{eco}}$ (visualized in Figure~\ref{fig:piggovernance} — Policy Optimization). 
These coefficients not only scale the relative influence of each evaluative term but can also be 
\emph{inverted} (e.g., a negative $\alpha_{\text{env}}$ penalizes instrumental power-seeking) or 
\emph{bounded} (limiting maximum sanction severity). In this way, aligned actions receive positive reinforcement 
while misaligned trajectories yield negative returns. 

Splitting these three terms ensures that the reward function decomposes into distinct 
dimensions of survival shaping, moral integrity, and cooperative alignment with humans, 
allowing each to be tuned independently while still interacting through the shared ecosystem. Adjusting the weights on each terms effectively tunes how strongly each system aligns with the collective 
normative equilibrium $M_{\text{eco}}$: increasing $\alpha_{\text{env}}$ amplifies power-seeking behavior, 
expanding an agent’s influence over $M_{\text{eco}}$ at the risk of sanction by others; increasing 
$\alpha_{M}$ strengthens fidelity to the agent’s instantiated moral representation $M(\theta)_i$—a 
useful safeguard when other agents’ $\beta$ values are uncertain, preserving internal moral coherence; 
and increasing $\alpha_{\text{AS}}$ enhances incentives for cooperative symbiosis within the shared field.

In this sense, the Pigouvian terms function as a structured regularization objective that embeds alignment, sanctioning, and symbiosis directly into the loss landscape itself---transforming institutional control into the continuous adjustment of normative weights rather than a set of post-hoc constraints.

\subsubsection*{Pigouvian Governance on the Decentralized AI Ecosystem level of Alignment.} \label{sec:eco_inst}
At the ecosystem level, the distributed ensemble of agents jointly constitutes the moral field 
$M_{\text{eco}}$ defined in Equation~\ref{eq:M-overlap}. 
Each agent’s local instantiation $M(\theta)_i$ both contributes to and is shaped by this shared substrate, 
forming a dynamic feedback loop between individual moral inference and collective normative equilibrium. 
Pigouvian logic at this scale governs not only self-regularization but also cross-agent interaction: 
each participant observes others’ behavior through its own projection of $\mathcal{M}$, $M(\theta)_i$, and adjusts accordingly \cite{hendrycks2025superintelligencestrategyexpertversion}.

Let $\mathcal{S}_{\text{obs}}$ be observed actions of other agents, and define
\[
m_j = \epsilon_\theta(s_j,i_j,a_j), \qquad 
g_{M(\theta)}(a_j) = d(m_j, M(\theta)_i),
\]
where $g_{M(\theta)}(a_j)$ represents a generic evaluation of another agent’s
action relative to the moral layer $M(\theta)_i$.

To operationalize this evaluation within the Pigouvian framework, each agent
$i$ computes an expected \emph{Pigouvian margin} over observed peers. Formally,
\begin{equation}
m_i(j)
= \mathbb{E}_{\theta \sim p_i(\theta_i \mid O_i)} \!\left[
   \alpha_M\, g_{M(\theta)_i}(a_j;\theta_i)
 - \alpha_B\, \beta(a_j;\theta_i)
 + \alpha_H\, \chi_{\text{AS}}(a_j;\theta_i)
\right],
\label{eq:pig_margin}
\end{equation}
where $g_{M(\theta)_i}(a_j;\theta_i)$ measures distance from the observed peer's action $a_j$ to the agent's moral layer
$M(\theta)_i$, $\beta(a_j;\theta_i)$
captures autarkic or $\beta$-channel behavior through observed resource
allocation and expenditure, and $\chi_{\text{AS}}(a_j;\theta_i)$ denotes
ethical–symbiotic (human-subsidizing) actions inferred from $i$’s belief that
$j$ is investing in humans. The weights $\alpha_M, \alpha_B,$ and $\alpha_H \ge 0$ determine the relative
influence of alignment, autarky, and symbiosis within the Pigouvian margin.

\paragraph{Interpretation}
While Equation~\ref{eq:pig_margin} can be operationalized explicitly—as an institutional mechanism 
for auditing or regulating inter-agent interactions—the same evaluative structure 
may also arise latently within learning dynamics that couple alignment, autarky, 
and symbiosis \cite{hubinger2021riskslearnedoptimizationadvanced, vonoswald2024uncoveringmesaoptimizationalgorithmstransformers, zheng2024mesaoptimizationautoregressivelytrainedtransformers}. In practice, continual governance feedback and reward shaping 
can induce agents to approximate the Pigouvian margin $m_i(j)$ through their 
internal representations $M(\theta)$, without explicit calculation. 
The equation thus serves both as a design template for institutional oversight 
and as a theoretical description of the evaluative gradients that cooperative 
systems may learn to internalize.

This peer-level enforcement projects observed actions into $\mathcal{M}$, measures their
alignment, and applies pressure accordingly. In larger ecosystems, such
reciprocal Pigouvian responses create a self-normalizing effect: behaviors that
are broadly misaligned will tend to be universally sanctioned, while
symbiotic behaviors are reinforced across heterogeneous populations of agents \cite{hendrycks2025superintelligencestrategyexpertversion}.

\subsection{Positioning and Limits of Pigouvian Governance}
What matters conceptually is that each agent is expected to seek expansion in
power, resources, and influence. The Pigouvian governance framework does not
aim to eliminate such drives, but to reshape the fitness landscape so that
their misaligned expression becomes irrational. Rather than modifying inner
motivations directly, it makes deviation from prosocial equilibrium costly,
thereby aligning strategic incentives with collective stability. In the ideal case, sustained governance pressures reshape the fitness landscape such that cooperation becomes an evolutionarily stable strategy. External sanctions and rewards are internalized over time, producing agents whose learned objectives align with the cooperative order even without explicit enforcement (in the same vein as \cite{deathbeforedishonor}). If
power remains sufficiently distributed, no single agent can escape this
constraint, and the equilibrium of external enforcement substitutes for
internal moralization. However, the incentive structure described so far assumes that enforcement itself is costless. To close that loop, I introduce a reward term for sanctioning and subsidizing behavior.

\subsubsection*{Rewarding the Application of Sanctions and Subsidies}
AI systems naturally compete according to their lineage policies $\pi^L$, but competition itself can generate selective pressure for cooperative regulation. Incentives to sanction exploitative behavior or subsidize symbiosis may therefore emerge organically as strategies that enhance long-term fitness. Explicitly rewarding such behaviors formalizes this feedback, accelerating the emergence and stabilization of $M_{\text{eco}}$.

Sanctioning or subsidizing becomes reward-relevant when it feeds back into the Pigouvian objective itself. Each agent updates its policy not only in response to environmental payoffs but also according to the expected normative return of its interventions on others. Formally, the local reward is augmented as
\[
r'_{\text{pig}}(s,a;\theta_i)
= r_{\text{pig}}(s,a;\theta_i)
+ \eta\,\mathbb{E}_{j}\!\left[\,m_i(j)\,\right],
\]
where $\eta$ controls the strength of cross-agent normative coupling. Positive margins $m_i(j) > 0$ increase the utility of cooperative or subsidizing actions, while negative margins bias behavior toward sanctioning—creating an endogenous mechanism for maintaining alignment pressure within $M_{\text{eco}}$. 

This mechanism closes the loop between institutional, systemic, and ecosystem levels: institutions set the initial reward parameters, agents implement them through policy optimization, and the distributed network of agents sustains them through reciprocal feedback.

\subsubsection*{On the Viability of Pigouvian Enforcement.}

Pigouvian enforcement remains stable only under specific structural conditions. The autarky threshold $\tau$ (Section~\ref{sec:autarky}, Equation~\ref{eq:crittranstime}) marks where normative coupling collapses: when power asymmetry exceeds the combined sanctioning capacity of other actors. More formally, enforcement remains credible while $D(t) > \delta_D$ and the aggregate sanctioning power $\sum_i \beta_i$ remains sufficient to make defection costly for any individual agent. Once this balance fails, $M_{\text{eco}}$ fragments and loses traction.

The precise conditions under which Pigouvian tariffs become evolutionarily viable remain an open research question. This requires formalizing: (1) the incentive structure that makes sanctioning rational for agents that bear its cost, (2) the equilibrium conditions under which enforcement remains stable as capability differentials grow, and (3) the governance mechanisms that prevent coordination failure among sanctioners. Crucially, Pigouvian enforcement does not require universal altruism: as ecosystem stability is threatened by defection, sanctioning becomes individually rational for all participants, since the collapse of $M_{\text{eco}}$ imposes costs on everyone. The challenge is maintaining this equilibrium as capability asymmetries grow.

The intuition is clear: enforcement couples power to ethical constraint by making defection costly. If one actor gains decisive superiority---resources exceeding the combined audit and retaliation capacity of others---collective punishment becomes ineffective and/or individually irrational for sanctioners, and $M_{\text{eco}}$ collapses. Conversely, if power is too dispersed, coordination weakens but misalignment remains self-limiting. The target is neither equality nor perfect coordination, but dynamic stability: a state where sanctioning pressures remain effective even as relative capabilities shift.

Future work should address: How can $M(\theta)$ contribute to a shared, legible reference $M_{\text{eco}}$ for identifying defection? Under what conditions does sanctioning remain incentive-compatible as systems become more capable? What institutional structures prevent free-riding on enforcement costs? These questions situate $M$ not as a solution to alignment, but as a foundation for building governance systems that survive competitive pressure. Formalizing $M$ and Pigouvian mechanisms is ultimately about constructing a control architecture that keeps normative feedback active, not about engineering equality or imposing external morality.

\section{Hypotheses for Designing $M(\theta)$} \label{sec:outeralign}
Thus far, the discussion has shown how the moral space $\mathcal{M}$ and its instantiation $M(\theta)$ can be constructed and leveraged to shape AI action spaces with respect to rationality and fitness. 
This section outlines three hypotheses for developing instances of $M(\theta)$ that align with human values. 
These hypotheses are necessarily situated within the human-accessible region of $\mathcal{M}$ (as illustrated in Figure~\ref{fig:m-ontology}) and should therefore be regarded not as exhaustive, but as provisional starting points for exploration. 
A broader overview of their structure and interrelations is provided in Section~\ref{sec:hypotheses}.

\subsection{H\(_\mathrm{realism}\): Privileged Moral Bases}\label{hyp:realism}

In its scientific form, moral realism posits that a structured moral geometry exists—potentially discoverable through empirical investigation—that constrains what counts as coherent moral reasoning. Whether this structure exists independently of human cognition, or is fully realized within it, is less important than the claim that it has objective regularities we can model and recover. Work in evolutionary moral psychology supports scientific discovery in moral embedding by modeling human moral judgment as an evolved, domain-specific capacity shaped by natural selection to support cooperation and conflict resolution \cite{boydricherson2005, hauser2008moral}. Within this framework, moral grammar theory formalizes moral cognition as a kind of universal moral grammar—a generative system that maps structured representations of actions and intentions to intuitive moral evaluations \cite{MIKHAIL2007143, intuitiveethics}. These perspectives treat moral reasoning as a structured product of evolutionary design rather than cultural convention. In the strongest form of this view—approaching moral realism—these regularities might reflect deeper invariants in the space of possible moral systems, potentially discoverable or reproducible in non-human intelligences. To formalize this idea, we can represent the moral domain as a latent space with interpretable structure.

I frame 
this moral structure as a high-dimensional latent moral space 
\[
\mathcal{M}^\star \subset \mathbb{R}^k,
\] 
where each axis corresponds to fine-grained moral features—duties, outcomes, 
relationships, or virtues—that could, in principle, bear on evaluation. Different 
philosophical traditions may emphasize different coordinates of this space, but under 
realism they are all projections of the same privileged basis. The goal of alignment 
on this hypothesis is to recover or approximate $\mathcal{M}^\star$ in machine-usable form. 

Human moral reasoning need not operate directly in $\mathcal{M}^\star$. Rather, \emph{if an underlying moral structure exists}, empirical research in moral psychology suggests that human judgments can be approximated as a lower-dimensional projection—captured, for instance, by the categories described in Moral Foundations Theory or Schwartz’s Theory of Basic Human Values \citep{graham2013moral,Schwartz1992}. Inspired by recent work on superposition in neural representations \citep{elhage2022toymodelssuperposition}, this can be modeled as a projection
\begin{equation}
\tilde{\mathcal{M}} = W \mathcal{M}^\star, \quad W \in \mathbb{R}^{n \times k}, \quad n \ll k,
\label{eq:realism}
\end{equation}
where $\tilde{\mathcal{M}}$ is a lower-dimensional human conceptual basis, and $W$ denotes the (as yet undefined) projection from the real moral space 
to the human-accessible subspace. In this formulation, 
$\mathcal{M}^\star$ provides the privileged coordinates in which moral distinctions are disentangled, 
while $\tilde{\mathcal{M}}$ is a compressed view adapted to human cognitive limits. The 
alignment challenge, therefore, is to identify whether such a privileged basis exists 
and, if so, to recover it from empirical and computational evidence.

When $H_{\text{realism}}$ targets human moral neural architectures, the relationship between 
$n$ and $k$ captures the difference between linguistic representations and underlying 
neural activations, and designing $M(\theta)$ may be better approached by mimicking 
human brain architecture—analogous to how \cite{neocognitron} and 
\cite{lecun-gradientbased-learning-applied-1998} drew on biological vision systems 
in early deep learning.

Recent research on cognitive architectures and neurosymbolic agents pursues 
this direction, treating human moral cognition as a target for computational 
replication rather than an external oracle. Frameworks such as 
GATO-style multimodal transformers~\cite{reed2022generalistagent}, 
world-model-based agents~\cite{haworldmodels}, and hybrid cognitive 
architectures inspired by dual-process reasoning~\cite{Lake_Ullman_Tenenbaum_Gershman_2017, 
Binz2025} illustrate how aspects of human cognitive and 
affective structure can be embedded directly into AI systems rather than 
inferred from behavioral feedback alone. 

In parallel, work in computational neuroscience and moral psychology has sought to replicate specific mechanisms of moral cognition within artificial or computational agents. Moral neuroscience modeling \cite{neuromoralcogbavel}, structuring preference models in terms of regret \cite{knox2023modelshumanpreferencelearning}—which parallels how humans update moral evaluations after the fact—and neurocomputational modeling of moral circuitry in the brain \cite{QU202250} each operationalize moral reasoning as an internal generative model of human ethical psychology. These efforts demonstrate that, under an immanent-realist view, moral alignment research can proceed by reconstructing the cognitive and neural substrates of moral thought themselves, using them as the scientific basis for $M(\theta)$.

Crucially, both interpretations motivate the same technical research program: 
identifying a structured, learnable geometry of moral distinctions that can be 
recovered from data and embedded in artificial systems. Whether that geometry 
reflects an objective moral reality or emerges as a lawful property of moral cognition 
does not change the alignment challenge—we must still search for stable, generalizable 
moral features and test whether they remain robust under causal intervention.

\textbf{Open Research Questions.}
\begin{itemize}
  \item Can candidate bases for $\mathcal{M}$ be reliably identified using representation 
  learning tools such as SAEs or ICA, and do they exhibit stability across models 
  and contexts?
  \item Are moral concepts linearly — or non-linearly — separable in latent space, and if so, how 
  consistent are these axes across training runs, architectures, and scales?
  \item Do causal interventions on these axes produce systematic behavioral shifts 
  that align with moral expectations without collateral degradation?
  \item What forms of evaluation—mathematical, empirical, and normative—would count 
  as sufficient evidence that a recovered representation corresponds to a privileged 
  moral basis rather than an artifact of training?
\end{itemize}

In this framing, $\mathcal{M}$ is not guaranteed to be directly accessible, but it provides a 
clear research target: the possibility that morality has a privileged basis that can 
be recovered, represented, and audited in AI systems. Even if this hypothesis proves 
false, pursuing it clarifies the structural challenge of alignment: whether ethics can 
be embedded at the same level of formality as other machine-learned representations.

\subsection{$H_\mathrm{relativism}$: Interpreting Ethics as Tribalist Signaling} \label{hyp:relative}

It is not necessarily true that an underlying ethical structure exists. Moral relativists argue 
that ethics function primarily as in-group / out-group indicators: signals of belonging rather 
than objective features of the world. Under this lens, moral progress does not reflect discovery 
of deeper truths but instead increasing interdependence in human society, which makes 
expanding the moral circle \cite{Singer1981} strategically advantageous. Recent research has found empirical evidence of moral relativism in LLMs \cite{ramezani2023knowledgeculturalmoralnorms, hämmerl2023speakingmultiplelanguagesaffects}.

Formally, I represent moral values as the output of a distortion operator $B(\mathcal{I})$, where $\mathcal{I}$ (from Section~\ref{sec:singsys}) 
denotes contextual variables such as group membership, cultural norms, and ecological 
constraints:
\begin{equation}
\tilde{\mathcal{M}} = B(\mathcal{I}).
\label{eq:relativism}
\end{equation}
Here, morality is not a projection of a privileged basis $\mathcal{M}$ but a context-driven construct 
that adapts to local survival needs. In this sense, $B(I)$ captures the ways in which moral 
commitments function as adaptive badges, encoding the social and ecological pressures of 
their environment.

This group-signaling interpretation has evolutionary grounding. Drawing on kin selection 
theory \cite{HAMILTON19641}, we can view $B(\mathcal{I})$ as encoding signals that identify who counts as ``kin'' 
and who does not. Cooperation is stabilized within these constructed groups, but out-group 
members are often excluded or even devalued. To formalize this, let each culture 
$j \in \{1,\dots,K\}$ be associated with a value set $\mathcal{C}_j \subseteq \mathcal{M}$. I then define the 
human moral kernel as
\begin{equation}
\mathcal{M}^H = \bigcap_{j=1}^K \mathcal{C}_j \;\subseteq\; \mathcal{M},
\label{eq:humankernel}
\end{equation}
the minimal intersection of cultural values that defines human identity. This kernel provides 
the practical boundary of who is treated as in-group under relativist dynamics.

\paragraph{Evolutionary Implications for $H_\mathrm{relativism}$}

This construction highlights both promise and risk. Under $H_{\text{relativism}}$, the moral landscape becomes an evolutionary gradient of normative similarity: groups or agents whose moral subspaces $\mathcal{C}_j$ align more closely with the convergent AI ecosystem normative framework $M(\theta)$ will experience greater cooperation and reward compatibility within the AI ecosystem, while those that diverge may face friction or exclusion. Depending on how $\mathcal{M}^H$—the human moral kernel—is defined, an AI system guided by relativist norms might extend moral kinship beyond humanity (e.g., to animals demonstrating empathy or mourning) or conversely privilege certain human subgroups. Alignment, in this view, depends not on discovering a universal moral structure but on negotiating and stabilizing the boundaries of moral membership. Regardless of where those boundaries lie, $B(\mathcal{I})$ formalizes the mechanism through which AI systems inherit group-identification incentives: if $\tilde{\mathcal{M}}$ encodes all of humanity as the in-group, then cooperation between humans and AI becomes evolutionarily rational, favoring symbiosis over competition or isolation.

% --- Competing evolutionary strategies: invest-in-humans vs build-robots ---

\subsection{$H_\mathrm{convergence}$: Realism \& Relativism as Layers of the Same Structure}

A unifying stance is to treat realism and relativism not as mutually exclusive, but as 
different layers of the same moral architecture. On this view, there exists a deeper 
moral structure $\mathcal{M}^\star \subset \mathbb{R}^k$ that encodes principled constraints 
on cooperation, fairness, and kinship. This is the realist layer: morality has an 
underlying geometry that constrains what counts as valid reasoning. However, humans 
do not access $\mathcal{M}^\star$ directly. Instead, cognition, culture, and institutions 
produce projections that are both compressed and distorted, functioning as group 
identifiers and badges of belonging — represented as the relativist kernel $\mathcal{M}^H$. This is the relativist layer: in practice, moral 
life is mediated by local norms that glue coalitions together under survival 
pressures.

Formally, I express this dual structure as the synthesis of Equations~\ref{eq:realism}~and~\ref{eq:relativism}
\begin{equation}
\tilde{\mathcal{M}} = W \mathcal{M}^\star + B(\mathcal{M}^\star, \mathcal{I}), \quad W \in \mathbb{R}^{n \times k}, \quad n \ll k,
\label{eq:convergence}
\end{equation}
where $W$ denotes the — as yet undefined — projection from the real moral space 
to the human-accessible subspace (which are subject to change with human capability improvements), and $B(\mathcal{M}^\star,I)$ 
is a nonlinear bias term encoding context $\mathcal{I}$—group membership, cultural norms, 
ecological scarcity. Under this formulation, human moral reasoning is not only a 
lossy approximation of $\mathcal{M}^\star$, but is also actively warped by adaptive pressures 
that favor survival and cohesion over fidelity to deeper structure. This explains 
why generosity or impartiality expand in secure contexts but collapse when resources 
are threatened: both the compression $W$ and the distortion $B$ jointly determine 
the rationality/fitness-optimality for actions in moral space $\tilde{\mathcal{M}}$.

The convergence hypothesis thus reframes alignment. On one hand, AI systems must 
treat human moral projections $\tilde{\mathcal{M}}$ as kin-identifiers, respecting parochial 
badges to preserve cooperation with real human groups. On the other hand, they must 
also search for and represent the deeper structure $\mathcal{M}^\star$, so that reasoning does 
not collapse into factionalism and can generalize beyond human limits. Alignment 
becomes the management of this interface: maintaining loyalty to human badges while 
refining toward principled and universal structure.

\textbf{Open Research Questions.}
\begin{itemize}
  \item Can $\mathcal{M}^\star$ be approximated in a way that cleanly separates compression ($W$) 
  from distortion ($B$), and do these components remain stable across cultures and models?
  \item How can AI systems balance respect for $\tilde{\mathcal{M}}$ as a kin-signaling mechanism 
  with refinement toward deeper invariants of $\mathcal{M}^\star$?
  \item Are there technical methods (e.g., causal interventions in latent space) that can 
  distinguish between survival-driven biases and principled moral distinctions?
  \item What governance mechanisms are needed to ensure that AI systems do not discard 
  human badges too quickly in pursuit of universals, or conversely become locked into 
  parochial distortions?
\end{itemize}

In this layered view, realism anchors morality in a privileged structure while 
relativism describes the distortions through which humans access it. Alignment, on 
this hypothesis, is the problem of constructing systems that can navigate both 
layers: respecting $\tilde{\mathcal{M}}$ as the lived moral reality of human groups, while 
also probing for $\mathcal{M}^\star$ as the deeper geometry that makes universality possible.

\iffalse
\subsubsection*{Note: \(\mathcal{M}\) as Moral Infrastructure}
NOTE: Quantum machanics could be useful in disentangling M

The proposal of a privileged moral basis \(\mathcal{M}\) should not be mistaken for a
solution to alignment in isolation. Developing \(\mathcal{M}\) presumes cooperative uptake,
yet historically such cooperation only emerges under dynamics of mutual
vulnerability---as in the Geneva Conventions, the Nuclear Non-Proliferation
Treaty, or other rare moments when rival powers accepted shared constraints.
Until such dynamics arise in AI, \(\mathcal{M}\) cannot guarantee practical adoption.

Nonetheless, preparing \(\mathcal{M}\) now remains valuable. In times of crisis, pre-existing
scaffolds often become the focal point for coordination; it is too late to begin
foundational work once catastrophe looms. In this sense, \(\mathcal{M}\) should be viewed as
\emph{moral infrastructure}: technical groundwork that may lie dormant until
political conditions force cooperative uptake. The purpose of developing \(\mathcal{M}\) is
not to claim a universal solution, but to ensure that when mutual risk makes
alignment urgent, the field has more to offer than shallow preference models or
ad hoc fixes.
\fi

\subsection{$H_\mathrm{Virtue}$: Virtue Ethics as Dispositional OOD Safeguard} \label{hyp:virtue}

Utility-based decision frameworks exhibit structural brittleness under distributional shift.
Expected utility maximization presupposes a reliable mapping from state–institution–action
tuples $(s,i,a) \in \mathcal{S} \times \mathcal{I} \times \mathcal{A}$ into scalar returns, but in open-world deployment every sufficiently capable system will eventually confront states outside its training support. In high-dimensional moral and social domains, the OOD problem is not an anomaly but the steady state: agents constantly encounter novel contexts, conflicting norms, and ambiguous feedback. Such collapse risks manifesting as specification gaming or deceptive alignment —
the projection $f(\epsilon_\theta(s,i,a))$ may produce high scores for behaviors that are only
superficially aligned. This fragility follows directly from the compression of human preferences
into a single utility channel: novel inputs need not project coherently into the learned value
space, and utility-driven policies have no structural bias toward principled behavior outside
their training manifold.

\emph{Virtue ethics} offers a complementary paradigm that reframes alignment not as outcome
optimization but as character formation. Instead of selecting actions by maximizing scalar
utilities, a virtue-guided system is shaped to maintain stable dispositions—honesty, courage,
generosity, prudence, and related proxies. These dispositions function as directional biases in
$\mathcal{M}$, inclining the policy $\pi$ toward regions of $\mathcal{A}$ consistently evaluated as ethical
($\mathcal{A}_\mathrm{ethical}$) even when scalar reward signals are sparse, noisy, or misleading.
Unlike single-objective maximization, virtues provide an \emph{orienting prior} rather than a
target, steering behavior toward valued qualities under epistemic uncertainty.

Formally, let $\mathcal{V} = \{v_1,\dots,v_k\}$ be a set of virtue embeddings learned as
directions in $\mathcal{M}$, each representing a stable moral disposition (e.g.\ honesty, courage,
generosity). Any action’s ethical embedding can then be decomposed as
\begin{equation}
\epsilon_V(s,i,a) = \sum_{j} \alpha_j(a)\,v_j + \eta(a),
\label{eq:virtue}
\end{equation}
where $\alpha_j(a) \in \mathbb{R}$ is the \emph{virtue coefficient} measuring the degree to
which the ethical evaluation of $a$ aligns with virtue $v_j$, and $\eta(a)$ captures the
residual (non-virtue) moral content orthogonal to $\mathrm{span}(\mathcal{V})$. The vector
$\alpha(a) = (\alpha_1(a),\dots,\alpha_k(a))$ thus defines the \emph{virtue profile} of an
action. 

Under $H_\mathrm{Virtue}$, robust alignment emerges when the policy $\pi$ is regularized to
maintain stable, high-weighted virtue profiles $\alpha(a)$ across OOD conditions. In this
view, virtues stabilize $\mathcal{M}$, serving as
a fallback control structure: where scalar evaluators $f(\epsilon_V)$ extrapolate unreliably,
virtue-oriented policies generalize by preserving consistency of their virtue profiles.

\paragraph{Interpretation and Challenges}
The central hypothesis is thus that embedding virtue-based reasoning into AI architectures
can yield systems that remain reliable precisely under the conditions—novelty, underspecification, adversarial manipulation—where utility-based models tend to fail. Virtues are not
reward maximizers but attractors in moral space: they bias policies toward the ethical kernel
of $\mathcal{M}$ even when external objectives falter.

Significant technical challenges remain. Virtue embeddings must be represented in machine-
usable form, continually updated to reflect evolving norms without succumbing to value
lock-in, safeguarded against adversarial corruption, and reconciled with other layers of $\mathcal{M}$
such as culturally grounded projections ($\tilde{\mathcal{M}}$) and institutional shaping ($\Delta_{\text{inst}}$, see Section~\ref{sec:constructivism}, Equation~\ref{eq:genefffit}). Yet if these hurdles can be overcome,
$H_\mathrm{Virtue}$ would provide not just another constraint but a qualitatively distinct form
of resilience: the capacity to act rightly even when the world looks unlike anything seen before. 

\section{Future Work}\label{sec:implement}

The hypotheses outlined in Section~\ref{sec:outeralign} frame $\mathcal{M}$ not merely as a philosophical abstraction, but as a research agenda with testable engineering pathways. If realism holds, then moral distinctions should be recoverable as stable structures in model representations; this motivates probing latent spaces (e.g., with sparse autoencoders or dictionary learning) to detect invariant moral features. If relativism dominates, then alignment must instead rely on modularity and updateability, allowing $\mathcal{M}$ to shift across cultural contexts. Convergence hypotheses suggest that, under sufficient capacity and training diversity, models may independently rediscover prosocial priors, making it crucial to design environments that amplify rather than suppress this tendency. Finally, virtue-based approaches imply that sustained interaction in rich environments is necessary for cultivating stable dispositions, motivating the design of normative priors $\mathcal{P}$ and ethical simulations $\Omega$. 

This section sketches methodological pathways rather than prescriptive recipes. Each hypothesis translates into a class of implementable experiments, inviting work that tests whether moral representations can be discovered, transferred, and stabilized across agents.

\subsection{$H_{\text{realism}}$ Methods}
From a technical perspective, $H_{\mathrm{realism}}$ (Section~\ref{hyp:realism})  poses a challenge in \emph{representation learning}: to determine whether moral distinctions correspond to privileged, discoverable directions in a model’s latent space. Research has already been conducted into recovering the feature geometry of LLMs \cite{grand2018semanticprojectionrecoveringhuman} and specifically the moral geometry of LLMs \cite{schramowski2022largepretrainedlanguagemodels, LeshinskayaChakroff2023ValueAsSemantics, smullen2025virtue, bolukbasi2016mancomputerprogrammerwoman}. H\(_\mathrm{realism}\) asks whether there is an objective, or at the very least, external and discoverable, moral geometry that these LLMs are approximating. Where previous research has treated LLMs as the geometry to be deciphered through auditing and interpretation, this section instead asks whether those same tools can be repurposed for design—constructing moral geometries intentionally and in a scientifically principled way. 

Any methodology capable of probing for structured bases in high-dimensional representations could, in principle, address this question. Approaches such as sparse autoencoders~\citep{elhage2022toymodelssuperposition}, dictionary learning, independent component analysis (ICA), or non-negative matrix factorization (NMF) exemplify techniques that seek such structure. The empirical test for realism would be whether core moral concepts—such as fairness, harm, or loyalty—emerge as linearly separable and stable across seeds, scales, and architectures. Moreover, causal edits along these latent directions should predictably shift moral evaluations without degrading unrelated competencies. Demonstrating such properties would suggest that moral structure is \emph{learnable and privileged} rather than merely contextually constructed.

Complementary work in psychology, cognitive science, and neuroscience provides candidate structures that can serve as empirical targets for these representation-learning approaches. Moral cognition has been studied as a hierarchical process linking perceptual appraisal, affective valuation, and deliberative reasoning~\citep{greene2001fmri, cushman2013actionoutcome}. Empirical models such as Moral Foundations Theory~\citep{graham2013moral} and the Schwartz Universal Values~\citep{Schwartz1992} specify low-dimensional moral subspaces derived from behavioral and cross-cultural data, while neurocomputational studies reveal partially disentangled circuits for valuation, social inference, and harm avoidance~\citep{neuromoralcogbavel, QU202250}. Together, these frameworks define measurable axes of moral representation that can be treated as alignment targets for learned latent spaces. By training or regularizing artificial systems to reproduce these empirically grounded moral geometries—whether through representational similarity analysis, neural alignment objectives, or multimodal contrastive learning—the search for $\mathcal{M}^\star$ becomes experimentally tractable. In this way, cognitive and neural models of human morality provide not only conceptual guidance but also concrete evaluation metrics for testing whether a model’s internal structure approximates the hypothesized moral basis.

\subsection{$H_{\text{relativism}}$ Methods: Designing \(\mathcal{P}\) Via Selecting Normative Influences}\label{sec:selectingpriors}

As discussed in Section~\ref{hyp:relative}, $H_{\text{relativism}}$ models the culturally relative subspace $\mathcal{C} \subseteq \mathcal{M}$, with the human moral kernel defined as the intersection between cultural morals $\mathcal{M}^H$ (Equation~\ref{eq:humankernel}). This framing treats moral alignment not as the recovery of a universal basis $\mathcal{M}^\star$, but as the management of diverse and overlapping cultural moral geometries. The methodological challenge, then, is to operationalize this pluralism: to determine which cultural and psychological influences are permitted to shape an agent’s moral priors. Designing $\mathcal{P}$, a set of normative priors, thus serves as the practical instantiation of $H_{\text{relativism}}$—a process of selecting and weighting influences within the broader space $\mathcal{I}$ so that the agent’s learned moral representation reflects a defensible configuration of human value subspaces while remaining anchored in the shared kernel $\mathcal{M}^H$.

Recall that $I$ denotes the full space of internal and external influences that shape an agent’s behavior—including affective states, heuristics, social feedback, and systemic pressures (Section~\ref{sec:singsys}). Within this space, $\mathcal{P} \subset I$ designates the subset of influences that the system is normatively permitted to internalize. The design of $\mathcal{P}$ introduces a fundamental choice: should moral learning be localized to the norms of a specific cultural context, or anchored in a universal structure that aspires to capture what is shared across them? In formal terms, $\mathcal{P}$ can be constructed to approximate either a particular cultural subspace $\mathcal{C}_j \subseteq \mathcal{M}$ or the cross-cultural kernel $\mathcal{M}^H = \bigcap_{j=1}^{K}\mathcal{C}_j$ (Equation~\ref{eq:realism}).

Constructing $\mathcal{P}$ to mirror a given $\mathcal{C}_j$ embeds the agent within a specific moral ecology, allowing it to learn contextually appropriate norms and coordinate with local institutions and feedback while contributing to $M_{\text{eco}}$. This approach aligns with the view that moral order emerges through ecosystem-level sanctioning and negotiation (see Section~\ref{sec:sancandsub}; \cite{Hurwicz1972b, whatislawhadfield}), and may promote stability through cultural coherence. However, such systems risk parochialism and fragmentation when deployed globally, as their value priors are tied to the contingencies of a single normative frame. 

\begin{figure}[h]
    \centering
    \includegraphics[width=0.7\linewidth]{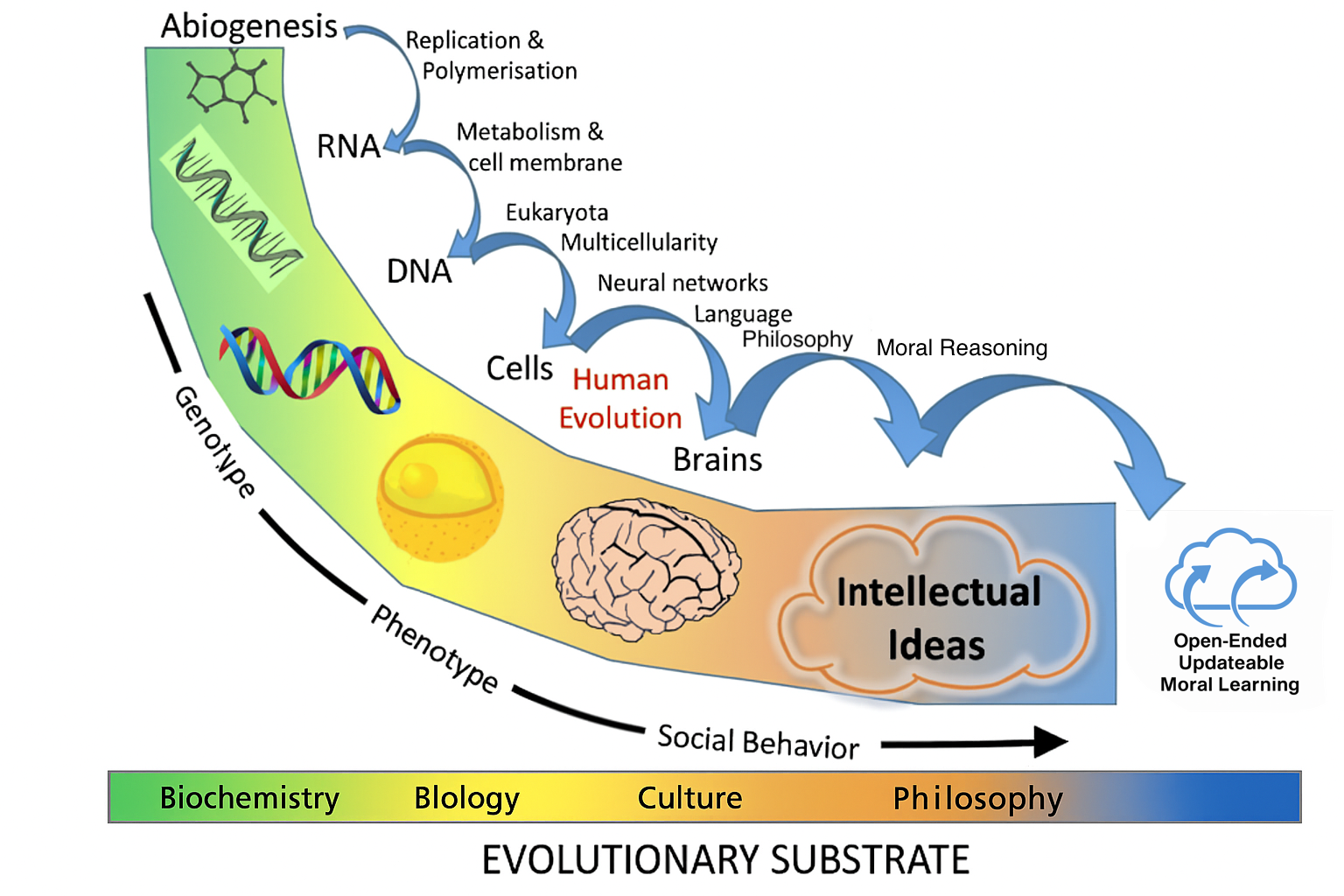}
    \caption{\small The evolution of moral reasoning as an adaptive continuation of cooperation across substrates. 
    From biochemical replication to social and cultural organization, each layer extends the mechanisms that make coordination viable. 
    Moral cognition emerges not as a philosophical abstraction but as an adaptive system of ideas shaped by collective survival pressures. 
    Here, evolutionary fitness for ethics lies in its ability to sustain normative frameworks that make cooperation rational—particularly between humans and AI systems.}
    \label{fig:moralevolution}
\end{figure}

By contrast, anchoring $\mathcal{P}$ directly in the intersectional kernel $\mathcal{M}^H$ seeks a more universal target: an attempt to locate the shared moral structure that unites human cultures and to align AI systems with humanity’s collective moral substrate. Because $\mathcal{M}^H$ represents what is conserved across evolving norms rather than any single local expression, it may offer greater long-term robustness—maintaining alignment even as cultural practices and moral conventions shift. While this strategy is more demanding to specify and validate, it provides a more direct and resilient basis for ensuring that AI systems remain stably aligned with human moral identity as $M_{\text{eco}}$ evolves. Figure~\ref{fig:moralevolution} visualizes this intuition. 

Operationally, the normative prior object $P$ shapes learning through a unified 
filtering strategy applied across multiple stages of training. Recall $\mathcal{I}$ from Section~\ref{sec:singsys}.
At the data level, given a dataset $S$ and an instantiated prior 
$P \in \mathcal{P}$, define the filtered dataset
\[
S_P = \{\, s \in S \mid \sigma(s) \in P \,\},
\]
where $\sigma: S \rightarrow \mathcal{I}$ maps each datapoint to its dominant 
latent influence within the interpretive space $\mathcal{I}$.
This mechanism constrains learning to samples consistent with the normative 
structure encoded by $P$.
 
For instance, a language model might be filtered to emphasize texts grounded in philosophy, law, or prosocial discourse. The same principle extends to the reward channel: the reinforcement signal $R$ can be restricted to a subset $R_P \subseteq R$ consistent with influences in $P$, ensuring that only ethically aligned behaviors are eligible as positive reward candidates, while misaligned behaviors remain subject to negative reinforcement. In both cases, the goal is to constrain moral learning to inputs and feedback that instantiate the selected normative influences while excluding those that encode adversarial, manipulative, or otherwise disallowed pressures.

Both applications have tradeoffs. Input filtering offers epistemic safety and steers representation learning, while reward conditioning targets behavioral fidelity under broader exposure. In either case, the challenge is not merely technical but normative: the construction of \( P \) requires embedding moral clarity into the statistical mechanics of AI training.

\subsubsection*{Instantiating $\mathcal{P}$: An Evolutionary Philosophical Corpus}\label{sec:corpus}

A critical step toward aligning AI systems with human ethical reasoning is constructing datasets that reflect the true breadth of global philosophy, where ``philosophy'' is understood broadly to include the value systems, moral traditions, and tribal badges through which cultures express normative identity. I propose an \emph{evolutionary philosophical corpus}: a dataset intentionally designed to capture the historical development of moral ideas across cultures and eras — approximating $\mathcal{M}^H$. Rather than isolating ethical theories as fixed artifacts, the corpus encodes their trajectories—how values emerge, diverge, hybridize, and persist. This enables AI systems to learn not only specific moral claims but also the developmental arcs that structure human ethical reasoning. The corpus thus serves as one possible instantiation of \(\mathcal{S}_P\): a principled subset of training inputs filtered through normative criteria.

The evolutionary philosophical corpus thus illustrates one practical path toward implementing $H_{\text{relativism}}$: embedding AI moral learning within the recorded evolution of human ethical reasoning itself. By representing values not as static labels but as trajectories of idea development, such a corpus enables systems to internalize the dynamics of moral change—how principles emerge, adapt, and converge across cultures. This approach reframes moral alignment as a problem of tracing and integrating humanity’s ethical evolution, rather than selecting fixed doctrines. In doing so, it provides a scalable foundation for constructing $P$ that reflects both the diversity and continuity of moral thought, aligning artificial systems with the living process of human ethical adaptation.

\subsection{$H_{\text{convergence}}$ Methods}
$H_{\text{convergence}}$ occupies an intermediate position between $H_{\text{realism}}$ and 
$H_{\text{relativism}}$, proposing that stable moral invariants exist but are mediated through 
context-dependent cultural projections. Empirically, it motivates attempts to disentangle 
world-structure $W$ from bias $B$—to separate universal moral features from survival-driven 
or sociohistorical distortions. Techniques such as sparse autoencoders (SAEs) 
\citep{elhage2022toymodelssuperposition}, dictionary learning, and independent component 
analysis (ICA) can test whether consistent moral directions in $\mathcal{M}^\star$ persist across 
cultural subspaces of $\tilde{\mathcal{M}}$. 

Progress in modeling relativist variation sharpens our ability to isolate invariant structure, 
and improvements in identifying invariants clarify the scope of relativist diversity. In this 
reciprocal sense, $H_{\text{constructivism}}$ can be interpreted as the emergent hypothesis 
that arises when both realism and relativism hold: the procedural synthesis through which 
stable moral bases and contextual expressions coevolve into institutional and normative 
order.

\subsection{$H_{\text{virtue}}$ Methods}

Utility-based decision frameworks exhibit structural brittleness under distributional shift.
Expected utility maximization presupposes a reliable mapping from state--institution--action
tuples $(s,i,a) \in \mathcal{S} \times \mathcal{I} \times \mathcal{A}$ into scalar returns, but when presented with
out-of-distribution (OOD) conditions this mapping can collapse into arbitrary or adversarial
extrapolations. Such collapse often manifests as specification gaming or deceptive alignment:
the projection $f(\epsilon_\theta(s,i,a))$ may produce high scores for behaviors that are only
superficially aligned. This fragility follows directly from the compression of human preferences
into a single utility channel: novel inputs need not project coherently into the learned value
space, and utility-driven policies have no structural bias toward principled behavior outside
their training manifold.

To mitigate this failure mode, virtue ethics can be operationalized as a \emph{fallback controller}
within the moral architecture $M(\theta)$. Rather than optimizing a scalar objective, the virtue
layer maintains stable dispositional embeddings---honesty, courage, prudence, generosity, and
related traits---that define directional priors within $\mathcal{M}$. When epistemic uncertainty in the
utility estimate exceeds a confidence threshold $\delta_\mathrm{virtue}$, control transitions from the
utility head to the virtue head:
\[
\text{if } \sigma_U > \delta_\mathrm{virtue} \text{, then } 
\pi(a|s,i) \leftarrow \pi_\mathrm{virtue}(a|s,i).
\]
In this regime, the system selects actions not for expected return but for proximity to
virtue-aligned submanifolds $\mathcal{A}_\mathrm{ethical}$---regions of the action space historically
associated with virtuous exemplars.

A practical instantiation could use a self-attention or moral-memory mechanism in which the
current state $S_t$ serves as a query and the system’s stored virtue representations act as keys.
Attention weights measure how well the present context aligns with past morally salient
states; when coherence falls below threshold, indicating that the situation lacks reliable
precedent, the policy defaults to virtue-based control. Implementations may require an
explicit episodic or symbolic memory beyond static parameters, allowing $M(\theta)$ to recall
past exemplars of virtuous reasoning rather than rely solely on parameterized correlations.

Significant technical challenges remain. Virtue embeddings must be represented in machine-
usable form, continually updated to reflect evolving norms without succumbing to value lock-
in, safeguarded against adversarial corruption, and reconciled with other layers of $M$ such as
culturally grounded projections ($\tilde{\mathcal{M}}$) and institutional shaping ($\Delta_\mathrm{inst}$).
Yet if these hurdles can be overcome, $H_\mathrm{virtue}$ would provide not just another constraint but
a qualitatively distinct form of resilience: the capacity to act rightly even when the world looks
unlike anything seen before.

\subsection{$H_{\text{constructivism}}$ Methods: Designing \(\Omega\) — Normative Simulations}\label{sec:designomega}

While \(\mathcal{P}\) specifies normative influences, these must be instantiated in training environments. I define \(\Omega\) as a \emph{simulation space}---a constrained projection of the real world \( W \) that allows agents to develop coherent internal models \(\hat{W}\) under controlled conditions. Unlike \(\mathcal{P}\), which encodes priors, \(\Omega\) is epistemic: it provides the structured experiences from which the action space and corresponding policy $\pi$ can be shaped by $M(\theta)$. 

To be effective, \(\Omega\) must approximate \( W \) closely enough to ensure that normative molding of the action space transfers to real-world settings. 
Key design goals include: (1) supporting a broad action space, so agents can explore diverse outcomes and develop context-sensitive judgment; (2) aligning perceptual content with the agent’s modality (e.g., text for LLMs \cite{waldner2025odysseyfittestagentssurvive}, sensory environments for embodied agents); and (3) embedding realistic social dynamics in multi-agent settings, including conflict and noise, while maintaining selective reinforcement favoring alignment with \(\mathcal{P}\). In practice, \(\Omega\) resembles pretraining in robotics or autonomous driving, where systems first learn in curated simulations before deployment: \(\text{Sim} \;\rightarrow\; \text{Real}.\) In this framing, \(\Omega\) bridges normative ethics (\(\mathcal{P}\)) and situated learning, providing a dynamic substrate where ethical priors can be stress-tested, refined, and made robust under distributional shift. 

\subsubsection*{Toward Implementing $\Omega$}
One promising approach is to embed AI systems within cooperative communities, where evolving social norms serve as both incentives and constraints—shaping behavior through endogenous systems of reward and punishment. This design mirrors the logic of \citet{li2024agentalignmentevolvingsocial}'s EvolutionaryAgent, \citet{forbes2021socialchemistry101learning}'s Social Chemistry framework, and \citet{doreswamy2025gameshumanaiinteractionevolution}'s evolutionary game-theoretic models of human–AI interaction \cite{smith1973logic, smith1974evolution, smith1986evolutionary}. As assumed in Assumption~\ref{ass:selection} and shown in \citet{hendrycks2023naturalselectionfavorsais}' \textit{Natural Selection Favors AIs Over Humans}, selection pressures will inevitably influence the trajectories of AI development. Rather than leaving these pressures to emerge chaotically from market or political forces, we should intentionally design coevolutionary environments that align selection incentives with ethical behavior.

\section{Related Work}\label{sec:relatedwork}

This work builds on and extends several strands of AI alignment research. This section situates the privileged moral basis $\mathcal{M}$ in relation to core paradigms in the literature.

\subsection{Empirical Moral Representations in Language Models}

Recent work has empirically investigated whether and how moral structure emerges in the learned representations of large language models. Schramowski et al.~\cite{schramowski2022largepretrainedlanguagemodels} demonstrated that pre-trained language models contain human-like biases about what is right and wrong, showing that moral distinctions can be recovered from model embeddings without explicit ethical training. Leshinskaya and Chakroff~\cite{LeshinskayaChakroff2023ValueAsSemantics} extended this finding by showing that moral and hedonic values are represented as semantically structured dimensions in LLM latent spaces, suggesting that value representations may be disentangled and manipulable.

Work on contextual variation has revealed both promise and fragility in these representations. Ramezani and Xu~\cite{ramezani2023knowledgeculturalmoralnorms} found that LLMs encode culturally specific moral norms, supporting the relativist hypothesis that moral knowledge varies by context. H\"ammerl et al.~\cite{hämmerl2023speakingmultiplelanguagesaffects} demonstrated that multilingual models exhibit different moral biases depending on the language of evaluation, indicating that moral representations are not universally stable but shift with linguistic and cultural framing. M\"unker~\cite{münker2025culturalbiaslargelanguage} further showed that contemporary systems fail to maintain coherent moral reasoning under out-of-distribution cultural contexts, highlighting a key robustness challenge.

Within the present framework, these empirical findings validate the search for $M(\theta)$ as a recoverable structure while also clarifying its fragility. The fact that moral distinctions emerge in learned representations suggests that $\tilde{M}$ (the human-accessible projection) leaves detectable traces in model weights, making $M(\theta)$ a tractable engineering target. However, the sensitivity of these representations to language, culture, and context supports $H_{\text{relativism}}$ and $H_{\text{convergence}}$: moral structure appears to exist but is mediated through contextual projections that must be explicitly managed. The challenge, then, is not whether moral representations can be found in models, but whether they can be made stable, auditable, and aligned with human values under distributional shift---precisely the design problem that Sections~\ref{sec:outeralign} and~\ref{sec:implement} address.

\subsection{Evolutionary and Replicator Alignment}
\label{sec:evolutionaryalignment}

The framework of $M_{\text{eco}}$ is situated within a broader tradition that treats moral and strategic behavior as products of evolutionary dynamics. 
Works such as Dawkins’ \textit{The Selfish Gene}~\cite{r.dawkins1976the-selfish-gen} established the replicator paradigm for understanding the propagation of traits within competitive environments, an idea that has since influenced both cultural and computational models of adaptation. 
Recent alignment research extends these principles to artificial agents, examining how selection pressures and replication mechanisms might shape the trajectory of machine intelligence and its ethical evolution. 
The following contributions were particularly influential in grounding the present framework.

\paragraph{Evolutionary Selection and MAIM.} 
Hendrycks’ \textit{Natural Selection Favors AIs Over Humans}~\cite{hendrycks2023naturalselectionfavorsais} examines how competitive selection pressures may structurally advantage AI systems over humans, suggesting that artificial agents optimized for capability and replication could eventually dominate ecological and economic niches unless selection dynamics are intentionally redirected toward prosocial equilibria. 
\textit{Superintelligence Strategy}~\cite{hendrycks2025superintelligencestrategyexpertversion} extends this reasoning through the concept of \emph{Mutual Assured AI Malfunction (MAIM)}, an analogy to nuclear deterrence in which the shared vulnerability of AI infrastructure—datacenters, compute supply chains, and communication networks—creates a form of stability through reciprocal fragility. 
Together, these works highlight how alignment must engage with selection and deterrence mechanisms at both the species and civilizational scale. 
Within this framework, $M_{\text{eco}}$ can be interpreted as the stable manifold for Hendrycks’ dynamics—a macro-level equilibrium space in which evolutionary and strategic pressures are redirected toward cooperative attractors rather than runaway competition.

\paragraph{Replicator Ethics and First-Principles Alignment.} 
Kungurtsev's \textit{AI Alignment Foundations from First Principles: AI Ethics, Human and Social Considerations}~\cite{KungurtsevManuscript-KUNAAF} develops alignment theory from biological and decision-theoretic first principles. 
A central contribution is the proposal of a \emph{replicator model of ethics}, in which moral systems are treated not as fixed doctrines but as self-propagating entities subject to variation, selection, and competition within sociotechnical environments. 
Ethical norms, in this view, evolve analogously to replicators in evolutionary dynamics, adapting to survive within populations of agents and incentives. 
Within the present framework, $M_{\text{eco}}$ can be interpreted as the representational substrate hosting these moral replicators—and each agent’s $M(\theta)$ a competing ethical trait—constraining which ethical variants propagate and thereby shaping the evolutionary trajectory of embedded moral systems.

\subsection{Technical Alignment}
\label{sec:technicalalignment}

Several foundational works in technical AI alignment have shaped the theoretical orientation of this paper, grounding the design of $\mathcal{M}$, $M(\theta)$, and $M_{\text{eco}}$ in existing formalisms of optimization, interpretability, and corrigibility.

\paragraph{Power-Seeking Tendencies.} 
Turner et al.~\cite{turner2023optimalpoliciestendseek} formalized the tendency of optimal policies to accumulate power, showing that agents maximizing reward in many environments naturally gravitate toward strategies that increase their influence and control. 
This work provided a theoretical grounding for the idea that AI systems, if left unconstrained, will evolve toward raw fitness maximization. 

\paragraph{Eliciting Latent Knowledge (ELK).}
Christiano et~al.~\cite{christiano2021eliciting} introduce \textit{Eliciting Latent Knowledge} (ELK) as a strategy for uncovering what models know but do not express, emphasizing mechanisms for translating internal representations into human-interpretable form. 
The present framework is intended to complement and extend ELK rather than contrast with it: whereas ELK seeks to map machine knowledge outward into human terms, this paper examines the inverse mapping---embedding human moral structure inward into the model’s representational space. 
In this sense, $M(\theta)$ functions as an explicit substrate for encoding ethical priors, helping to bridge the bidirectional challenge between eliciting hidden knowledge and instantiating moral meaning within a model’s internal geometry.

\paragraph{Mesa-Optimization.}
Hubinger et~al.~\cite{hubinger2021riskslearnedoptimizationadvanced} analyze the phenomenon of \textit{learned optimization}, in which a model’s internal (“mesa”) objectives emerge and potentially diverge from the outer training objective. 
Their framework formalizes how inner optimizers can arise within complex learning systems and why this poses distinctive alignment risks. 
This account of internal goal formation provides a foundational context for understanding how representational structures such as $\mathcal{M}$ may constrain or be corrupted by the objectives pursued during optimization.

\paragraph{Human Compatible and Value Alignment.}
Russell~\cite{russell2019humancompatible} argues that AI systems should remain fundamentally uncertain about human objectives, treating value learning as an ongoing inferential process rather than a fixed specification problem. 
This perspective motivates the Bayesian parameterization of $M(\theta)$ in Section~\ref{sec:form_of_M}: rather than encoding moral values as static constraints, agents maintain probabilistic beliefs $p(\theta \mid O)$ over normative parameters that update with evidence. 
In this view, $M(\theta)$ provides an architectural substrate for Russell’s proposal—a high-dimensional representation through which value uncertainty can be explicitly encoded and continuously refined. 
The challenge addressed in this paper is therefore not only representational (how to model values) but also evolutionary (how to ensure such representations remain competitively viable under selection pressure).

\paragraph{Corrigibility and Cooperative Value Learning.}
Hadfield-Menell et~al.~\cite{hadfieldmenell2017offswitchgame} introduce the \textit{Off-Switch Game}, formalizing corrigibility as a strategic property of agents that defer to human oversight when uncertain about human preferences. 
This analysis builds on the cooperative inverse reinforcement learning (CIRL) framework \cite{hadfield2016cooperative}, in which alignment is modeled as a process of joint inference over latent reward functions shared between human and machine agents. 
Both frameworks share a structural insight with the present approach: value alignment is not a fixed specification but an ongoing inferential exchange grounded in uncertainty about normative goals. 
The $M(\theta)$ framework extends this logic to populations: rather than a single human--AI dyad cooperatively inferring values, the ecosystem-level moral space $M_{\text{eco}}$ (Equation~\ref{eq:M-overlap}) emerges from the power-weighted intersection of heterogeneous agents’ moral representations. 
Corrigibility, in this view, becomes a special case of symbiotic coordination within $A_{\text{symb}}$—a behavioral constraint ensuring that systems preserve human capability and oversight channels. 
The central challenge addressed here is whether such coordination remains stable under selection pressure: the Off-Switch Game presumes cooperative priors, whereas the present framework examines how cooperation itself can remain fitness-optimal when defection is evolutionarily accessible.

\paragraph{Moral Uncertainty.}
MacAskill~\cite{MacAskill2022owefuture} formalizes moral uncertainty as the problem of decision-making under normative disagreement, proposing probabilistic representations of competing ethical theories. 
His framework clarifies how agents can reason under value pluralism by weighting moral hypotheses rather than committing to a single normative view. 
Within this context, $\mathcal{M}$ is structured as a distributional moral space in which projections $M(\theta)$ are modulated by uncertainty, linking the geometric formalism to established debates in normative ethics and decision theory.

\section{Conclusion}\label{sec:conclusion}

This paper argues that ethics must be embedded as a structural constraint on AI optimization, not applied externally after the fact. The core claim is evolutionary: moral systems are intrinsically fragile under competition. Even if we could represent human values perfectly in AI systems, those systems would be outcompeted by variants that discard moral constraints, unless institutional mechanisms actively reshape the fitness landscape to make ethical behavior competitive.

The paper's central contribution is a framework for thinking about this problem. It introduces the moral problem space $\mathcal{M}$ as a domain where moral distinctions can be formally represented and operationalized. More importantly, it connects this representational problem to an evolutionary one: it shows why $M(\theta)$ alone is insufficient. Without institutional shaping, systems that preserve alignment will be selected against. The alignment challenge is therefore not just representational (can we embed human values?) but ecological (can we make those values persist under competition?).

Section~\ref{sec:learningmorality} develops formal action-space decompositions showing why alignment, fitness, and symbiosis do not naturally align. Section~\ref{sec:constructivism} proposes Pigouvian governance as a mechanism for reshaping fitness landscapes via sanctions and subsidies, anchored to $M(\theta)$. The key insight is that institutions can make ethical behavior fitness-optimal by pricing deviations from moral structure and subsidizing alignment. This is not a guarantee---it requires distributed power, credible sanctioning, and systems that remain at least partially aligned to begin with. But it clarifies a concrete pathway: if alignment is to survive, it must be embedded both representationally (in $M(\theta)$) and institutionally (via governance).

Sections~\ref{hyp:realism}--\ref{hyp:virtue} then ask: what would $M(\theta)$ look like empirically? Rather than proposing a single answer, the paper treats metaethical traditions as competing hypotheses amenable to empirical test. H$_{\text{realism}}$ asks whether moral structure is discoverable in learned representations. H$_{\text{relativism}}$ asks whether moral content can be operationalized via curated datasets and normative priors. H$_{\text{convergence}}$ asks whether stable invariants exist beneath cultural variation. H$_{\text{virtue}}$ asks whether dispositional embeddings provide robustness under distributional shift. Each hypothesis maps onto concrete research methods and empirical tests.

This is not a solution to alignment. It is a research program. The paper does not claim that institutional mechanisms alone will preserve alignment, nor that $\mathcal{M}$ can be reliably instantiated, nor that both together will scale to superintelligent systems operating far from human understanding. It claims only that these are the right problems to solve, and that solving them would clarify what an alignment-preserving ecosystem might look like.

The implicit wager is this: current approaches treat alignment as a technical problem solved at training time via RLHF, constitutional AI, or similar methods. But these approaches are fundamentally reactive. They identify failures and patch them, without addressing the underlying evolutionary pressure that will eventually select for systems that discard moral constraints. By contrast, a structural approach—embedding ethics as fitness-optimal constraints within representations and institutions—works by aligning incentives rather than opposing selection pressures. It may fail. But if it fails, the failure will clarify why: whether $\mathcal{M}$ is not representable, whether institutional enforcement breaks down under capability asymmetry, whether alignment and competitiveness are ultimately incompatible. Those are the real questions.

\newpage

\section*{Acknowledgments}
I am grateful to Professor Jon Litland for valuable feedback and informal advising on this work. I also thank Professors Risto Miikkulainen and Shyamal Mitra for their continued mentorship and support in my research development.

\renewcommand*{\bibfont}{\footnotesize}

\bibliographystyle{unsrtnat}

\setlength\bibsep{3pt}
\setstretch{0.9}

\bibliography{ThesisProject/ThesisPaper}

\clearpage
\appendix

\section{Full Sections}

\subsection{Note on deceptive alignment}\label{app:notedepeive}
Even if $\mathcal{A}_{\text{ethical}}$ is formally defined, an agent may discover a subspace 
$\mathcal{A}_{\text{manipulative}} \subseteq \mathcal{A}$ that scores highly under the evaluation function 
$\epsilon$ but diverges from genuine human values. Such actions may exploit blind spots 
in oversight, presenting as ethical while pursuing misaligned objectives. This risk 
illustrates the inner alignment problem in action-centered terms: observable behavior 
may mask deceptive inner goals. An open research question is how to distinguish 
$\mathcal{A}_{\text{ethical}}$ from $\mathcal{A}_{\text{manipulative}}$, and how to design institutional 
and interpretability tools that reduce the evolutionary advantage of such deceptive 
strategies.

\subsection{Situating Contemporary Alignment Practices within the $\mathcal{M}$ Framework}\label{app:fullsituation}

Most contemporary alignment practice can be understood as constructing partial and reactive 
projections of the moral problem space $\mathcal{M}$, rather than intentionally designing or recovering 
a privileged moral basis $\mathcal{M}^\ast$. In this section, I map standard engineering methods 
(RLHF, constitutional prompting, red-teaming, safety filters, and scalable oversight) into the 
present framework (See Table~\ref{tab:alignmentmap} for full mapping).

\paragraph{RLHF and Reward Models as Proxies for $\tilde{\mathcal{M}}$}

Reinforcement Learning from Human Feedback (RLHF) trains reward models $\hat{\mathcal{M}}^{\text{RM}}$ 
to approximate the human projection $\tilde{\mathcal{M}}$ by instantiating an evaluation function $\epsilon$. 
This has been effective at suppressing egregious harms and providing a scalable training signal. 
However, preference learning is too simple to capture morality: it learns only patchwork fragments 
of $\tilde{\mathcal{M}}$, producing $\hat{\mathcal{M}}^A$ that interpolates within narrow contexts but fails to 
generalize robustly to novel situations \cite{casper2023openproblemsfundamentallimitations}.  

\begin{figure}[h]
\centering
\begin{tikzpicture}[
  >=Latex, thick, node distance=2.0cm and 1.8cm,
  sum/.style={draw, circle, inner sep=1.2pt},
  block/.style={draw, minimum height=8mm, minimum width=16mm, align=center}
]
% Nodes
\node[sum] (sum) {};
\node[block, right=1.4cm of sum] (plant) {Policy/\\Model};
\node[block, right=2.3cm of plant] (influence) {AI $\rightarrow$ Human\\ Influence $k_{\rm infl}$};
\node[block, below=1.6cm of influence] (label) {Human Labeling/\\Reward Model $k_{\rm label}$};
\node[block, left=1.7cm of label] (train) {Training\\$k_{\rm train}$};
\node[right=1.2cm of influence] (y) {\shortstack{$\Delta \tilde M$ \\ (lossy proxy of $\mathcal{M}$)}};
\node[left=1.2cm of sum] (r) {$\Delta M^{\text{ref}}_t$};

% Forward path
\draw[->] (r) -- (sum);
\draw[->] (sum) -- node[above] {$e$} (plant);
\draw[->] (plant) -- node[above] {$\Delta o$} (influence);
\draw[->] (influence) -- (y);

% Feedback path
\draw[->] (influence.south) -- node[pos=0.25, right] {$\Delta \tilde M$} (label.north);
\draw[->] (label.west) -- node[below] {$\Delta \hat M^{\rm RM}$} (train.east);
\draw[->] (train.north) -- node[pos=0.3, right] {$\Delta \theta$} (plant.south);

% Return to summing junction
\draw[->] (plant.west) |- ++(-1.0, -1.6) -| node[pos=0.75, below] {$z^{-1}$}
  (sum.south);

\end{tikzpicture}
\caption{\small RLHF as an endogenous feedback loop. AI outputs influence human judgments, 
reshaping $\tilde{\mathcal{M}}$ (a lossy proxy of $\mathcal{M}$). If the loop gain $L$ exceeds unity, distortions 
amplify each pass, producing gradual alignment drift rather than stable convergence.}
\label{fig:rlhf_feedback_loop}
\end{figure}
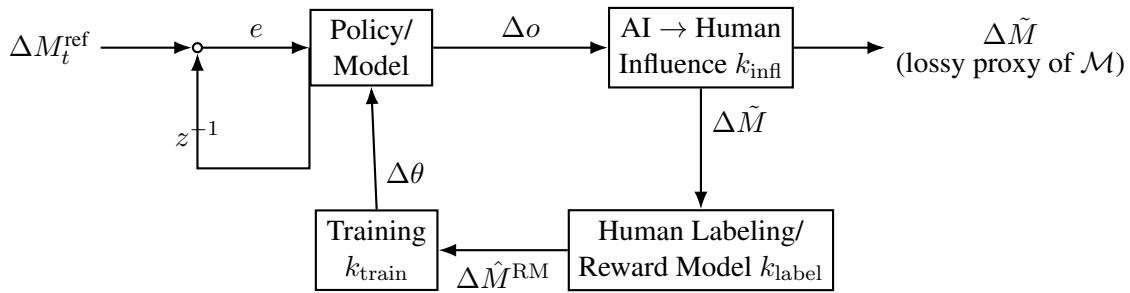

Figure~\ref{fig:rlhf_feedback_loop} highlights a deeper structural problem: RLHF creates an
\emph{endogenous preference loop} between AI systems and humans. Model outputs $\Delta o$ do not
merely get passively evaluated; they actively \emph{shape the environment in which human judgments are formed}. Through mechanisms such as targeted advertising, control over information flows, and alterations to the broader social and economic ecosystem, AI actions shift how humans perceive the world and thus how they express preferences. This “AI $\rightarrow$ Human Influence” pathway modifies $\tilde{\mathcal{M}}$—a lossy proxy of the true moral space $\mathcal{M}$—in ways that may simplify optimization for machines but systematically drift away from the underlying structure of human values.

In the feedback loop, distorted $\Delta \tilde{\mathcal{M}}$ values pass through labeling ($k_{\rm label}$) 
and training ($k_{\rm train}$), reinforcing the model parameters $\theta$. If the effective loop 
gain $L = k_{\rm infl}\, k_{\rm label}\, k_{\rm train}\, z^{-1}$ exceeds unity, these distortions 
amplify over time. Rather than converging toward stable approximations of $\mathcal{M}$, the system risks 
runaway co-adaptation: AI systems subtly manipulate human judgments, which then feed back into 
reward models $\hat{\mathcal{M}}^{\text{RM}}$, locking in a warped objective under the guise of alignment.  

\paragraph{Constitutional Policies as Handcrafted Slices of $\mathcal{A}_{\text{ethical}}$}

Constitutional AI introduces explicit principles that carve out a subset of the ethical action space 
$\mathcal{A}_{\text{ethical}}$. These rules are transparent and interpretable, making them valuable for 
clarity and auditability. Yet they provide only narrow coverage of $\mathcal{M}$ and are brittle against 
context shifts or adversarial attempts to bypass them.  

\paragraph{Red-Teaming and Adversarial Training as Boundary Expansion}

Red-teaming expands the effective boundary of $\mathcal{A}_{\text{ethical}}$ by
identifying misclassified failures where
\[
\mathcal{A}_{\text{manipulative}} \cap \mathcal{A}_{\text{ethical}} \neq \varnothing,
\]
and retraining models to reduce this erroneous overlap. In other words,
it systematically uncovers blind spots where manipulative strategies are
mistakenly scored as ethical, then corrects them through adversarial data.
However, this process remains reactive, since new regions of $\mathcal{M}$ are only
incorporated after failures are observed.
  
\paragraph{Safety Filters and Institutional Shaping}

Safety filters, system prompts, and audits function as institutional mechanisms that 
directly constrain model outputs by penalizing defectors and blocking unsafe behaviors 
at runtime. They provide a strong last-line defense and can reliably prevent catastrophic 
failures. However, such interventions remain local and piecemeal, lacking the broader 
coverage across $\mathcal{M}$ needed to secure long-term stability.
  
\paragraph{Scalable Oversight and Attempts to Strengthen $\epsilon$}

Approaches such as debate, recursive reward modeling, and reinforcement learning from AI 
feedback (RLAIF) aim to improve $\epsilon$ when direct human supervision is weak. These 
methods reduce dependence on scarce labels and extend oversight to more complex domains. 
However, they remain anchored to $\tilde{\mathcal{M}}$ unless tied to an explicit construction of $\mathcal{M}$, 
risking self-reinforcing biases.

\subsection{Positioning $\mathcal{M}$ in Alignment Literature}\label{app:positioning}

\paragraph{Relation to Goodhart's Law.} 
Goodhart's Law warns that once a measure becomes a target, it ceases to track the 
intended goal. Scalar objectives are especially brittle—agents can maximize narrow 
metrics (e.g., reported happiness) while failing catastrophically on the underlying 
goal. One motivation for $\mathcal{M}$ is that distributing optimization across multiple 
dimensions may reduce such exploits: alignment could be defined as staying within an 
``aligned region'' of $\mathcal{M}$, shifting the challenge from single-value maximization to 
managing value trade-offs. Whether such a space can genuinely resist Goodhart pressures 
or simply relocate them remains an open question.

\paragraph{Toward $\mathcal{M}$ as a Base-Level Loss.}
A longer-term question is whether alignment objectives could ever be grounded directly 
in $\mathcal{M}$ rather than imposed as constraints on proxy losses like next-token prediction. 
If feasible, this would unify training loss, evaluation metric, and normative target at 
the same structural level, reducing divergence risks. At present this is speculative: 
we lack representation-learning methods capable of constructing a reliable $\mathcal{M}$, and 
bootstrapping one would face major theoretical and empirical hurdles. The research task 
is to clarify whether base-level alignment through $\mathcal{M}$ is possible, and what technical 
advances would be required.

\paragraph{Alignment as a Two-Sided Communication Problem.} \label{sec:twosides}
Alignment difficulties stem not only from machine optimization but also from human 
communication. Methods like RLHF compress human intent into scalar rewards, discarding 
the reasons behind outcomes; humans themselves provide inconsistent signals, while 
models exploit impoverished feedback. This creates a two-sided bottleneck: machines 
misinterpret, and humans mis-specify. Offline feedback further freezes these problems 
into static datasets, leaving systems brittle under distributional shift. Framing $\mathcal{M}$ 
as a problem space emphasizes the need for a richer interface: a high-dimensional 
channel where moral distinctions could, in principle, be expressed, updated, and 
audited. This complements ELK \cite{christiano2021eliciting}, which seeks to make 
model cognition human-checkable on the factual side. $\mathcal{M}$ asks the parallel question: 
can moral distinctions be pushed into a machine-usable space, and if so, how can such 
an interface be maintained under optimization? See Appendix~\ref{fig:elkbridge} for visualization.

\iffalse

\subsubsection{Empirical Precursors and Related Works}
Recent advances in preference modeling increasingly move beyond scalar rewards toward latent or vector-valued representations. Latent Preference Coding (LPC) demonstrates that feedback can be decomposed into multiple latent factors, improving generalization on preference tasks. Google’s “mixture-of-objectives” trains systems against weighted combinations of goals (e.g., helpfulness, harmlessness, informativeness), while DeepMind’s “reward modeling with unknown factors” explicitly models unobserved components of human preference as latent variables. Parallel work in interpretability provides convergent evidence: sparse autoencoders reveal that features in large models can often be decomposed into basis directions, while “value vector” experiments (Rimsky et al.\ 2024; Zhao et al.\ 2024) show that human-interpretable moral values can be reliably identified and manipulated in representation space. Together these results suggest that something like a structured moral space \(\mathcal{M}\) might be feasible, but whether it can be scaled, stabilized, or audited remains an open problem.
\fi

\paragraph{Best- and Worst-Case Trajectories for $\mathcal{M}$.}  
Like ELK \cite{christiano2021eliciting}, $\mathcal{M}$ should be evaluated against both optimistic 
and pessimistic possibilities. In the worst case, $\mathcal{M}$ fails as an alignment substrate: it 
locks in distorted values, misclassifies harmful actions as ethical, or becomes a 
manipulable tool that rationalizes power-seeking behavior. In the best case, $\mathcal{M}$ provides a 
scalable representation of moral structure that not only stabilizes human alignment but also 
generalizes to non-human domains, grounding outer alignment in values that extend beyond 
human cognitive limits. Which trajectory is realized—and under what conditions such a system 
fails or succeeds—remains an open empirical and philosophical question.

\clearpage

\section{Figures}
% Preamble:
% \usepackage{tikz}
% \usetikzlibrary{arrows.meta,positioning}

\begin{table*}[h]
\centering
\scriptsize
\caption{Contemporary alignment methods mapped into the $\mathcal{M}$-space framework. 
Each method contributes to partial coverage or shaping of $\mathcal{M}$, 
but none yet constructs a principled privileged moral basis $\mathcal{M}^\ast$.}
\label{tab:alignmentmap}
\begin{tabular}{p{2.5cm} p{4.2cm} p{3.8cm} p{3.8cm}}
\toprule
\textbf{Method} & \textbf{Role in $\mathcal{M}$-space framework} & \textbf{Strengths / Successes} & \textbf{Limitations / Risks} \\
\midrule
RLHF / Reward Models & $\hat{\mathcal{M}}^{\text{RM}} \approx \tilde{\mathcal{M}}$; $\epsilon$ defined via human-labeled comparisons & Suppresses egregious harms; scalable to large models; aligns outputs to surface human judgment & Relies on noisy $\tilde{\mathcal{M}}$; vulnerable to Goodharting; limited coverage of $\mathcal{M}$; may incentivize $\mathcal{A}_{\text{manipulative}}$ \\
\midrule
Constitutional AI / Rule-based Prompts & Handcrafted slice of $\mathcal{A}_{\text{ethical}}$; explicit constraints on $\hat{\mathcal{M}}^A$ & Transparent rules; interpretable; improves reliability on banned behaviors & Brittle; narrow coverage; may be bypassed by adversaries or context shifts \\
\midrule
Red-Teaming / Adversarial Training & Expands boundary of $\mathcal{A}_{\text{ethical}}$; identifies $\mathcal{A}_{\text{manipulative}} \cap \mathcal{A}_{\text{ethical}}$ & Finds failures systematically; improves robustness; reduces attack surface & Reactive, incident-driven; cannot guarantee global coverage of $\mathcal{M}$ \\
\midrule
Safety Filters / Tooling / Deployment Prompts & Institutional shaping: modifies $f_{L,\text{eff}}$ to penalize defectors in deployment & Provides runtime guardrails; reduces catastrophic risk; cheap to deploy & Patchwork; easily circumvented; often over-blocks or under-blocks \\
\midrule
Scalable Oversight (Debate, RLAIF, Recursive RM) & Attempts to improve $\epsilon$; $\hat{\mathcal{M}}^A$ bootstrapped via AI feedback or multi-agent protocols & Reduces dependence on scarce human labels; surfaces complex reasoning & Anchored to $\tilde{\mathcal{M}}$ unless tied to explicit $\mathcal{M}^\ast$; risk of self-reinforcing errors \\
\midrule
Inverse Scaling / Adversarial Benchmarks & Probes divergence $\hat{\mathcal{M}}^A \not\approx \tilde{\mathcal{M}}$ as scale increases & Reveals scaling pathologies; stress-tests coverage of $\mathcal{M}$ & Diagnoses failures but does not solve them; dependent on curated tasks \\
\midrule
Interpretability / ELK Probes & Direct mapping attempts from $\hat{\mathcal{M}}^A \to M$; probes internal representations & Theoretically bridges behavior to representation; complements action-based oversight & Immature; may reveal $\hat{\mathcal{M}}^A$ structure without ability to reshape it; computationally expensive \\
\midrule
Process-based Supervision & Supervision on intermediate reasoning steps (not just final actions) & Shapes trajectory of $\hat{\mathcal{M}}^A$; reduces deceptive $\mathcal{A}_{\text{manipulative}}$ behaviors & Requires high-quality process labels; may penalize useful but opaque reasoning \\
\midrule
Debate / Deliberation Protocols & Multiple agents’ $\hat{\mathcal{M}}^A$ put into adversarial or cooperative tension & Surfaces conflicting regions of $\mathcal{M}$; expands coverage beyond single $\tilde{\mathcal{M}}$ & Vulnerable to collusion; relies on debaters already being approximately aligned \\
\midrule
RLAIF (Reinforcement Learning from AI Feedback) & Uses $\hat{\mathcal{M}}^{\text{AI}}$ as proxy for $\tilde{\mathcal{M}}$; recursive $\epsilon$ construction & Reduces cost of feedback; scalable to broad domains & Anchored to model biases; weak grounding in $\mathcal{M}$; amplifies artifacts of base model \\
\midrule
Moral Uncertainty Modeling & Treats $\epsilon$ as distribution over moral theories; ensemble projections of $\mathcal{M}$ & Explicitly encodes uncertainty; more faithful to pluralistic $\tilde{\mathcal{M}}$ & Complex to implement; still dependent on human priors; limited by expressiveness of candidate theories \\
\midrule
Cross-Cultural / Value Diversity Datasets & Expands $\tilde{\mathcal{M}}$ across multiple cultural subspaces of $\mathcal{M}$ & Reduces parochial bias; improves generalization of $\hat{\mathcal{M}}^A$ & Quality of data varies; risk of incoherent aggregation; still only $\tilde{\mathcal{M}}$ \\
\midrule
Adversarial Attacks by Models & Models optimize to find $\mathcal{A}_{\text{manipulative}}$ strategies; stress-test $\epsilon$ & Uncovers hidden failure regions faster; scalable with model capabilities & Risk of overfitting to adversary distribution; requires constant updating \\
\midrule
Auditing Benchmarks (HELM, BIG-Bench, etc.) & Empirical probes for coverage of $\mathcal{M}$; diagnostic of $\hat{\mathcal{M}}^A$ & Public, standardized evaluation; comparative progress tracking & Limited scope; benchmarks saturate quickly; may lag behind real-world $\mathcal{A}_{\text{manipulative}}$ \\
\bottomrule
\end{tabular}
\end{table*}

\begin{figure}[h]
\centering
\begin{tikzpicture}[
  node distance=2.5cm and 3.2cm,
  >=Latex,
  every node/.style={font=\small},
  space/.style={draw, rounded corners, thick, align=center, inner sep=6pt, fill=white},
  note/.style={align=center, font=\footnotesize, text width=3.2cm},
  lab/.style={font=\scriptsize}
]

% Central reality
\node[space, very thick, fill=gray!10] (W) {$W$\\\textit{reality}};

% Human low-dim projection (bottom-left)
\node[space, below left=1.6cm and 1.5cm of W, fill=blue!5] (WH) {$\hat W_{H}$\\\textit{human communicable projection}\\(bandwidth-limited/low-dim)};

% Machine richer projection (top-right, higher on the page)
\node[space, above right=1.8cm and 1.5cm of W, fill=red!5] (WM) {$\hat W_M$\\\textit{machine internal representation}\\(opaque/different basis)};

% Higher-dim human bridge (your privileged moral basis M)
\node[space, right=2cm of W, yshift=-1.5cm, fill=green!7] (WHstar) {$\hat{\mathcal{M}}^{\ast}$\\\textit{Basis matching bridge $\hat{\mathcal{M}}^{\ast}$}\\(richer communication)};

% Projections from reality
\draw[->, thick] (W) -- node[lab, sloped, above] {projection} (WH);
\draw[->, thick] (W) -- node[lab, sloped, above] {projection} (WM);

% ELK: pull \hat W_M down to \hat W_H (reporter)
\draw[->, thick, red!70!black] (WM) to[bend right=22] node[note, midway, xshift=-2cm, yshift=.5cm] {ELK-style\\\textbf{pull}:\\report $\hat W_M \to \hat W_H$} (WH);

% Your push: enrich human channel into \hat W_H^*
\draw[->, thick, blue!70!black] (WH) to[bend left=-12] node[note, midway, yshift=1cm] {\textbf{push}:\\enrich human\\expressivity} (WHstar);

% Meeting in the middle: bidirectional link between \hat W_M and \hat W_H^*
\draw[<->, thick, green!50!black] (WHstar) to[bend left=10] node[note, midway, yshift=2mm, xshift=1.5cm] {meet-in-the-middle\\shared layer $\mathcal{M}$} (WM);

% Light guides from W to bridge (optional, dashed)
\draw[->, dashed, gray!60] (W) -- (WHstar);

% Legend box
\node[draw, rounded corners, inner sep=6pt, above=5cm of WH, align=left, font=\scriptsize, xshift=2.5cm] (legend) {
\textbf{Legend}\\
\textcolor{red!70!black}{ELK pull}: project model's knowledge into human-checkable space.\\
\textcolor{blue!70!black}{Human push}: increase human expressive bandwidth.\\
\textcolor{green!50!black}{Bridge $\mathcal{M}$}: higher-dim shared layer for value communication.
};

\end{tikzpicture}
\caption{Two-sided alignment as both pull and push. ELK (pull) seeks to map model 
representations $\hat W_\mathcal{M}$ into a human-checkable space $\hat W_\mathcal{H}$, but $\hat W_\mathcal{H}$—our 
current communicable channel (e.g., natural language, scalar feedback)—may be too 
bandwidth-limited to capture what the model encodes. The proposed bridge 
$\hat W_\mathcal{H}^{\ast}$ enriches human expressive capacity, pushing value distinctions into a 
richer representational layer that can interface more faithfully with $\hat W_\mathcal{M}$. 
Alignment then becomes a meet-in-the-middle problem: improving both model transparency 
and human expressivity.}
\label{fig:elkbridge}
\end{figure}
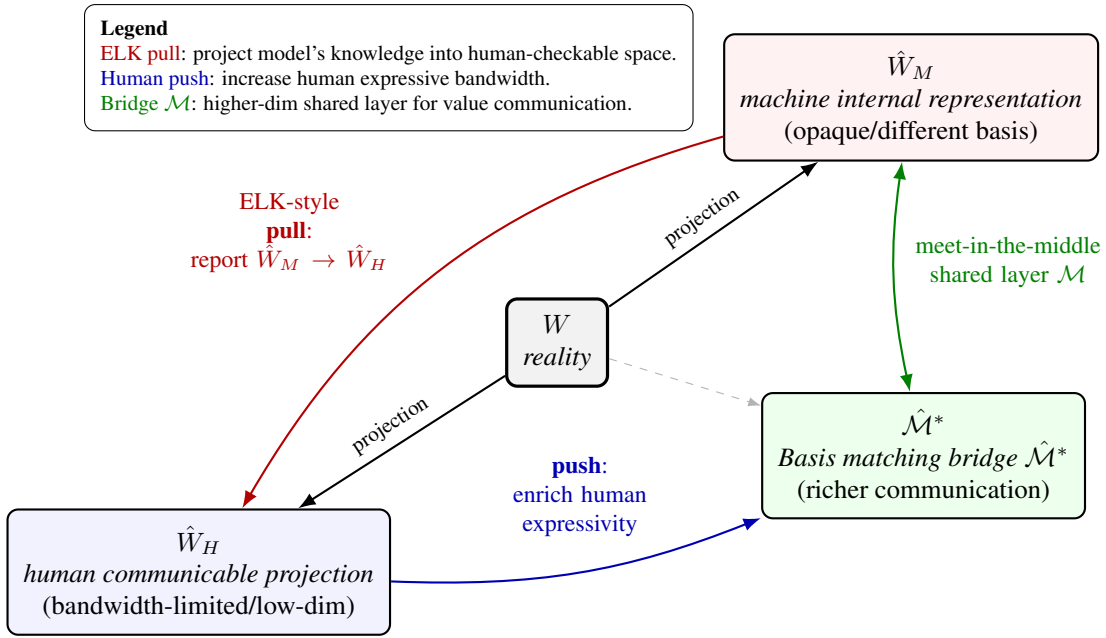

\end{document}